\documentclass[11pt]{article}
\pdfoutput=1
\usepackage{jcapmod}
\usepackage{booktabs}
\usepackage{mathtools}
\usepackage{microtype}
\usepackage{setspace} 
\usepackage[english]{babel}
\usepackage{amsmath,amssymb,amsbsy,amstext,amsthm,xcolor}
\usepackage{graphicx}
\usepackage{amsfonts}
\usepackage{verbatim}
\usepackage{float}
\usepackage[utf8]{inputenc}
\usepackage{caption}
\usepackage{subcaption}
\usepackage{enumitem}
\RequirePackage{color}

\usepackage{colortbl}
\definecolor{green2}{cmyk}{0, 1, 0.5, 0}
\definecolor{lightgreen}{cmyk}{0.2, 0, 0.2, 0.2}
\definecolor{lightgray}{cmyk}{0.1,0.2,0,0.1}
\definecolor{lightgray2}{cmyk}{0.4,0.4,0,0.8}
\definecolor{black}{cmyk}{1.0,1.0,1.0,1.0}

\allowdisplaybreaks[1]

\usepackage{colortbl}

\definecolor{lightgreen}{cmyk}{0.2, 0, 0.2, 0.2}
\definecolor{lightgray}{cmyk}{0.1,0.2,0,0.1}
\definecolor{lightgray2}{cmyk}{0.1,0.1,0,0.1}

\makeatletter
\newlength{\apb@width}
\newcommand{\autoparbox}[2][c]{\settowidth{\apb@width}{#2}\parbox[#1]{\apb@width}{#2}}

\makeatother

\numberwithin{equation}{section}

\def\beq{\begin{equation}}
\def\eeq{\end{equation}}

\def\bea{\begin{eqnarray}}
\def\eea{\end{eqnarray}}
\def\eg{{\it e.g.~}}

\def\ie{{\it i.e.~}}
\def\d{{\rm d}}

\def\d{{\rm d}}

\def\nn{\nonumber}

\def\sgm{\sigma}

\def\del{\partial}
\def\Mp{M_{\rm pl}}

\def\fr{\frac}

\def\0{{\boldsymbol 0}}

\def\fr{\frac}

\begin{document}

\begin{titlepage}

\setcounter{page}{1} \baselineskip=15.5pt \thispagestyle{empty}

\bigskip\

\vspace{1cm}
\begin{center}

{\fontsize{20}{28}\selectfont  \sffamily \bfseries {Synthetic Gravitational Waves from a Rolling Axion Monodromy}}

\end{center}

\vspace{0.2cm}

\begin{center}
{\fontsize{13}{30}\selectfont Ogan \"Ozsoy
$^{\clubsuit}$$^{\spadesuit}$
}
\end{center}

\begin{center}

\vskip 8pt
\textsl{
$\clubsuit$ Institute of Theoretical Physics, Faculty of Physics, University of Warsaw, ul. Pasteura 5, Warsaw, Poland,\\
$\spadesuit$ CEICO, Institute of Physics of the Czech Academy of Sciences, Na Slovance 1999/2, 182 21, Prague.}
\vskip 7pt

\end{center}

\vspace{1.2cm}
\hrule \vspace{0.3cm}
\noindent
In string theory inspired models of axion-like fields, sub-leading non-perturbative effects, if sufficiently large, can introduce steep cliffs and gentle plateaus onto the underlying scalar potential. During inflation, the motion of a spectator axion $\sigma$ on this potential becomes temporarily fast, leading to localized amplification of one helicity state of gauge fields. In this model, the tensor and scalar correlators sourced by the vector fields exhibit localized peak(s) in momentum space corresponding to the modes that exit the horizon while the roll of $\sigma$ is fast. Thanks to the gravitational coupling of gauge fields with the visible sector and the localized nature of particle production, this model can generate observable gravitational waves (GWs) at CMB scales while satisfying the current limits on scalar perturbations. The resulting GW signal breaks parity and exhibit sizeable non-Gaussianity that can be probed by future CMB B-mode missions. Depending on the initial conditions and model parameters, the roll of the spectator axion can also generate an observably large GW signature at interferometer scales while respecting the bounds on the scalar fluctuations from primordial black hole limits. In our analysis, we carefully investigate bounds on the model parameters that arise through back-reaction and perturbativity considerations to show that these limits are satisfied by the implementations of the model that generate GW signals at CMB and sub-CMB scales. \vskip 10pt
\hrule

\vspace{0.6cm}
 \end{titlepage}

 \tableofcontents
 
\newpage

\section{Introduction}

The observations on the Cosmic Microwave Background (CMB) radiation strongly suggests that the universe went through an early phase of accelerated expansion called inflation \cite{Guth:1980zm,Linde:1981mu,Linde:2005ht}. Apart from its success in addressing the puzzles of hot Big Bang cosmology, inflation provides us an explanation for the quantum mechanical origin of large scale cosmological fluctuations that are observed to be nearly Gaussian and adiabatic with a small red-tilt \cite{Ade:2013ydc,Ade:2015xua,Akrami:2019izv}.  Another robust prediction of inflation is the production of gravitational waves (GWs) which can be probed or constrained through the B-mode polarization of the CMB. This signal is conventionally parametrized by the ratio between the GW power spectrum and the scalar power spectrum-- denoted by $r$ -- which is currently restricted to $r < 0.063$ \cite{Array:2015xqh,Akrami:2018odb}.  This limit is expected to be improved by upcoming CMB polarization measurements such as PIXIE \cite{Kogut:2011xw}, LiteBIRD \cite{Hazumi:2019lys} and CMB-S4 \cite{Abazajian:2016yjj} which aim at the ambitious sensitivity goal of $\sigma(r)  \approx 10^{-3}$  where $\sgm(r)$ denotes uncertainty on $r$. 

In single field models of inflation, it is often considered that a detection of primordial B-modes of CMB fluctuations would provide us the energy scale of inflation. This direct relationship is typically expressed as
 \beq\label{EI}
H_{\rm inf} \simeq 2.5 \times 10^{-5} \left(\fr{r}{0.068}\right)^{1/2} \Mp,
\eeq
characterizing the dependence of vacuum fluctuations of the metric on the expansion rate $H_{\rm inf}$ during inflation or equivalently to the  inflationary energy scale $E_{\rm inf} = (3 H_{\rm inf}^2 \Mp^2)^{1/4}$. This relation alone makes the measurement B-modes an important scientific objective of current and upcoming CMB probes \cite{Kamionkowski:2015yta} and therefore, it is important to reconsider the validity and scope of \eqref{EI}. In principle, since GWs can be produced by any energetically viable contribution to the energy momentum tensor, it is possible to invalidate this result by simply considering additional matter fields sources of GWs that exhibit different parametric dependence on $H_{\rm inf}$. For example, this can be achieved by additional field configurations that are not in their vacuum state \cite{Cook:2011hg,Senatore:2011sp}.

However, introducing additional sources of GWs come with a price as the sector that sources GWs also interacts with the scalar perturbations at least gravitationally\footnote{See \eg \cite{Garcia:2019icv} for a detailed study on stochastic particle production in a spectator scalar sector and \cite{Garcia:2020mwi} for interesting features this production may impart on the correlators of curvature perturbation at cosmological scales.}  or stronger in the case where the sources are directly coupled to the sector responsible for the generation of density perturbations. This situation in general results with a decrease in the observed value of $r$ otherwise leads to large non-gaussian statistics for the scalar fluctuations, particularly if we insist on a large component of tensor perturbations produced by secondary sources \cite{Barnaby:2010vf,Barnaby:2012xt,Mirbabayi:2014jqa}.

An efficient mechanism\footnote{Other scenarios that can generate observable GWs during inflation include the amplification of chiral tensor modes through non-abelian gauge fields \cite{Dimastrogiovanni:2012ew,Adshead:2013qp,Namba:2013kia,Obata:2014loa,Obata:2016tmo,Maleknejad:2016qjz,Adshead:2017hnc}, by spectator fields with reduced sound speed \cite{Biagetti:2013kwa,Biagetti:2014asa,Fujita:2014oba}, modification of tensor dispersion relation \cite{Cannone:2014uqa,Cannone:2015rra}, varying sound speed of tensor fluctuations \cite{Cai:2015dta,Cai:2016ldn}, breaking of space diffeomorphisms \cite{Bartolo:2015qvr} and transient non-attractor phase(s) during inflation \cite{Mylova:2018yap,Ozsoy:2019slf}. Another mechanism that can lead observable GWs has been studied in \cite{Choi:2015wva,Fujita:2018zbr,Kawasaki:2019hvt} where a rolling dilaton field coupled to gauge fields through $f(\sgm) F^2$ \cite{Ratra:1991bn} is considered.} that can generate observable GWs from secondary sources utilizes the motion of a rolling scalar field X (an inflaton or a spectator scalar) as a dynamo to amplify abelian gauge fields, which in turn act as a source for GWs. A natural candidate for the sector X is thus an axion-like field because i) due to their approximate shift symmetry \cite{Freese:1990rb} axions are light and thus can roll a significant amount of time during inflation ii) as a result of the shift symmetry, they are expected to interact with gauge fields through a dimension five operator\footnote{Shift symmetric scalars can also couple to fermions through dimension five operators. See \cite{Adshead:2015kza,Adshead:2018oaa,Domcke:2018eki,Adshead:2019aac} for the phenomenological consequences of such coupling during inflation.}: 
\beq\label{Lint}
\mathcal{L}_{\rm int} = \fr{\alpha_c}{4f} X F\tilde{F},
\eeq
where $F$ is field-strength tensor, $\tilde{F}$ is its dual and $\alpha_{\rm c} / f$ controls the size of the coupling with X, $f$ being the axion decay constant. The coupling \eqref{Lint} of vector fields with $X$ leads to an exponential enhancement in gauge field modes, giving rise to an inflationary dynamics with a rich set of phenomenological consequences including, inflation on a steep potential \cite{Anber:2009ua}, magneto-genesis during inflation \cite{Prokopec:2001nc,Anber:2006xt,Caprini:2014mja,Fujita:2015iga,Adshead:2016iae}, large scalar \cite{Barnaby:2010vf,Barnaby:2011vw}, tensor \cite{Cook:2013xea,Agrawal:2018mrg} and mixed \cite{Fujita:2018vmv,Dimastrogiovanni:2018xnn} non-Gaussianity, parity violation in the CMB \cite{Sorbo:2011rz,Shiraishi:2013kxa,Adshead:2013qp}, at interferometers \cite{Crowder:2012ik} and production of primordial black holes \cite{Linde:2012bt,Bugaev:2013fya,Garcia-Bellido:2016dkw}.

In the presence of the coupling in \eqref{Lint}, the influence of gauge field sources on the scalar sector can be minimized by identifying the sector $X$ as a hidden scalar sector $X=\sgm$ that only interacts gravitationally with inflaton \cite{Barnaby:2012xt,Mukohyama:2014gba,Ozsoy:2014sba}. However, even in this case, the roll of the spectator ($\dot{\sgm}\neq 0$) allows for a mass mixing between $\phi$ and $\sgm$ which results with a channel that can feed into the correlators of curvature perturbation through the conversion of $\delta \sgm$ to the inflation fluctuations $\delta \phi$: $\delta A + \delta A \to \delta \sgm \to \delta \phi \propto \mathcal{R}$ \cite{Ferreira:2014zia}. The amplitude of $\delta \phi$ fluctuations sourced through this channel is proportional to number of e-folds during which $\sgm$ is rolling. As a result, in order to avoid excess power in the scalar correlators, the spectator $\sgm$ should roll no more than several e-folds in order to simultaneously grant for observable tensors at the level of $r \lesssim 10^{-3}$ and scalar fluctuations consistent with CMB observations  \cite{Ozsoy:2017blg}.

In \cite{Namba:2015gja}, a model of a spectator axion-like field that can roll transiently over its standard cosine potential, $V_\sgm (\sgm) \propto \Lambda^4 \left( 1 -\cos(\sigma/f)  \right)$ is considered. In this model, the shape of the potential allows for a very small velocity $\dot{\sgm}$ at early and late times, \ie when $\sgm$ is close to maximum ($\sgm = \pi f$) and minimum ($\sgm=0$) of $V_\sgm (\sgm)$, and a relatively fast motion in between where $\dot{\sgm}$ increases. This transient motion in turn generates a scale dependent enhancement of scalar fluctuations $\delta \phi \propto \mathcal{R}$ through the gauge fields where only modes that leave the horizon when $\dot{\sgm} \neq 0$ are excited, allowing us to keep the production of sourced scalar fluctuations under control with respect of sourced GW production at various cosmological scales \cite{Namba:2015gja}.

In this work, we propose an alternative mechanism that is capable of producing scale dependent, observable GWs at CMB and sub-CMB scales while keeping scalar fluctuations at observationally viable levels. In particular, we consider a string-inspired model where the spectator scalar $\sgm$ is identified with a non-compact axion field, \eg axion monodromy \cite{Silverstein:2008sg,McAllister:2008hb,Flauger:2014ana}. In this framework, discrete shift symmetry of the axion is broken by a monomial term in its potential,
\beq\label{Vs}
V_\sgm(\sigma) = \mu^3 \sgm+ \Lambda^4 \left[ 1 - \cos\left(\fr{\sigma}{f}\right) \right],
\eeq
which features characteristic axion oscillations with a period $f^{-1}$, superimposed on the monomial term. For sub-leading but sizable modulations $\Lambda^4 \lesssim \mu^3 f$ (which we refer to bumpy regime in what follows), the second term in \eqref{Vs} introduces plateau-like regions in the potential connected by steep cliffs (See \eg Figure \ref{fig:pot}). In each step like region, the roll of $\sgm$ give rise to very small  field velocity in the plateaus, whereas $\dot{\sgm}$ transiently peaks when $\sgm$ rolls over the cliff(s) connecting the plateaus. The amount of e-folds where $\dot{\sgm}$ is significant is  given by $\Delta N \sim \mathcal{O}(H^2 f / \mu^3)$ where $\mu^3/f$ is roughly the mass square $m^2_{\rm axion}$ of the $\sgm$ in its global minimum\footnote{$\sgm$ should settle to its global minimum (see Section \ref{SS}) long before the end of inflation to remove the effect of spectator fluctuations on the curvature perturbation. The roll of $\sgm$ to its global minimum ($\sgm = 0$) can be captured by replacing $\mu^3 \sgm \to \mu^3 f ([1+\left({\sgm}/{f}\right)^2]^{1/2}-1)$ in \eqref{Vs} such that the potential interpolates between $\mu^3\sgm$ and $(\mu^3/f)\sgm^2$ for large and small field values respectively.}. Therefore, similar to the pure periodic potential (where $\mu \to 0$) studied in \cite{Namba:2015gja}, each step-like feature of the bumpy potential can give rise to a transient, relatively fast-roll evolution for the spectator $\sgm$. In particular, we will show that in the presence of the coupling \eqref{Lint}, such a motion can generate suitable conditions for the production observable GWs while keeping the level of scalar fluctuations at acceptable levels imposed by the observations at CMB and sub-CMB scales. In this context, the mechanism we consider in this work constitutes one of the few existing examples in the literature that is capable of producing GWs of non-vacuum origin from Abelian gauge field sources, across a wide range of cosmological scales.

The non-compact nature of the spectator axion model we study here can provide a rich phenomenology at both CMB and/or sub-CMB scales for a broad range of initial conditions: in contrast to pure periodic potential studied in \cite{Namba:2015gja}, $\sigma$ can probe multiple wiggles in its scalar potential, generating multiple sourced signals through its coupling \eqref{Lint} to vector fields\footnote{An interesting possibility is production of observable signals at both CMB and sub-CMB scales either by the presence of enhanced tensor (GWs) and/or scalar fluctuations (e.g primordial black holes).}. Another appealing aspect of this framework is the insensitivity of phenomenological implications to the choice of initial conditions of $\sgm$ (\ie to the initial $\sgm$ and $\dot{\sgm}$): the presence of flat plateaus in the bumpy regime and Hubble drag induced by the inflaton ensures that the spectator axion will quickly settle to a quasi slow-roll regime\footnote{See \eg \cite{Parameswaran:2016qqq, Ozsoy:2018flq} in the context of canonical single field inflation.} suitable for the efficient particle production in the gauge field sector and its subsequent sourcing of scalar and tensor fluctuations\footnote{For a pure periodic potential studied in \cite{Namba:2015gja}, background dynamics of $\sgm$ is limited to a field range of $\Delta \sgm = \pi f$ in which all the phenomenological results of the spectator axion-gauge field model are obtained. In this case, initial field value should be chosen with care, \ie close to the maximum of $V_\sgm (\sgm) \propto \Lambda^4 \left( 1 -\cos(\sigma/f)  \right)$ so as to allow for small enough $\dot{\phi} \propto V'_\sgm$ that can lead to sufficient e-fold of evolution in the quasi slow-roll regime, required to generate sufficiently large sourced scalar/tensor fluctuations by gauge fields. }.  

This work is organized as follows. In Section \ref{Sec2}, we describe the multi-field model we are considering and its background evolution together with the resulting gauge field production. In Section \ref{Sec3}, we review the dynamics of scalar and tensor fluctuations in the presence of gauge field sources.  In Section \ref{Sec4}, we present our results on sourced cosmological correlators and discuss their phenomenology at CMB and sub-CMB scales. In Section \ref{Sec5} contains our conclusions. We supplement our results with four appendices. In Appendix \ref{AppA}, we present the details on the background evolution of spectator $\sgm$ and compute the resulting gauge field mode functions in the WKB approximation. In Appendix \ref{AppB} and \ref{AppC}, we provide details on the computation of tensor and scalar correlators, respectively. In Appendix \ref{AppE}, we study limits on the parameter space of the model including back-reaction of the produced gauge quanta on the background dynamics and perturbativity of scalar/gauge field fluctuations to show that in the applications of the model that generates observable effects at CMB and sub-CMB scales these limits are satisfied. 

\section{The model}\label{Sec2}
We consider a model described by the following matter Lagrangian \cite{Barnaby:2012xt},
\beq\label{Lm}
\fr{\mathcal{L}}{\sqrt{-g}} = \fr{\Mp^2 R}{2} -\underbrace{\fr{1}{2}(\del\phi)^2 - V_{\phi}(\phi)}_\text{Inflaton Sector}-\underbrace{\fr{1}{2}(\del\sigma)^2 - V_{\sgm}(\sgm)-\fr{1}{4}F_{\mu\nu}F^{\mu\nu}-\fr{\alpha_{\rm c}\sgm}{4f}F_{\mu\nu}\tilde{F}^{\mu\nu},}_\text{Hidden Sector}
\eeq
where $R$ is the Ricci curvature, $\phi$ is the inflaton and the hidden sector includes the scalar $\sigma$, the gauge field $A_\mu$ and their interaction through the Chern-Simons term with its strength parametrized by the axion decay constant $f$ and the dimensionless number $\alpha_c$. In \eqref{Lm}, $V_\phi(\phi)$ and $V_\sgm(\sgm)$ are the potential of the inflaton and $\sgm$, whereas the gauge field strength tensor and its dual are defined by $F_{\mu\nu} = \del_\mu A_\nu - \del_\nu A_\mu$ and $\tilde{F}^{\mu\nu}\equiv \eta^{\mu\nu\rho\sigma} F_{\rho\sigma}/(2\sqrt{-g})$ where alternating symbol $\eta^{\mu\nu\rho\sigma}$ is $1$ for even permutation of its indices, $-1$ for odd permutations, and zero otherwise.
\subsection{Background evolution}
As indicated by the Lagrangian in \eqref{Lm}, we consider a setup two sectors only interact gravitationally and where the background energy density is dominated by the inflaton sector $\phi$ and the axion $\sgm$ is a spectator: \ie $\rho_\sgm \ll \rho_\phi$ where $\rho_X = \dot{X}^2/2+ V_X$ with $X =\{\phi,\sgm\}$. During inflation, assuming negligible back-reaction (See \eg Appendix \ref{AppE}) from gauge fields, this implies that 
\beq
3H^2\Mp^2 = \rho_\phi + \rho_\sgm ~~~\longrightarrow ~~~3H^2 \Mp^2 \simeq V_\phi(\phi). 
\eeq
Moreover, we will assume that the inflaton's potential $V_\phi(\phi)$ is very flat, such that we can treat Hubble rate as constant, \ie during the scales where the signal is generated through rolling $\sgm$.
\subsubsection{Bumpy regime for the spectator axion}
In the hidden axion sector $\sgm$, we consider a scenario that is based on an earlier observation of how sub-leading, non-perturbative effects can alter the dynamics of axions \cite{Parameswaran:2016qqq, Ozsoy:2018flq}. In low energy effective descriptions of string theory, the perturbative
axion shift symmetry is broken spontaneously by background vevs (e.g. fluxes) or non-perturbative effects (e.g. string instantons), leading to large field inflation models with
monomial \cite{McAllister:2008hb, Flauger:2014ana} or cosine (“natural inflation”) potentials \cite{Freese:1990rb}. As noted earlier in \cite{Parameswaran:2016qqq, Ozsoy:2018flq}, the sub-leading non-perturbative corrections – if sufficiently large – can superimpose oscillations onto the underlying potential. The size of these effects will depend on the vev’s of fluxes and
other moduli, which are already stabilized. Therefore, they may be small, large enough to
introduce new local minima and maxima into the potential, or anything in between. For concreteness, for the spectator scalar sector, we consider a model of axion monodromy with the potential\footnote{Potentials that shares similar features that we consider in this work can be found in \cite{Tatsuo,BLZ,Kallosh:2014vja}. For an investigation on primordial black hole and GW production from axion inflation that exhibit similar bumps in its scalar potential, see also \cite{Cheng:2016qzb,Cheng:2018yyr}.}  given in \eqref{Vs}.

The background dynamics of the spectator axion depends on the size of the non-perturbative corrections compared to the monomial term proportional to $\mu^3$ in the potential \eqref{Vs}, in particular on the ratio $\beta = \Lambda^4 / (\mu^3 f)$. In the limit $\beta \to 0$, non-perturbative corrections become negligible and we recover the usual smooth linear potential $V_\sigma \propto \sigma$.  For $\beta > 1$ however, one may introduce a large number\footnote{In fact, the number of extremum is approximately proportional to the value of $\beta$ for $\beta > 1$. For an interesting study of this case see \cite{Berges:2019dgr} in the context of axion-like scalar dark matter and \cite{Ballesteros:2019hus} in the context of primordial black hole dark matter from single field inflation.} of new stationary points (where $V'_{\sigma} =0$) into the smooth potential for a given range of field values. In this case, the classically rolling scalar field might eventually stuck in one of the minima depending on the initial conditions \cite{Banks:2003sx}. In this work, we would like to focus on the regime where non-perturbative effects in the scalar potential $V_\sigma $ are sizeable but subdominant, $\beta< 1$, without assuming $\beta \ll 1$. 
\begin{figure}[t!]
\begin{center}
\includegraphics[scale=0.87]{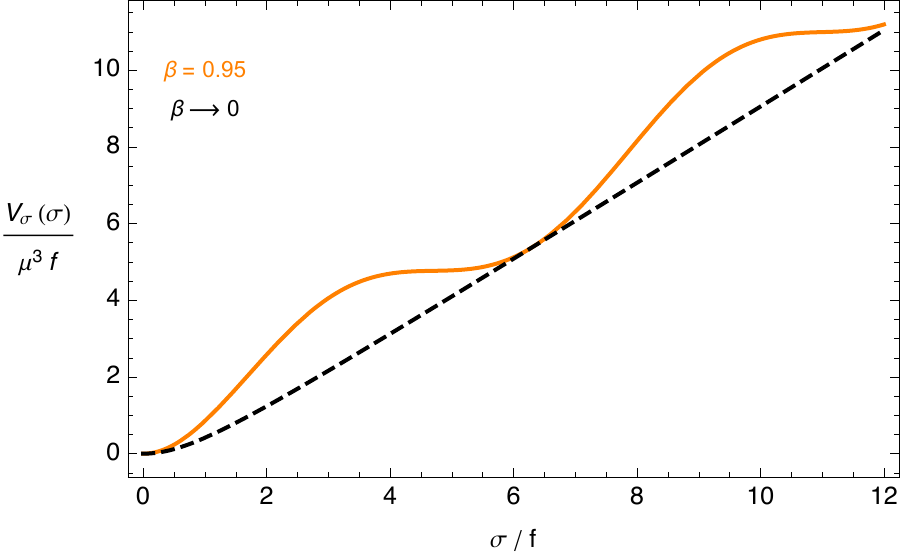}\includegraphics[scale=0.9]{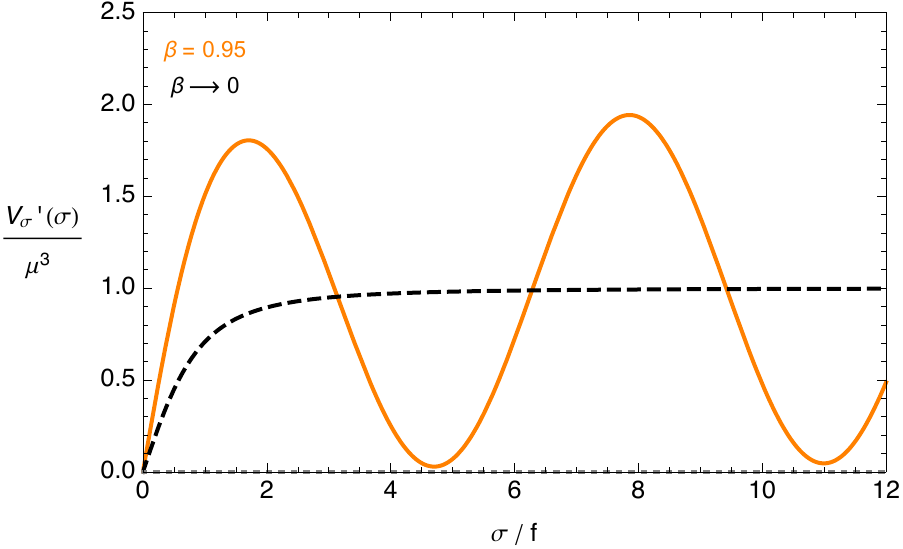}
\end{center}
\caption{The potential $V_\sigma$ and its slope $V'_\sgm$ for $\beta =0.95$ (Orange) and $\beta =0$ (black-dashed).\label{fig:pot}}
\end{figure}

In the bumpy regime ($\beta= 0.95 $), we illustrate the shape of the potential $V_\sgm(\sgm)$ and its slope in Figure \ref{fig:pot}. As we advertised before, we see that the potential exhibit plateau like regions followed by steep cliffs parametrized by large slopes $V'_\sgm/\mu^3 > 1$. On this potential, an initially displaced $\sgm$ rolls down in its wiggly potential, passing through the steep cliffs followed by flat plateaus to eventually settle on its global minimum at $\sgm =0$. With an aim to understand the gauge field production and its subsequent sourcing of GWs, it is enough to consider the evolution of $\sgm$ within a single bump that connects two plateaus with a cliff-like region in between. 

{\bf Background evolution of $\sgm$:}  In the slow-roll approximation $\ddot{\sgm}\ll 3H\dot{\sgm}$, the structure of the potential admits a simple analytical solution (see Appendix \ref{AppA}) for the field profile $\sgm$ within each bump --including two plateau regions separated by a cliff--  of the potential shown in Figure \ref{fig:pot}:
\beq\label{dsgm}
\fr{\dot{\sgm}}{2Hf} = -\fr{\delta}{1+ \ln\left[(\tau/\tau_*)^{\delta}\right]^2},
\eeq 
where we define the dimensionless parameter  $\delta \equiv (1+\beta)(\mu^3/6H^2f)$ with a constant Hubble rate $H$.  To ensure the validity of slow-roll solution \eqref{dsgm}, we require $\delta < 1$ (see Appendix \ref{AppA}). In \eqref{dsgm}, $\tau_*$ denotes the conformal time when $\dot{\sgm}$ in \eqref{dsgm} reaches its peak value, \ie when $\sgm$ rolls over the cliff regions in its potential. In the presence of the last term in the Lagrangian \eqref{Lm}, the roll of $\sgm$ provides a time dependent mass for the gauge field and amplifies its vacuum fluctuations. During inflation, this phenomenon is controlled by the dimensionless effective coupling $\xi = - {\alpha_c\dot{\sgm}}/{(2Hf)}$
which must be larger than unity in order to give rise to efficient particle production in the gauge field sector. Using \eqref{dsgm}, within each bump of the potential, $\xi$  \footnote{Note the minus sign difference in the definition of $\xi$ compared to the literature \cite{Anber:2009ua,Sorbo:2011rz}. However, this is just a matter of conventions. In this work, we work in a model where $\dot{\sgm} < 0$ and so $\xi > 0$.} can be re-written as
\beq\label{Joe}
 \xi (\tau)\equiv - \fr{\alpha_{\rm c}\dot{\sgm}}{2Hf} = \fr{\xi_*}{1+\ln\left[(\tau/\tau_* )^{\delta}\right]^2},
 \eeq
 where we defined $\xi_* = \alpha_{\rm c}\delta$ as the maximal value obtained by $\xi$ at $\tau = \tau_*$.
\subsection{Gauge field production from rolling $\sgm$}\label{gfp}
To study gauge field production, we focus on Coulomb gauge ($A_0 =0$) and decompose 
the gauge field $A_i$ in terms of the annihilation and creation operators as,
\beq\label{DGF}
\hat{A}_i(\tau, \vec{x}) = \int \fr{\d^3 k}{(2\pi)^{3/2}} ~ e^{i\vec{k}.\vec{x}}  \sum_{\lambda = \pm} \epsilon^{\lambda}_i(\vec{k}) \left[
A_\lambda(\tau,\vec{k})\hat{a}_\lambda(\vec{k}) + A^{*}_\lambda(\tau,-\vec{k})\hat{a}^{\dagger}_\lambda(-\vec{k})  \right],   
\eeq
where the helicity vectors obey $k_i \epsilon^{\pm}_i = 0$, $\epsilon_{ijk}~ k_j ~\epsilon^{\pm}_k = \mp i k \epsilon^{\pm}_i$, $\epsilon^{\pm}_i\epsilon^{\pm}_i = 0$, $\epsilon^{\pm}_i \epsilon^{\mp}_i =1$ and $(\epsilon^{\lambda}_i(\vec{k}))^{*} = \epsilon^{\lambda}_i(-\vec{k}) = \epsilon^{-\lambda}_i(\vec{k})$ and the annihilation/creation operators satisfy $\left[\hat{a}_\lambda(\vec{k}),\hat{a}^\dagger_{\lambda'}(\vec{k}')\right] = \delta_{\lambda\lambda'} \,\, \delta(\vec{k}-\vec{k}')$. 
Plugging the decomposition in \eqref{DGF} into the Lagrangian \eqref{Lm}, the mode functions $A_\lambda$ can be shown to obey 
\beq\label{MEA}
A_{\pm}''(x) + \left(1\pm \fr{2\xi}{x}\right)A_\pm(x) =0,
\eeq
where we defined $-k\tau = x$. We assume $\dot{\sgm} < 0$ or $\xi >0$, implying that only negative helicity modes $A_-$ will exhibit tachyonic instability in eq.\eqref{MEA} for modes satisfying $-k\tau < 2\xi$. For this reason, we will only consider $A_-$. For constant $\xi$, equation \eqref{MEA} can be solved exactly which is a case studied extensively in the literature \cite{Anber:2009ua}. In this work, we will focus on the case where $\xi$ evolves significantly as the spectator pseudo-scalar rolls through the cliffs before reaching on to the plateau regions in its scalar potential.
To understand the behavior of gauge field modes in this regime, we need to solve the following equation:
\beq\label{MEA2}
\fr{\d^2 A_{-}}{\d x^2} +\left(1 - \fr{2}{x}~ \fr{\xi_{*}}{1+\ln\left[(x_*/x )^{\delta}\right]^2} \right)A_{-} = 0,
\eeq
where we used \eqref{Joe} in \eqref{MEA}. For a general $\delta, \xi_*$ and $x_*$, it is not possible to find a closed form solution for eq. \eqref{MEA2}. However, we found that the growing mode of $A_{-}(\tau,k)$ can be captured very well by the following expressions at late times (See Appendix \ref{AppA}):
 \beq\label{Am}
A_{-}(\tau, k) \simeq\fr{1}{\sqrt{2k}}\left[\frac{-k\tau}{2 \xi(\tau)}\right]^{1 / 4} \tilde{A}(\tau, k), \quad A_{-}^{\prime}(\tau, k) \simeq \sqrt{\fr{k}{2}}\left[\frac{2 \xi(\tau)}{-k \tau}\right]^{1 / 4} \tilde{A}(\tau, k),
\eeq
where
\beq\label{tA}
 \tilde{A}(\tau, k) = N(\xi_*,x_*,\delta) ~\exp \left[- \fr{2\sqrt{2\xi_*}~ (-k\tau)^{1/2} }{\delta |\ln(\tau/\tau_*)| }\right], ~~~~~~ \tau/\tau_* < 1
\eeq
and we defined $x_* =- k\tau_*$ with $\tau_* = - (a_* H)^{-1}$ denoting the time at which $\xi$ reaches its peak value $\xi_*$ while $\sgm$ rolls through the cliffs. In \eqref{tA}, the time independent normalization factor $N(\xi_*,x_*,\delta)$ characterizes the dependence of the mode function amplitude on the background model parameters $\xi, x_*, \delta$. In this work, we will determine $N(\xi_*,x_*,\delta)$ by matching $A_-$ in \eqref{Am} to the full numerical solution of \eqref{MEA2} at late times, $-k\tau \ll 1$. We choose the arbitrary initial phase factor of $A_-$ to ensure that $N(\xi_*,x_*,\delta)$ is real and positive. As a result, the decomposition for gauge field in \eqref{DGF} becomes 
\beq\label{DGFs}
\hat{A}_i(\tau, \vec{x}) \simeq \int \fr{\d^3 k}{(2\pi)^{3/2}} ~ e^{i\vec{k}.\vec{x}}  \epsilon^{-}_i(\vec{k}) A_{-}(\tau,\vec{k})\left[
\hat{a}_{-}(\vec{k}) + \hat{a}^{\dagger}_{-}(-\vec{k})  \right], 
\eeq
where $A_{-}(\tau,k)$ is given by \eqref{Am} and \eqref{tA}. In analogy with Standard Model gauge fields,  we define ``Electric'' and ``Magnetic" fields as $\hat{E}_i = -a^{-2}~ \hat{A}'_i , ~\hat{B}_i = a^{-2}~\epsilon_{ijk}~\del_j \hat{A}_k$ to derive the following expressions Fourier space
\begin{align}\label{EBF}
\nn \hat{E}_i(\tau,\vec{k}) &= -(H\tau)^2 \sqrt{\fr{k}{2}} ~\epsilon_{i}^{-}(\vec{k}) \left(\fr{2\xi(\tau)}{-k\tau}\right)^{1/4} \tilde{A}(\tau,k)\left[\hat{a}_{-}(\vec{k})+~\hat{a}^{\dagger}_{-}(-\vec{k})\right],\\
\hat{B}_i(\tau,\vec{k}) &= -(H\tau)^2 \sqrt{\fr{k}{2}}~ \epsilon_{i}^{-}(\vec{k}) \left(\fr{-k \tau}{2\xi(\tau)}\right)^{1/4} \tilde{A}(\tau,k)\left[\hat{a}_{-}(\vec{k})+~\hat{a}^{\dagger}_{-}(-\vec{k})\right].
\end{align}

\section{Dynamics of primordial fluctuations}\label{Sec3}
The gravitational coupling between the inflaton and the hidden sector fields ($\sgm$ and $A_i$) induces
source terms in the equation of motion of the inflaton fluctuations. On the other hand, gauge fields inevitably couple to the metric and give rise to secondary contributions to tensor fluctuations in addition to those generated by quantum vacuum fluctuations of the metric.
In this section, we will analyze the dynamics of scalar and the tensor fluctuations in the presence of gauge field sources we studied in the previous section. 
 
In the spatially flat gauge, we first note the metric in the ADM form as 
\beq\label{MA}
\mathrm{d} s^{2}=a^{2}(\tau)\bigg\{-N^{2} \mathrm{d} \tau^{2}+ \left(\delta_{i j}+\hat{h}_{ij}(\tau,\vec{x})\right)\left(\mathrm{d} x^{i}+N^{i} \mathrm{d} 
\tau\right)\left(\mathrm{d} x^{j}+N^{j} \mathrm{d} \tau\right)\bigg\},
\eeq  
where $N(\tau,\vec{x}) = 1 +\delta N(\tau,\vec{x})$ and $N^{i}(\tau,\vec{x})$ are non-dynamical lapse and shift function respectively. In terms of its canonical mode functions $\hat{Q}_\lambda$, we decompose the metric as 
\beq
\label{tdc}\hat{h}_{ij}(\tau, \vec{x}) = \fr{2}{\Mp}\int \fr{\d^3 k}{(2\pi)^{3/2}}~ \mathrm{e}^{i\vec{k}.\vec{x}} \sum_{\lambda=\pm} \Pi^{*}_{ij,\lambda}(\vec{k}) \fr{\hat{Q}_{\lambda}(\tau,\vec{k})}{a(\tau)},
\eeq
where $\hat{h}_{ij}$ is the transverse, $\partial_i \hat{h}_{ij} = 0$ and traceless, $\hat{h}_{ii} =0$ metric perturbation and the polarization operators are defined as $\Pi^{*}_{ij,\pm} = \epsilon^{\pm}_i(\vec{k}) \epsilon^{\pm}_j(\vec{k}), ~\Pi_{ij,\pm} = \epsilon^{\mp}_i(\vec{k}) \epsilon^{\mp}_j(\vec{k})$, satisfying $\Pi^{*}_{ij,\lambda}\Pi_{ij,\lambda'} = \delta_{\lambda\lambda'}$. 
Besides two physical tensor modes, the Lagrangian \eqref{Lm} contains two scalar dynamical variables. To linear order in perturbations, using conformal time, we decompose these fluctuations as
\beq\label{sdc}
\hat{X}(\tau, \vec{x}) = X(\tau)+\int \frac{\d^{3} k}{(2 \pi)^{3 / 2}}~ \mathrm{e}^{i \vec{k} \cdot \vec{x}}~ \frac{\hat{Q}_{X}(\tau,\vec{k})}{a(\tau)},
\eeq
where $X=\{\phi,\sgm\}$ and we defined the canonical variables $\hat{Q}_{X} \equiv a(\delta \phi, \delta\sigma)^{T}$. Using the metric \eqref{MA} and the decompositions in \eqref{sdc} in the Lagrangian \eqref{Lm}, one can solve for the lapse and shift functions in terms of the dynamical scalar modes (See \eg \cite{Barnaby:2011vw, Ozsoy:2017blg}). In this way, we found that the action for physical scalar fluctuations $\hat{Q}_X$ is given by
\beq\label{ss2}
S\left[\hat{Q}_\phi,\hat{Q}_\sgm\right]=\frac{1}{2} \int \d \tau \d^{3} k\Bigg\{\hat{Q}_{a}'^{T} \hat{Q}'_{a}-\hat{Q}^{T}_{a} \bigg[ k^{2}~\delta_{a b}+M_{a b}^{2}\bigg] \hat{Q}_{b} + 2\hat{Q}^{T}_a ~ \hat{J}_a (\tau, \vec{k})\Bigg\},
\eeq
where $M^2_{ab}$ is the effective mass term for canonical fluctuations, including mass mixing between $\hat{Q}_\phi$ and $\hat{Q}_\sgm$ and is given by 
\beq
M^{2}_{ab} = - (aH)^2  \bigg[ (2-\epsilon) \delta_{ab} - (3 - \epsilon)~2\sqrt{\epsilon_{a}\epsilon_{b}} - \fr{V_{,ab}}{H^2} - \left(\fr{\sqrt{2\epsilon_a}V_{,b}+\sqrt{2\epsilon_b}V_{,a}}{H^2 \Mp}\right)\bigg],
\eeq
where $\epsilon = \epsilon_\phi + \epsilon_{\sigma}$, $\epsilon_b \equiv \dot{\varphi}_b^2/(2H^2\Mp^2)$ and $V_{,b} \equiv \partial V/\partial\varphi_b$ with $\varphi_b=(\phi,\sgm)^T$ and $V(\phi,\sgm) = V_\phi + V_\sgm$. On the other hand, the source term that is induced by the presence of gauge fields is given by  $\hat{J}_a \simeq (0, \hat{J}_\sgm(\tau,\vec{k}))^{T}$ \footnote{Through the gravitational interactions, both fluctuations $\hat{Q}_\phi$ and $\hat{Q}_\sgm$ obtain Planck suppressed couplings to the gauge fields and may in principle receive contributions of $\delta A + \delta A\to \delta \phi$ and $\delta A + \delta A\to \delta \sgm$ type. However, as shown in \cite{Barnaby:2012xt,Ozsoy:2017blg}, these contributions are negligible compared to the process $\delta A+ \delta A \to \delta \sgm \to \delta \phi$ and thus can be safely ignored.}:
\beq
\hat{J}_\sgm(\tau, \vec{k}) \equiv \frac{\alpha_{\mathrm{c}} a(\tau)^{3}}{f} \int \frac{\mathrm{d}^{3} x}{(2 \pi)^{3 / 2}}~\mathrm{e}^{-i \vec{k} \cdot \vec{x}}~ \hat{E}_i(\tau,\vec{x})  \hat{B}_i(\tau,\vec{x}).
\eeq
Similarly, for each polarization, tensor fluctuations $Q_\lambda$ has
\beq\label{SQl}
S\left[\hat{Q}_{\lambda}\right]=\frac{1}{2} \int \d \tau \d^{3} k\left\{\hat{Q}'_{\lambda}\hat{Q}_{\lambda}'-\left[k^{2}- \fr{a''(\tau)}{a(\tau)}\right] \hat{Q}^2_{\lambda} + 2\hat{Q}_\lambda ~ \hat{J}_\lambda (\tau, \vec{k})\right\},
\eeq
where the source induced by gauge fields given by the following Fourier transform
\beq\label{JT}
\hat{J}_{\lambda}(\tau,\vec{k})
 \equiv - \fr{a(\tau)^3}{\Mp}\Pi_{ij,\lambda} (\vec{k})\int \fr{\d^3 x}{(2\pi)^{3/2}} ~{\mathrm e}^{-i\vec{k}.\vec{x}} \bigg[ \hat{E}_i (\tau, \vec{x}) \hat{E}_j (\tau, \vec{x})+ \hat{B}_i (\tau, \vec{x}) \hat{B}_j (\tau, \vec{x})\bigg].
\eeq
Next, we study the scalar and tensor modes in the presence of sources, \ie $\hat{J}_\phi$ and $\hat{J}_\lambda$. 
\subsection{Scalar Fluctuations}\label{SS}
Defining the second slow-roll parameter by $\eta_b = \Mp^2 V_{,bb}/V$, the total mass matrix, $M_{ab}^2$ can be written as
\beq\label{mm}
M_{ab}^{2} \simeq-\frac{1}{\tau^{2}}\left(\begin{array}{cc}
{2+9 \epsilon_{\phi}+3 \epsilon_{\sigma}-3 \eta_{\phi}} & {6 \sqrt{\epsilon_{\phi} \epsilon_{\sigma}}} \\
{6 \sqrt{\epsilon_{\phi} \epsilon_{\sigma}}} & {2+9 \epsilon_{\sigma}+3 \epsilon_{\phi}-3 \eta_{\sigma}}
\end{array}\right),
\eeq
where we kept leading terms in slow-roll. Using \eqref{mm} and \eqref{ss2} (See footnote 12), the equation of motion for the canonical scalar fluctuations read as
\begin{align}
\label{uphi}&\left(\frac{\partial^{2}}{\partial \tau^{2}}+k^{2}-\frac{2}{\tau^{2}}\right) \hat{Q}_{\phi} \simeq \frac{6}{\tau^{2}} \sqrt{\epsilon_{\phi} \epsilon_{\sigma}} ~\hat{Q}_{\sigma}\\
\label{usgm}&\left(\frac{\partial^{2}}{\partial \tau^{2}}+k^{2}-\frac{2}{\tau^{2}}\right) \hat{Q}_{\sigma} \simeq\fr{\alpha_{\rm c}a(\tau)^3}{f}\int\fr{\d^3 p}{(2\pi)^{3/2}}~ \hat{E}_{i}(\tau, \vec{k}-\vec{p}) ~\hat{B}_{i}(\tau,\vec{p}).
\end{align}
In the following, we focus on the production of $\delta \sgm$ from the gauge field and its subsequent sourcing of $\delta \phi$, namely the process $\delta A + \delta A \to \delta \sgm \to \delta \phi$. To solve for $\hat{Q}_\phi$ we split it into its uncorrelated vacuum $\hat{Q}^{(v)}_\phi$ and sourced part $\hat{Q}^{(s)}_\phi$ corresponding to the homogeneous and particular solution of eq. \eqref{uphi}, respectively. The vacuum part can be decomposed as $\hat{Q}_{\phi}^{(v)}(\tau, \vec{k})=Q_{\phi}^{(v)}(\tau, k)\,a(\vec{k})+Q_{\phi}^{(v)\,*}(\tau, k)\,a^{\dagger}(-\vec{k})$ where the solution that reduces to Bunch-Davies vacuum in the far past $-k\tau \gg 1$ is given by
\beq\label{vs}
Q_{\phi}^{(v)}(\tau, k)=\frac{\mathrm{e}^{-i k \tau}}{\sqrt{2 k}}\left(1-\frac{i}{k \tau}\right).
\eeq
On the other hand, the solution to the sourced part $\hat{Q}_{\phi}^{(s)}(\tau, k)$ can be found by first solving eq. \eqref{usgm} and then plugging the resulting solution as a source in the eq. \eqref{uphi}, \ie
\beq\label{QS}
\hat{Q}_{\phi}^{(s)} (\tau,k)=6 \sqrt{\epsilon_{\phi}} \int \d \tau^{\prime} G_{k}\left(\tau, \tau^{\prime}\right) \frac{\sqrt{\epsilon_{\sigma}\left(\tau^{\prime}\right)}}{\tau^{\prime 2}} \int \d \tau^{\prime \prime} G_{k}\left(\tau^{\prime}, \tau^{\prime \prime}\right) \hat{J}_{\sigma}\left(\tau^{\prime \prime}, \vec{k}\right),
\eeq
where $G_k(\tau,\tau')$ is the retarded Green's function\footnote{In \eqref{QS}, we neglect scale dependence of the Green's functions $G_k$ that might arise from  $\mathcal{O}(\eta_\sigma,\epsilon_\sgm)$ corrections in \eqref{mm}. We note that at the time ($N=N_*$) where the dominant contribution to $\hat{Q}_{\phi}^{(s)}$ takes place, $\eta_\sigma = \Mp^2 (V''_\sigma/V) \propto N-N_* \to 0$ and $\epsilon_{\sgm,*} \propto (f/\Mp)^2 \ll 1$ as can be verified from the discussion presented in Appendix \ref{AppA} and \ref{AppE}.}
for the operator $\partial_\tau^2 + k^2 -2/\tau^2$: 
\beq\label{gf}
G_{k}\left(\tau, \tau^{\prime}\right)=\Theta\left(\tau-\tau^{\prime}\right) \frac{\pi}{2} \sqrt{\tau \tau^{\prime}}\left[J_{3 / 2}(-k \tau) Y_{3 / 2}\left(-k \tau^{\prime}\right)-Y_{3 / 2}(-k \tau) J_{3 / 2}\left(-k \tau^{\prime}\right)\right],
\eeq
where $J_{\nu}$ and $Y_{\nu}$ denote Bessel functions of real argument.  In Appendix \ref{AppC}, we will compute in detail the scalar correlators that arise in the presence of the sourced contribution in \eqref{QS}.

{\bf The comoving curvature perturbation:} For the multi-sector model we consider, $\mathcal{R}(\tau,\vec{k})$ can in principle obtain contributions from all fields involved in the Lagrangian \eqref{Lm}. However, below we show that the standard expression $\mathcal{R} = (H/a\dot{\phi})Q_\phi$ valid in single field inflation still provides a very good approximation for the computation of late time correlators in our model. We begin by defining comoving curvature perturbation $\mathcal{R}$ in spatially flat gauge: $\mathcal{R} = - ({H}/{(\rho+p)}) \,\delta q$ \cite{Malik:2008im,Baumann:2009ds} where $\rho+p = \dot{\phi}^2+\dot{\sigma}^2$ is the sum of background energy density and pressure and $\delta q$ is the scalar momentum density in flat gauge. In terms of the perturbed energy momentum tensor, $\delta q$ is given by  $T^{0}_{\,\,\,i} = \partial_i \delta q$ where $T^{0}_{\,\,\,i} = g^{0\mu}(\partial_\mu\phi \partial_i \delta\phi +\partial_\mu\sigma\partial_i \delta\sgm+ g^{\rho\sigma} F_{\mu\rho}F_{i\sigma})$. Recalling the definition of the gauge field strength tensor, in real space we have $\partial_i \delta q = -\partial_i (\dot{\phi}\delta\phi+\dot{\sgm}\delta\sgm)+ a (\vec{E}\times\vec{B})$ at leading order in perturbations. Putting everything together, in Fourier space, total $\mathcal{R}$ is therefore given by 
\beq\label{Rtot}
\mathcal{R}(\tau,\vec{k})=\fr{H}{a(\dot{\phi}^{2}+\dot{\sigma}^{2})}\left(\dot{\phi}\, Q_\phi+\dot{\sigma}\,Q_\sgm-a \,\delta q_{(AA)}(\tau,\vec{k})\right),
\eeq
where the scalar momentum density due to the gauge fields is given by 
\beq\label{qAA}
\delta q_{(AA)}(\tau,\vec{k}) =- a\frac{i \hat{k}_{i}}{k} \epsilon_{i j k} \int \frac{\d^{3} q}{(2 \pi)^{3 / 2}} E_{j}(\tau,\vec{k}-\vec{q}) B_{k}(\tau,\vec{q}).
\eeq
In this work, independent of how many wiggles $\sgm$ probes on its potential, we assume that $\sgm$ settles to its global minimum much before the end of inflation where $\dot{\sgm} \to 0$ which allow us to neglect the term linear in $Q_\sgm$ for the computation of late time correlators of $\mathcal{R}$ in eq. \eqref{Rtot} \cite{Mukohyama:2014gba,Namba:2015gja}. In this case, at the end of inflation, the curvature perturbation directly induced due to gauge field fluctuations is given by
\beq\label{RAA}
\mathcal{R}_{(AA)}(\tau_{\rm end},\vec{k}) = \fr{H}{\dot{\phi^2}}a(\tau_{\rm end})\frac{i \hat{k}_{i}}{k} \epsilon_{i j k} \int \frac{\d^{3} q}{(2 \pi)^{3 / 2}} E_{j}(\tau_{\rm end},\vec{k}-\vec{q}) B_{k}(\tau_{\rm end},\vec{q}).
\eeq
To check if $\mathcal{R}_{(AA)}$ can significantly influence the late time correlators of $\mathcal{R}$, we studied the power spectrum of \eqref{RAA} in Appendix \ref{AppD} and found that it can be factorized as
\beq\label{PRAA}
\mathcal{P}_{\mathcal{R}_{(AA)}}(k) = \left[\epsilon_\phi \mathcal{P}_{\mathcal{R}}^{(v)}\right]^2 \left(\fr{\tau_{\rm end}}{\tau_*}\right)^6 {f}_{2,\mathcal{R}_{(AA)}}\left(\xi_*,\fr{k}{k_*},\delta\right),
\eeq
where $f_{2,\mathcal{R}_{(AA)}}$ is derived in \eqref{f2RAA} and it parametrizes the scale dependent enhancement of $\mathcal{P}_{\mathcal{R}_{(AA)}}$ due to gauge fields. In Appendix \ref{AppD}, we confirmed that the scale dependent part ${f}_{2,\mathcal{R}_{(AA)}}$ can never compete with the enormous suppression factor $(\tau_{\rm end}/\tau_*)^6 = e^{-6N_*}$ for all the phenomenological scenarios we consider in this paper where $N_* \geq 22$. We note that a similar conclusion applies for the higher point auto/cross correlators of $\mathcal{R}_{(AA)}$ because each $\mathcal{R}_{(AA)}$ contains a factor of
$a_{\rm end}^{-3}$ (See \eg \eqref{f2RAA}) representing the dilution of power in the gauge fields far in the IR (See \eg Figure \ref{fig:rhoA}). Therefore, we can safely adopt the standard relation for the purpose of calculating late time cosmological correlators involving $\mathcal{R}$:
\beq\label{CP}
\hat{\mathcal{R}}(\tau, \vec{k}) \simeq \frac{H}{a \dot{\phi}} \hat{Q}_\phi(\tau, \vec{k}) \simeq \frac{H \tau}{\sqrt{2 \epsilon_{\phi}} \Mp} \hat{Q}_{\phi}(\tau, \vec{k}).
\eeq
\subsection{Tensor Fluctuations}\label{ST}
To study the effects of gauge field amplification on the tensor power spectrum, we focus on the equation of canonical mode function $Q_\lambda$ which can be derived from \eqref{SQl} as
\beq\label{ctme}
\left(\partial^2_\tau + k^2 -\fr{2}{\tau^2}\right)\hat{Q}_\lambda(\tau,\vec{k}) =\hat{J}_{\lambda}(\tau,\vec{k}),
\eeq
with the following source term (see \eg \eqref{JT}),
\beq\label{ts}
\hat{J}_{\lambda}(\tau, \vec{k})=-\frac{a^{3}(\tau)}{M_{\mathrm{pl}}} \Pi_{i j, \lambda}(\vec{k}) \int \frac{\mathrm{d}^{3} p}{(2 \pi)^{3 / 2}}\left[\hat{E}_{i}(\tau, \vec{k}-\vec{p}) \hat{E}_{j}(\tau,\vec{p})+\hat{B}_{i}(\tau,\vec{k}-\vec{p}) \hat{B}_{j}(\tau,\vec{p})\right],
\eeq
where we used Fourier transforms of $\vec{E}$ and $\vec{B}$ fields to write the source as a convolution in momentum space. Similar to the case with scalar fluctuations, equations of motion for $\hat{Q}_\lambda$ in \eqref{ctme} is solved by separating $\hat{Q}_\lambda$ into a vacuum mode, $\hat{Q}^{(v)}_\lambda$, \ie solution to the homogeneous part of \eqref{ctme} and the sourced mode $\hat{Q}^{(s)}_\lambda$. Assuming, $a \simeq -1/(H\tau)$, the vacuum mode is given by 
\begin{align}\label{VM}
\nn \hat{Q}_{\lambda}^{(v)}(\tau, \vec{k}) &= Q_{\lambda}(\tau, k)~ \hat{a}_{\lambda}(\vec{k})+ Q_{\lambda}^{*}(\tau, k)~ \hat{a}_{\lambda}^{\dagger}(-\vec{k}), 
\\ Q_{\lambda}(\tau, k) &=\frac{\mathrm{e}^{-i k \tau}}{\sqrt{2 k}}\left(1-\frac{i}{k \tau}\right),
\end{align}
where $\hat{a}_{\lambda}^{\dagger}$ creates a graviton with helicity $2\lambda$. On the other hand, the sourced contribution can be written formally as
\beq\label{QSL}
\hat{Q}_{\lambda}^{(s)}(\tau, \vec{k}) = \int^{\tau} \d\tau'~ G_k(\tau,\tau')~ \hat{J}_\lambda(\tau',\vec{k}),
\eeq
where the retarded Green's function in this case is also given by \eqref{gf}. 
\section{Phenomenology of Cosmological Correlators}\label{Sec4}
The roll of the spectator scalar $\sgm$ through the cliffs of its wiggly potential produces gauge field fluctuations that can be considered as a source of inflaton and metric fluctuations through the corresponding inverse decay processes: $\delta A + \delta A \to \delta \sgm \to \delta\phi$ and $\delta A + \delta A \to h_\lambda$. Building upon our discussion on the sourced scalar (Section \ref{SS}) and tensor perturbations (Section \ref{ST}) in the previous section, we calculate cosmological correlators of curvature perturbation $\mathcal{R}$ and metric perturbation $h_\lambda$ in Appendix \ref{AppB} and \ref{AppC}. In the following subsection we present our results and study their phenomenological implications.
\subsection{Scalar and tensor correlators}\label{S4p1}
The total power spectrum and bispectrum of tensor and scalar curvature perturbation are defined as in \eqref{DTPS}, \eqref{tbsdef}, \eqref{DRPS} and \eqref{rbsdef}. All the cosmological correlators in this model can be written as a superposition of uncorrelated vacuum and sourced parts as we discussed in the previous section. Therefore, for power and bispectra\footnote{We note that due to the spectator nature of the axion and the fact that typical displacement of $\sigma$ is comparable to frequency $f$ of modulations in the $\Lambda^4\lesssim \mu^3 f$ regime of the potential \eqref{Vs}, resonant effects in the power and bispectra can not appear in the model under consideration, in contrast to the original axion monodromy models studied in \cite{Flauger:2009ab,Flauger:2010ja}.}, we have
\begin{align}
 \nn\mathcal{P}_{\mathcal{R}}(k)& =  \mathcal{P}^{(v)}_{\mathcal{R}}(k) +  \mathcal{P}^{(s)}_{\mathcal{R}}(k), ~~~   \mathcal{P}_{\lambda}(k) =  \mathcal{P}^{(v)}_{\lambda}(k) +  \mathcal{P}^{(s)}_{\lambda}(k),\\
 \mathcal{B}_{\mathcal{R}}(k)& =  \mathcal{B}^{(v)}_{\mathcal{R}}(k) +  \mathcal{B}^{(s)}_{\mathcal{R}}(k), ~~~   \mathcal{B}_{\lambda_1\lambda_2\lambda_3}(k) =  \mathcal{B}^{(v)}_{\lambda_1\lambda_2\lambda_3}(k) +  \mathcal{B}^{(s)}_{\lambda_1\lambda_2\lambda_3}(k).
\end{align}
The scalar and tensor vacuum bispectrum is below the present observational limits \cite{Maldacena:2002vr,Acquaviva:2002ud}, and thus only sourced modes are of our interest,  $\mathcal{B}^{(v)}_{\mathcal{R}} \to 0$, $\mathcal{B}^{(v)}_{\lambda_1\lambda_2\lambda_3} \to0$. In contrast to the vacuum fluctuations of the metric, only $-$ of the helicity of sourced metric fluctuations are amplified significantly in the presence of vector field sources, making only $\mathcal{P}^{(s)}_{-}$ contribution relevant. Similarly, due to the parity violating nature of gauge field production, $\mathcal{B}^{(s)}_{---}$ will appear as the dominant contribution to the tensor non-Gaussianity. 

At leading order in slow-roll the vacuum power spectrum of scalar and tensor fluctuations are given by
\beq\label{PSV}
 \mathcal{P}^{(v)}_{\mathcal{R}}(k) = \fr{H^2}{8\pi^2 \epsilon_\phi \Mp^2}, ~~~\mathcal{P}^{(v)}_{\lambda}(k) = \fr{H^2}{\pi^2\Mp^2},
\eeq
implying the standard relation for the vacuum tensor to scalar ratio $r_{v} = 16 \epsilon_\phi$.

 All the non-standard features of scalar and tensor perturbations are encoded in the modes sourced by vector fields, namely $\mathcal{P}^{(s)}_{\mathcal{R}}, \mathcal{P}^{(-)}_{\lambda}, \mathcal{B}^{(s)}_{\mathcal{R}}$ and $\mathcal{B}^{(s)}_{---}$. In the model we are considering, as $\sgm$ traverses the step like regions in its wiggly potential (see Figure \ref{fig:pot}), the effective coupling $\xi$ \eqref{Joe} between vector fields and $\sgm$ obtains a bump in time direction. The gauge field modes that crosses the horizon around the time where $\xi$ reaches its peak value will be maximally amplified in a localized manner in momentum space. For the correlators of $\mathcal{R}^{(s)}$ and $h_\lambda^{(s)}$ sourced by the vector fields, this directly translates into a localized bump in momentum space. The height of this scale dependent signal depends on the maximum value $\xi_*$ achieved by $\xi$ whereas the width depends on the number of e-folds $\dot{\sgm}$ significantly differs from zero, $\Delta N \simeq \delta ^{-1}$, implying its dependence on the mass of the axion in its global minimum, \ie $\delta \propto \mu^3 /(f H^2) \simeq m^2_{\rm axion}/H^2$. For larger $\delta$, $|\dot{\sgm}|$ will reach its maximum faster before it reduces to very small values in the plateau regions in the potential. This implies that, fewer $k$ modes of gauge fields will be influenced by the roll of $\sgm$, reducing the width of the bump in the cosmological correlators\footnote{In \cite{Namba:2015gja}, larger values of $\delta$ is considered as a favorable way of reducing the effect of the sourced scalar fluctuations with respect to GWs. Focusing on the same parameter choices, we found that the velocity profile of the spectator field we consider here (See eq. \eqref{dsgm}) is rather more spiky compared to model considered in \cite{Namba:2015gja}, leading to more GW production for the same amount of enhanced scalar fluctuations, implying a slightly improved situation from the perspective of constraints on scalar fluctuations at various cosmological scales. This result can be confirmed by comparing our Table \ref{tab:fit1} with Table 1 and 2 of \cite{Namba:2015gja} or by simply comparing the width of the gauge field normalization factor we found in eq. \eqref{fitN} with the ones appearing in Appendix B of \cite{Peloso:2016gqs}.}.

As one can anticipate from the discussion above, the sourced power spectra and bispectra obtains the following functional dependence on the model parameters,
\begin{align}\label{SC}
\nn\mathcal{P}^{(s)}_{\mathcal{R}}(k) &= \left[\epsilon_\phi \mathcal{P}^{(v)}_\mathcal{R}(k)\right]^2 f_{2,\mathcal{R}}\left(\xi_*, \fr{k}{k_*},\delta\right),\\\nn
\mathcal{P}^{(s)}_{\lambda} (k)&= \left[\epsilon_\phi \mathcal{P}^{(v)}_\mathcal{R}(k)\right]^2 f_{2,\lambda}\left(\xi_*, \fr{k}{k_*},\delta\right), \\\nn
\mathcal{B}^{(s)}_{\mathcal{R}}(k_1,k_2,k_3) &= \frac{\left[\epsilon_{\phi} \mathcal{P}_{\mathcal{R}}^{(v)}(k)\right]^{3}}{k_{1}^{2} k_{2}^{2} k_{3}^{2}} f_{3, \mathcal{R}}\left(\xi_{*}, \fr{k_1}{k_*},\fr{k_2}{k_*},\fr{k_3}{k_*}, \delta\right)\\
\mathcal{B}^{(s)}_{\lambda \lambda \lambda} (k_1,k_2,k_3) &= \frac{\left[\epsilon_{\phi} \mathcal{P}_{\mathcal{R}}^{(v)}(k)\right]^{3}}{k_{1}^{2} k_{2}^{2} k_{3}^{2}} f_{3, \lambda}\left(\xi_{*}, \fr{k_1}{k_*},\fr{k_2}{k_*},\fr{k_3}{k_*}, \delta\right),
\end{align}
where the dimensionless functions $f_{i,j}$ with $i=2,3$ and $j = \{\mathcal{R},+,-\}$ at the right hand parametrizes the dependence of the sourced correlators on the model parameters. The functions $f_{3,j}$ encode the full dependence of the bispectrum on the external momenta $k_i$, $i=1,2,3$. Similar in spirit to the model considered in \cite{Namba:2015gja} where a localized bump in the cosmological correlators present, we studied the shape of the scalar and tensor bispectrum using the formulas we derived in \eqref{f3l} and \eqref{f3R}. In this way, we found that both bispectra can be approximated by an equilateral shape when the signal is maximal (\ie at $k_i = \mathcal{O}(1) k_*$). In addition to the 2-pt functions, in this work we will therefore study phenomenology of 3-pt correlators by focusing on the functions $f_{3,j}$ for equal momenta. 

\begin{figure}[t!]
\begin{center}
\includegraphics[scale=0.63]{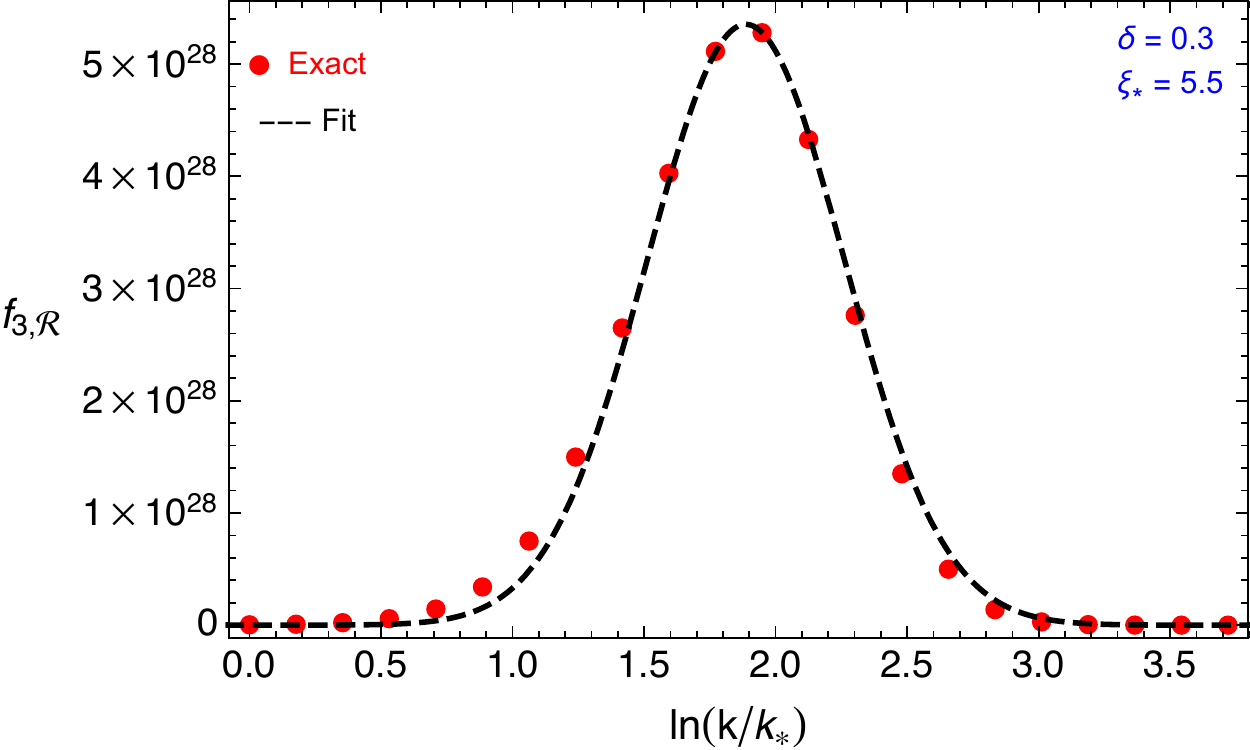}~\includegraphics[scale=0.64]{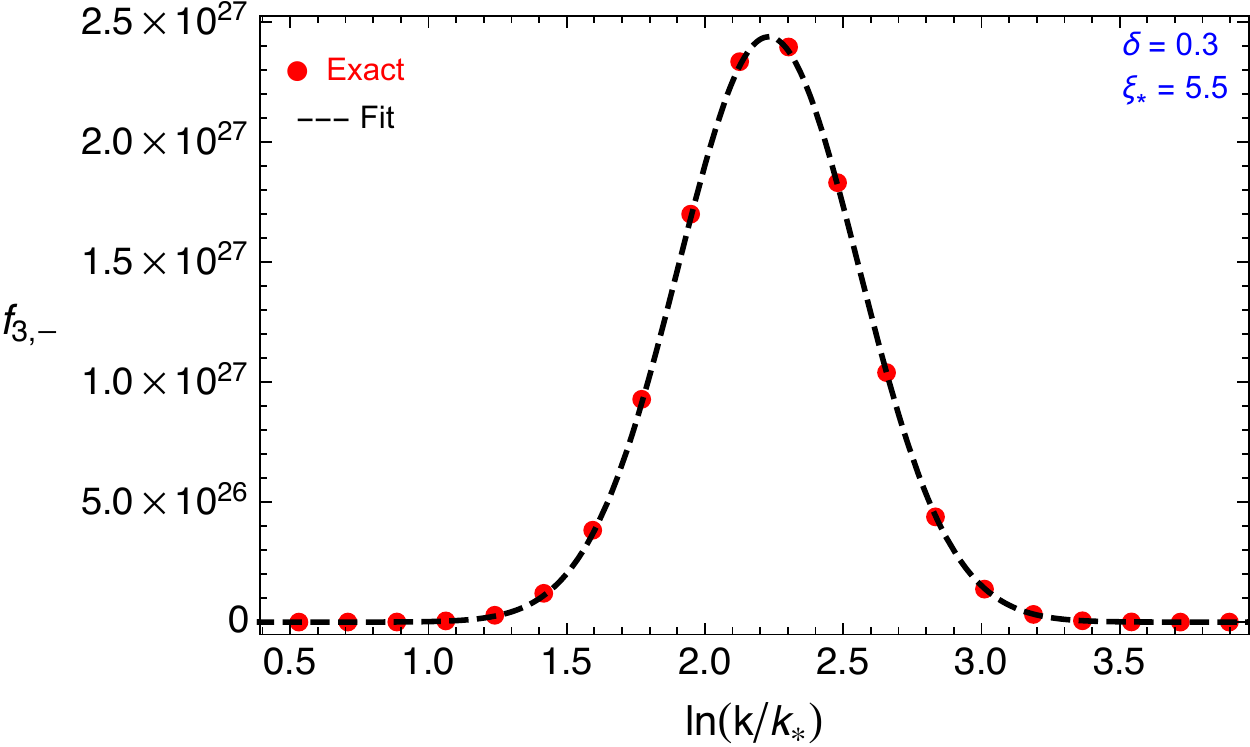}
\end{center}
\caption{ The red dots are obtained by exact numerical evaluation of $f_{3,\mathcal{R}}$ (Left) and $f_{3,-}$ (Right) while their approximate forms are obtained using eq. \eqref{fpheno} (black dashed lines) making use of Table \ref{tab:fit1}. The location of the peaks in both functions $f_{3,\mathcal{R}}$/$f_{3,-}$ appear at $k = \mathcal{O}(1) k_*$ and are due to the acceleration of $\sgm$ during its rollover from the cliff like regions in its potential.\label{fig:f3Rfit}}
\end{figure}

By studying the integrals defined in Appendix \ref{AppB} and \ref{AppC} for fixed $\xi_*$ and $\delta$ numerically, we found that the functions $f_{i,j}$ can be well described by a log-normal distribution in momentum space,
\beq\label{fpheno}
f_{i, j}\left(\frac{k}{k_{*}}, \xi_{*}, \delta\right) \simeq f_{i, j}^{c}\left[\xi_{*}, \delta\right] \exp \left[-\frac{1}{2 \sigma_{i, j}^{2}\left[\xi_{*}, \delta\right]} \ln ^{2}\left(\frac{k/k_{*}}{x_{i, j}^{c}\left[\xi_{*}, \delta\right]}\right)\right].
\eeq
The information about the location, width and the height of the signal depends on the motion of $\sgm$ in its potential and is therefore characterized by $\xi_*$ and $\delta$ inside the functions $x^c_{i,j}, \sgm_{i,j}, f^c_{i,j}$. As it is clear from \eqref{fpheno}, $f_{i,j}$ is maximal at $k = k_* x^c_{i,j}$, where it evaluates to $f^c_{i,j}$ whereas $\sgm_{i,j}$ controls the width of this signal. For a given choice of $\xi_*$ and $\delta$, we derived approximate formulas for these functions by fitting the right hand side of eq. \eqref{fpheno} to reproduce the position, height and width of the sourced signal parametrized within the integrals defined in Appendix \ref{AppB} and \ref{AppC} for $f_{i,j}$ (See \eg \eqref{f2l}, \eqref{f3l}, \eqref{f2Rf} and \eqref{f3R}).
\begin{table}
\begin{center}
\begin{tabular}{| c | c | c | c |}
\hline
\hline
\cellcolor[gray]{0.9}$\{i,j\}$&\cellcolor[gray]{0.9}$\ln(|f^c_{i,j}|) \simeq $&\cellcolor[gray]{0.9}$x^c_{i,j} \simeq $ &\cellcolor[gray]{0.9}$\sgm_{i,j} \simeq $ \\
\hline
\cellcolor[gray]{0.9}$\{2,\mathcal{R}\}$&$ -15.13 + 10.09\,\xi_*+ 0.0389\,\xi_*^2$ & $6.63 -0.403\, \xi_* + 0.0856\, \xi_* ^2$&$0.89 -0.101\, \xi_* + 0.0066\, \xi_* ^2$ \\\hline
\cellcolor[gray]{0.9}$\{2,-\}$ & $-14.78 +9.91\, \xi_* + 0.0487\, \xi_* ^2$ & $7.78 - 0.166\,\xi_* + 0.0992\, \xi_*^2$&$0.83 -0.110\,\xi_* + 0.0070\,\xi_* ^2 $\\\hline
\cellcolor[gray]{0.9}$\{2,+\}$ & $-21.01 +9.94\, \xi_* +0.0469\, \xi_* ^2$ & $3.16 + 0.003\,\xi_* + 0.0401\,\xi_* ^2$ & $ 0.91 -0.091\,\xi_* + 0.0061\,\xi_* ^2$ \\\hline
\cellcolor[gray]{0.9}$\{3,\mathcal{R}\}$ & $-19.03 +15.18\,\xi_* + 0.0561\,\xi_*^2 $ & $6.21 -0.377\, \xi_* - 0.0814\,\xi_* ^2 $ & $0.68 -0.086\,\xi_* + 0.0055\, \xi_* ^2 $\\\hline
\cellcolor[gray]{0.9}$\{3,-\}$& $ -20.81 + 14.83\, \xi_* + 0.0773\, \xi_* ^2$ &$ 7.43 - 0.209\, \xi_* + 0.0996\, \xi_* ^2$ &$0.67 -0.095\, \xi_* + 0.0061\, \xi_* ^2 $\\
\hline
\hline
\end{tabular}
\caption{\label{tab:fit1} $\xi_*$ dependence of the height $f^{c}_{i,j}$, location $x^c_{i,j}$ and width $\sgm_{i,j}$ of \eqref{fpheno} for $\delta = 0.3$. Among the entries shown, only the first column of $\{3,-\}$ has a negative sign.}
\end{center}			
\end{table}
For $\delta = 0.3$, we found that these functions can be described by a smooth second order polynomial in $\xi_*$ in the interval $3 \leq \xi_*\leq 6.5$ as we present in Table \ref{tab:fit1}. To illustrate the accuracy of the expression in \eqref{fpheno}, we compare the exact and approximate form of $f_{3,\mathcal{R}}$ for a representative choice of model parameters in Figure \ref{fig:f3Rfit}.
\subsubsection{Scalar power spectrum and tensor to scalar ratio }\label{S4p1p1}
In this subsection, we study the phenomenology of the model at the level of 2-pt functions, particularly focusing on observables at CMB scales. For this purpose, we assume that during its motion, $\sgm$ traverses only a single cliff like regions in its bumpy potential such that its velocity peaks at the time when scales associated with CMB observations exit the horizon.

\begin{figure}[t!]
\begin{center}
\includegraphics[scale=0.64]{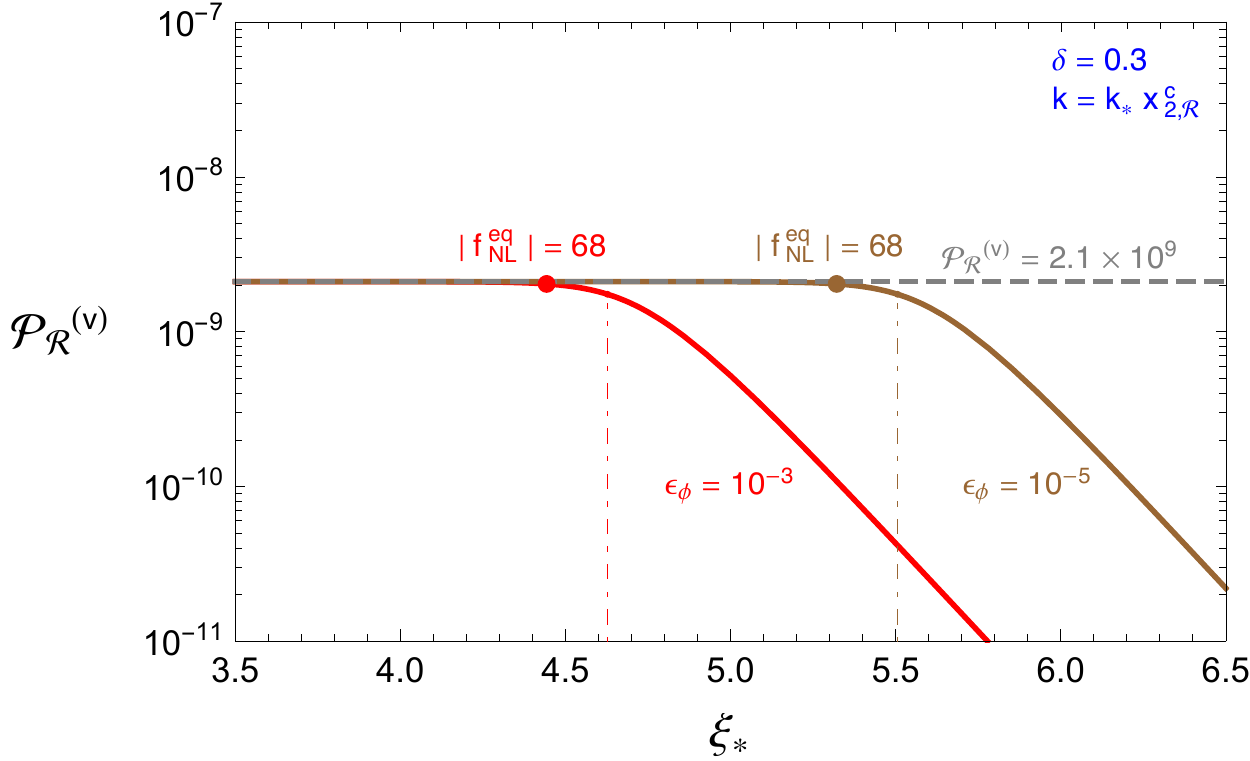}\includegraphics[scale=0.63]{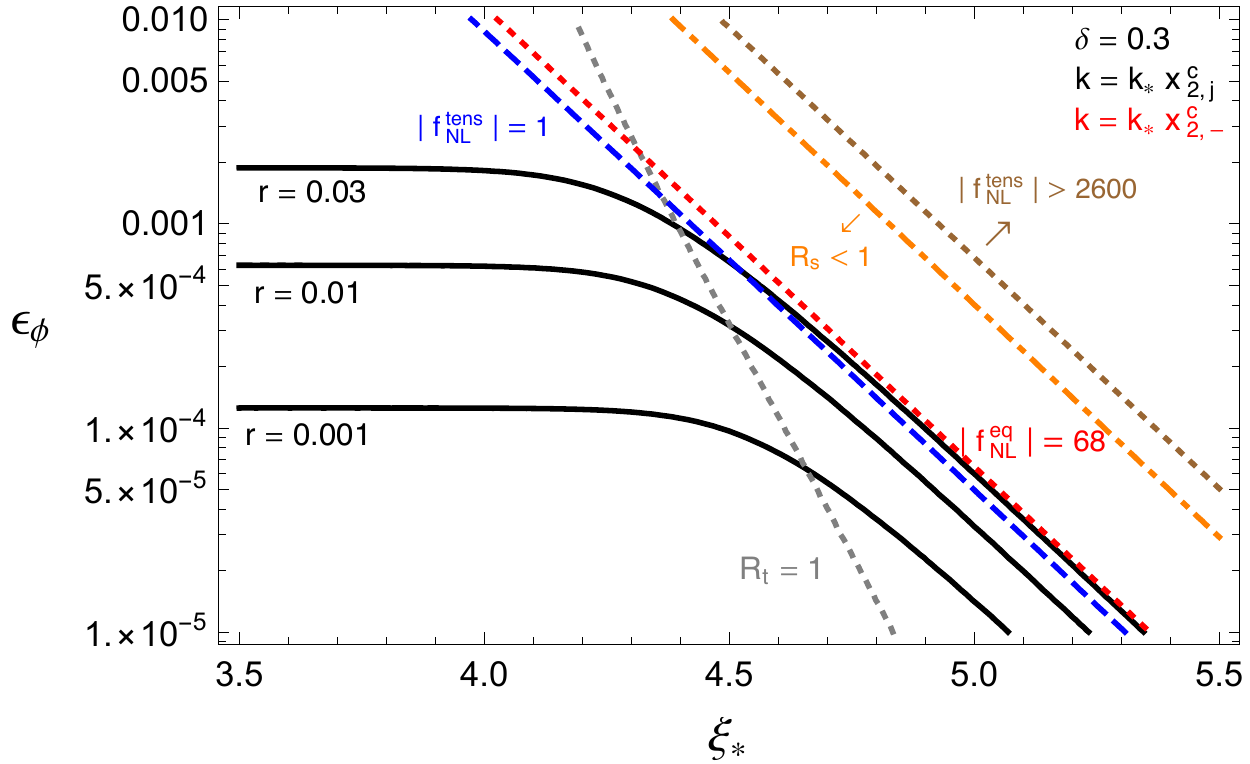}
\end{center}
\caption{Scalar power spectrum normalization in the $\mathcal{P}^{(v)}_\mathcal{R}-\xi_*$ using eq. \eqref{PSnorm} (Left): Dotted dashed lines indicate the location in terms of $\xi_*$ where $R_s =1$ in \eqref{Rts} and the coloured points indicate the location on the $\mathcal{P}^{(v)}_\mathcal{R} -\xi_*$ plane where $f^{\rm eq}_{\rm NL} =68$.  Curves of constant $r$ in the $\epsilon_\phi-\xi_*$ plane (Right): Sourced tensor spectrum dominates over the vacuum fluctuations on the right hand side of the dotted gray line where $R_t > 1$. Above the red dotted line,  $f^{\rm eq}_{\rm NL} > 68$ where eq. \eqref{fnl} is used. On the right panel, we also show $|f^{\rm tens}_{\rm NL}| = 1$ \eqref{fnltens} by the dashed blue line as an error of $\sigma(f^{\rm tens}_{\rm NL}) =1$ is expected for the upcoming CMB B-mode missions. Brown arrow indicates the parameter space that is ruled out by the current limits on tensor non-Gaussianity, $| f^{\rm tens}_{\rm NL} |< 2600$.\label{fig:r}} 
\end{figure}

{{\bf Normalization of the scalar power spectrum:}} The total scalar power spectrum is given by the sum of nearly scale invariant piece plus a sourced signal and should yield to the correct normalization $\mathcal{A}_s \simeq 2.1 \times 10^{-9}$ by Planck \cite{Akrami:2018odb}. In the $\mathcal{P}_\mathcal{R}^{(v)} - \xi_*$ plane, the power spectrum normalization is satisfied along the following curve,
\beq\label{PSnorm}
\mathcal{P}_\mathcal{R}^{(v)} = \fr{1}{2\, \epsilon_\phi^2\, f_{2,\mathcal{R}}(\xi_*,\delta)}\left[-1 + \sqrt{1+ 4\,\mathcal{A}_s \,\epsilon_\phi^2\, f_{2,\mathcal{R}}(\xi_*,\delta)}\right]
\eeq
\noindent
where we evaluated $f_{2,\mathcal{R}}(\xi_*,\delta) =  \exp({f^c_{2,\mathcal{R}}}(\xi_*,\delta))$ at the peak of the sourced signal. It is clear from \eqref{PSnorm} that if the second term inside the square root is much smaller than unity, we recover the standard result: $\mathcal{P}_{\mathcal{R}}^{(v)} = \mathcal{A}_s$. The value of $\xi_*$ where the sourced contribution becomes comparable to the vacuum one depends on the background model, in particular to the value of $\epsilon_\phi$ when the sourced contribution peaks. At fixed $\epsilon_\phi$, as $\xi_*$ increases, vacuum power spectrum should be exponentially decreased to avoid over production of scalar fluctuations. In general, for smaller $\epsilon_\phi $, it is easier to keep the sourced contribution to the total power spectrum sub-dominant compared to the vacuum fluctuations. We illustrate these facts in the left panel of Figure \ref{fig:r}.

{{\bf Tensor to scalar ratio:}} In the presence of sourced contribution, the tensor-to-scalar ratio modifies as
\beq\label{r}
r(k) = \fr{\sum_\lambda \mathcal{P}^{(v)}_\lambda(k) +  \mathcal{P}^{(s)}_\lambda(k) }{\mathcal{P}^{(v)}_\mathcal{R}(k) + \mathcal{P}^{(s)}_\mathcal{R} (k)} \simeq 16 \epsilon_\phi \left( \fr{1+ \fr{\epsilon_\phi}{16}\, \mathcal{P}^{(v)}_\mathcal{R}(k)\, f_{2,-}(k)}{1+ \epsilon_\phi^2\, \mathcal{P}^{(v)}_\mathcal{R}(k)\, f_{2,\mathcal{R}}(k)} \right), 
\eeq
where we have neglected the subdominant positive helicity mode as $f_{2,+} \ll f_{2,-}$. In \eqref{r}, the second term in both the numerator and denominator gives the ratio between the sourced and vacuum power spectrum for tensor/scalar fluctuations respectively: 
\beq\label{Rts}
R_{t} \equiv \fr{\epsilon_\phi}{16}\, \mathcal{P}^{(v)}_\mathcal{R}(k)\, f_{2,-}(k), ~~~ R_s \equiv  \epsilon_\phi^2\, \mathcal{P}^{(v)}_\mathcal{R}(k)\, f_{2,\mathcal{R}}(k).
\eeq 
It is immediately clear from these expressions that sourced tensor modes tend to become more dominant than the scalars at smaller values of $\epsilon_\phi$. This is the particular regime we are interested in because in this case vacuum tensor fluctuations remain to be small, $r_v = 16 \epsilon_\phi$ while the tensor power spectrum is mainly controlled by the sourced signal without over producing scalar fluctuations. We represent these facts on the right panel in Figure \ref{fig:r} where we show curves of constant $r$ (solid black lines), the ratio between the sourced and vacuum scalar power spectrum (orange dotted dashed lines) together with the line (dotted gray line) where the sourced spectrum of tensor fluctuations becomes comparable to the vacuum power spectrum. Notice that on the left hand side of this curve, \ie for smaller values of $\xi_*$, constant $r$ curves become $\xi$ independent, implying $r\simeq r_v = 16\epsilon_\phi$. On the right hand side of the $R_{t}$ line, \ie for greater values of $\xi_*$, $r \gg r_v$, especially towards smaller values of $\epsilon_\phi$ where $R_s \ll 1$. This is the parameter space we are interested in this work, as one can simultaneously realize $r \gg r_v$ and $R_s \ll 1$. In particular, in this regime, we found $r_*^{1/2} \simeq 2.8 \times 10^{-8} \epsilon_\phi\, e^{4.955\,\xi_*}$ at the peak of the sourced GW signal 
\noindent
where we have linearized the expression $f_{2,-}$ in $\xi_*$ using Table \ref{tab:fit1} and $ \mathcal{P}^{(v)}_{\mathcal{R}} = \mathcal{A}_s = 2.1 \times 10^{-9}$. Eliminating $\epsilon_\phi$ in favor of the Hubble rate $H_{\rm inf}$ using eq. \eqref{PSV}, we can relate $r$ to the energy scale of inflation at the peak of the sourced signal, analogously to the standard relation \eqref{EI} as,
\beq\label{rpeak}
\left(\fr{r_*}{0.063}\right)^{1/2} \simeq \left(\fr{H_{\rm inf}/\Mp}{2.5 \times 10^{-5}}\right)^2 e^{1.58 \pi (\xi_* - 4.05)}, ~~~~~~~\delta = 0.3.
\eeq
In contrast to the standard relation \eqref{EI} in single field inflation, \eqref{rpeak} implies that an observable GW spectrum sourced by the secondary sources is possible even for a low scale of inflation, as long as we compensate the reduction in $H_{\rm inf}$ with a sufficiently large $\xi_*$. 

In our analysis, we found that increasing $\delta$ opens up the available parameters space even further in the $\epsilon_\phi -\xi_*$ plane. This is because an increase in $\delta$ decreases the amount of e-folds ($\Delta N = \delta^{-1}$) $\sgm$ is rolling which in turn decreases the amplitude (characterized by $f^c_{i,j}$) and as well as the width (characterized by $\sgm_{i,j}$) of the produced signal as in the latter case, fewer gauge field modes will be excited to source cosmological correlators. However, the decrease in $\Delta N$ influences scalar correlators more than the tensor ones because apart from the reduction of the number of modes that are excited (decrease in the width $\sgm_{i,j}$), the time interval during which the conversion of $\delta \sgm$ (sourced by the $\delta A + \delta A \to \delta \sgm$) to $\delta \phi$ occurs will also decrease. 

It is worth emphasizing that the sourced GW signal we study here may be distinguished form its vacuum counterpart due to its scale dependence and its violation of parity ($f_{2,-}\gg f_{2,+}$). In particular, when the sourced GW signal dominates $R_t \gg 1$, we found that the running $\alpha_t$ of the tensor spectrum is given by \cite{Peloso:2016gqs},
\beq
\alpha_t (k) \equiv \fr{\d n_t(y)}{\d \ln k}\simeq -\fr{1}{\sigma_{2,-}^2}.
\eeq
Therefore, in contrast to the nearly scale invariant power spectrum, the presence of gauge field production gives rise to sourced GW power spectrum with a non-vanishing negative running that has a magnitude inversely proportional to the width of the peak. This running can be measured if the B-mode of the CMB is observed for a sufficiently large range of scales \cite{Namba:2015gja,Peloso:2016gqs}. 
\subsubsection{Scalar and tensor non-Gaussianity}\label{S4p1p2}
The $\{3,\mathcal{R}\}$ and $\{3, -\}$ entries of Table \ref{tab:fit1} clearly indicate that the sourced 3-pt correlators can be significantly large in our model. Considering scalar 3-pt correlators, the non-observation of scalar non-Gaussianity  \cite{Akrami:2019izv} thus impose further restrictions on the parameter space of the model. On the other hand, if the B-modes are observed by ongoing \cite{Ade:2015tva,Array:2015xqh,Ade:2017uvt} or proposed experiments like PIXIE \cite{Kogut:2011xw}, LiteBIRD \cite{Hazumi:2019lys} and CMB-S4 \cite{Abazajian:2016yjj}, the next important step is to reveal its origin. In this context, the presence of sizeable tensor non-Gaussianity can be considered as a distinguishing feature of our model, in particular as a source of primordial BBB correlator from non-vacuum excitations \cite{Shiraishi:2016yun}. In the following, we will therefore i) discuss the limits on the tensor-to-scalar ratio $r$ from scalar non-Gaussianity  ii) the observability tensor non-Gaussianity at CMB scales.

{{\bf Constraints on scalar non-Gaussianity:}} In the model under consideration, sourced primordial correlators exhibit a width in momentum space, thus making sourced signals manifest itself for a range of cosmological scales relevant for CMB measurements. Notice also from the Table \ref{tab:fit1} that compared to the sourced scalars  (with the exception of irrelevant $+$ helicity tensor 2-pt correlator), the peak of the tensor 2-pt correlator typically occurs at smaller scales in $k$ space as indicated by $x^c_{2,-} > x^c_{3,\mathcal{R}}$. On the other hand, the bulk of the constraints on the primordial bispectrum is carried by relatively small scales (\ie multipoles with $l >100$) compared to the corresponding observational window of scales ($l = 10-100$) for B-modes targeted by CMB probes.  For a sourced tensor signal occurring at $l \sim 100$, this implies that constraints from the non-observation of non-Gaussianity will be weaker, increasing the viability of the model in producing observable B-modes from secondary vector field sources. For sourced scalar and tensor signals with an appreciable offset in their respective peaks, one typically needs to carry a likelihood analysis, \ie similar to the one carried in \cite{Namba:2015gja}, to check the validity of the model when confronted with CMB data at the relevant scales.

In order to determine the level of tensor-to-scalar ratio $r$ allowed by observational limits on scalar non-Gaussianity, we will instead perform a preliminary check for the viability of the model by applying the constraints on scalar non-Gaussianity to the sourced cosmological correlators evaluated at their peaks, assuming the scales where the CMB data is relevant corresponds to the peak of these signals. Since the sourced scalar bispectrum is maximal for equilateral configurations of external momenta, we will use $f^{\rm eq}_{\rm NL}$ as an indicator of the constraints at the peak of the sourced signal and apply $2\sgm$ bound from CMB data: $ | f^{\rm eq}_{\rm NL} | < 68$ (at $k_p \simeq 0.05 \,\,{\rm Mpc^{-1}}$) \cite{Akrami:2019izv}. In particular we will impose this bound on the following expression derived in our model,
\beq\label{fnl}
f_{\mathrm{NL}}^{\mathrm{eq}}=\frac{10}{9} \frac{k^{6}}{(2 \pi)^{5 / 2}} \frac{\mathcal{B}_{\mathcal{R}}^{(s)}(k, k, k)}{\mathcal{P}_{\mathcal{R}}(k)^{2}},
\eeq
where $\mathcal{B}^{(s)}_{\mathcal{R}}$ is given by \eqref{SC} and $\mathcal{P}_\mathcal{R} = \mathcal{P}^{(v)}_\mathcal{R} + \mathcal{P}^{(s)}_\mathcal{R}$ recalling the expressions in eqs. \eqref{SC} and \eqref{PSnorm}. Using \eqref{fnl}, we present the restrictions imposed on the parameter space from $ | f^{\rm eq}_{\rm NL} | < 68$  in Figure \ref{fig:r}. From the left panel in Figure \ref{fig:r}, we observe that the bound on $ | f^{\rm eq}_{\rm NL} |$ is more restrictive than the normalization of the scalar power spectrum, as it (shown by red/brown dots) saturates on smaller $\xi_*$ on constant $\epsilon_\phi$ curves compared to the $\xi_*$ value where sourced scalar contribution becomes comparable to the vacuum counterpart (shown by red/brown dot dashed vertical lines). On the right panel of Figure \ref{fig:r}, the observational limit derived from scalar non-Gaussianity is shown by the red dotted line. We observe that in the model under investigation, a visible primordial GW spectrum with $r \simeq 10^{-2}$ can be generated without violating the bounds on CMB observations. It is important to note that, to derive the bound $ | f^{\rm eq}_{\rm NL} | < 68$ we have used \eqref{fnl} and evaluated $f_{2,\mathcal{R}}$ and $f_{3,\mathcal{R}}$ at wave numbers where the sourced contribution to the GWs peaks, \ie at $k = k_* \,x^c_{2,-}$ to properly take into account the offset between the peaks of sourced scalar and tensor fluctuations.

{\bf Tensor non-Gaussianity:} To quantify the strength of tensor non-Gaussianity, we will use the tensor analog of equilateral non-linearity parameter $f^{\rm tens}_{\rm NL}$ \cite{Ade:2015ava,Shiraishi:2019yux} 
\beq\label{fnltens}
f^{\rm tens}_{\rm NL} \equiv \fr{\mathcal{B}^{(s)}_{---}(k,k,k)}{2\sqrt{2}\,F^{\rm eq}_\mathcal{R}(k)}, 
\eeq
where we took into account a factor of $2\sqrt{2}$ that originates from the difference of our normalization convention of polarization tensors $\Pi_{ij}$ compared to the \cite{Ade:2015ava,Shiraishi:2019yux}, $F^{\rm eq}_{\mathcal{R}} \equiv \mathcal{B}^{(s)}_{\mathcal{R}}(k,k,k) / f^{\rm eq}_{\rm NL}$ which can be read from eq. \eqref{fnl}  and $\mathcal{B}^{(s)}_{---}$ is given in eq. \eqref{SC}. The constraint on the tensor bispectrum in the equilateral limit is also reported as a bound on $f^{\rm tens}_{\rm NL}$ \cite{Akrami:2019izv}:
\beq\label{tng}
-2600 < f^{\rm tens}_{\rm NL} < 3800,\quad\quad\quad\quad\quad(95 \%\, {\rm CL}, {\rm T \,only}).
\eeq
Since the tensor non-Gaussianity is negative $(f_{3,-} < 0)$, we will use the absolute value of the lower bound in \eqref{tng} to constrain the parameter space $\epsilon_\phi - \xi_*$ of the model. The resulting limits is shown by the dotted brown line in the right panel of Figure \ref{fig:r}. On the other hand, in order to determine the parameter space of the model that can be probed by tensor non-Gaussianity, we plot $\sigma (f^{\rm tens}_{\rm NL}) = 1$ line on the right panel in Figure \ref{fig:r}, which is expected to be the target sensitivity of LiteBIRD \cite{Matsumura:2013aja}. We observe that in addition to the B-modes at the level of $ r \simeq 10^{-2}$, observable tensor non-Gaussianity from vector field sources can be generated for a sizeable portion of the parameter space in our model.  From Figure \ref{fig:r} we also see that there is a small portion of the parameter space on the left of $R_t < 1$ where $r \simeq 10^{-2}$ and $f^{\rm tens}_{\rm NL} \gtrsim 1$, representing a scenario where $r_{s} < r_{v}$ with observable tensor non-Gaussianity. This regime is especially interesting because it allows us to obtain information on both quantum vacuum fluctuations of the metric and of the spectator fields (vector + axion) during inflation by combining tensor power and bispectrum. Similar to the case with tensor power spectrum, the resulting tensor bispectrum is parity violating (See \eg \cite{Cook:2013xea,Shiraishi:2013kxa}). Scanning the different regions of parameter space in our model where the roll of the $\sgm$ is faster around the cliffs, \ie $\delta >0.3$, it would also be interesting to further study the observability of parity violating tensor bispectrum. In this case, as the sourced contributions to the correlators are more spiky, an analysis similar to the one carried in \cite{Shiraishi:2016yun} is necessary to determine whether a significant signal to noise ratio for the bispectrum can be obtained or not\footnote{Private communication with Maresuke Shiraishi.}. Given the precision that will be achieved by future B-mode missions, such an investigation is particularly interesting in establishing vacuum vs. sourced nature of metric fluctuations.  

\subsection{Gravitational waves at interferometer scales}\label{S4p2}
\begin{figure}[t!]
\begin{center}
\includegraphics[scale=0.62]{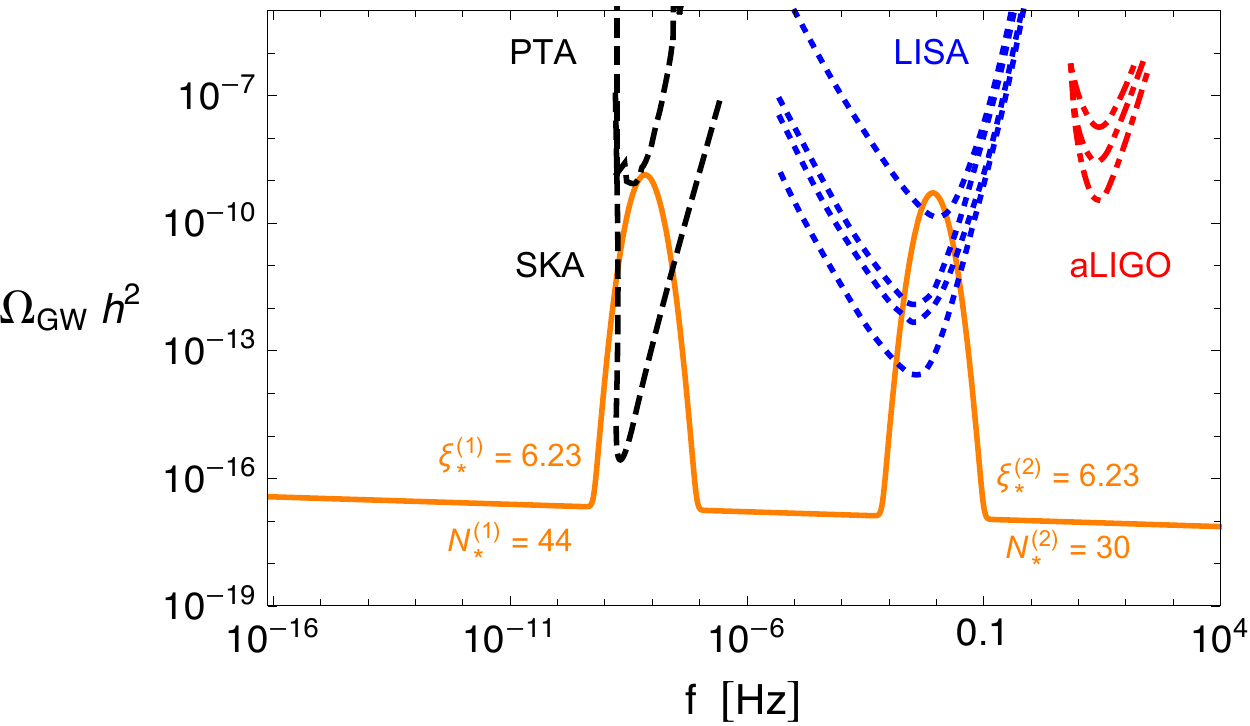}~~\includegraphics[scale=0.62]{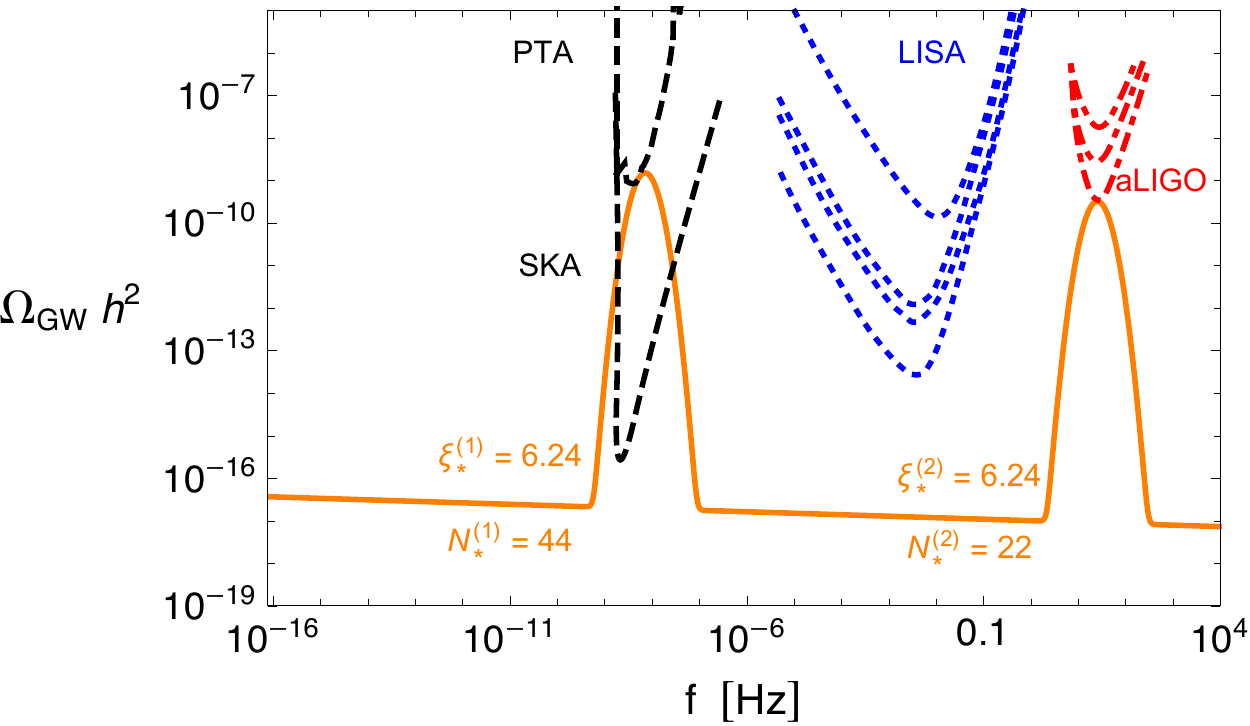}
\end{center}
\caption{ $\Omega_{\rm GW}\, h^2$ as a function of frequency in the two-field model \eqref{Lm}, where we assumed $\sgm$ field rolls through two successive cliff like regions in its potential, leading to a peaked signal at PTA-SKA and LISA scales (Left, Scenario 1) or at PTA-SKA and AdvLIGO scales (Right, Scenario 2). \label{fig:GW}}
\end{figure}

For suitable choices of initial conditions and model parameters, the model we are considering can also generate sufficiently large GW signal at interferometer scales without over producing scalars fluctuations. In this section, we will show that the model can generate observable GW signal at scales associated with PTA-SKA \cite{Arzoumanian:2015liz,Lentati:2015qwp,Shannon:2015ect}, LISA  \cite{Caprini:2015zlo,Bartolo:2016ami} and AdvLIGO \cite{TheLIGOScientific:2016wyq} experiments (see \eg Figure \ref{fig:GW}) without conflicting with the constraints on PBH abundance\footnote{For the constraints on PBH abundances we impose in this work, see Section 2 of \cite{Garcia-Bellido:2016dkw} and the references therein.} at sub-CMB scales.
\begin{table}[h!]
\begin{center}
\begin{tabular}{ l c c }
\hline
\hline

 \cellcolor[gray]{0.9}  ~~&~~    \cellcolor[gray]{0.9} 
  $f~~[~\rm Hz~]$ ~~&~~   \cellcolor[gray]{0.9}  $N_{\rm est}$\\

\hline
${\rm GW}~@~ {\rm AdvLIGO}$ ~~&~~ $~10-200$ ~~&~~ $18-22$ \\
${\rm GW}~ @~ {\rm LISA} $ ~~&~~~~~~~ $10^{-4} - 10^{-1}$ ~~&~~ $25-32$ \\
${\rm GW}~ @~ {\rm PTA}  $ ~~&~~ $~~~~~10^{-9} - 10^{-7}$ ~~&~~ $39-45$ \\
\hline
\hline
\end{tabular}
\caption{\label{tab:bparams} List of observational GW windows on inflation (Left column) and the corresponding sensitivity range in frequency, $f = k/ 2\pi$ (Middle column). Estimated number of e-folds (Right) before the end of inflation at which the corresponding scales exit the horizon, \ie $k= aH$, where we used eq. \eqref{Nvsf} assuming a constant Hubble rate during inflation and $N_{p} = 60$.\label{tab:GWW}}
\end{center}			
\end{table}

\begin{figure}[t!]
\begin{center}
\includegraphics[scale=0.6]{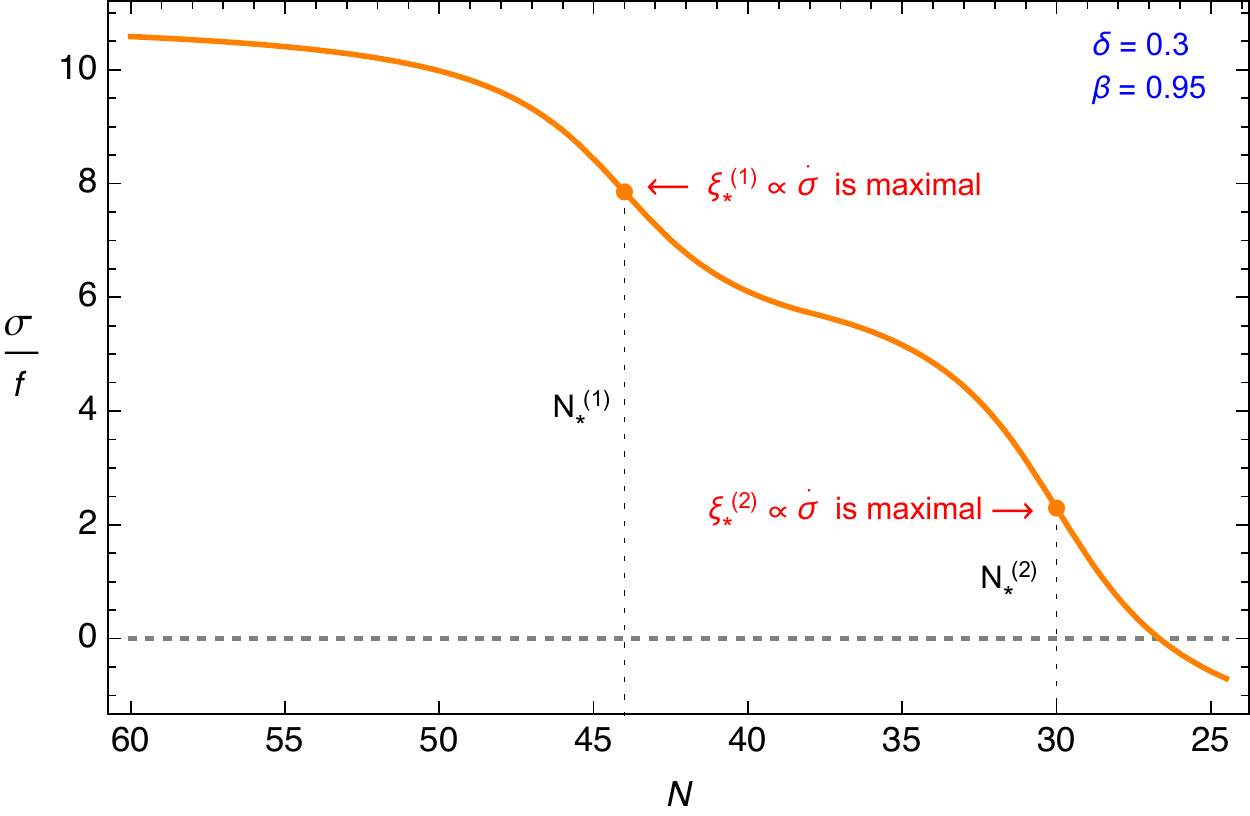}~~\includegraphics[scale=0.6]{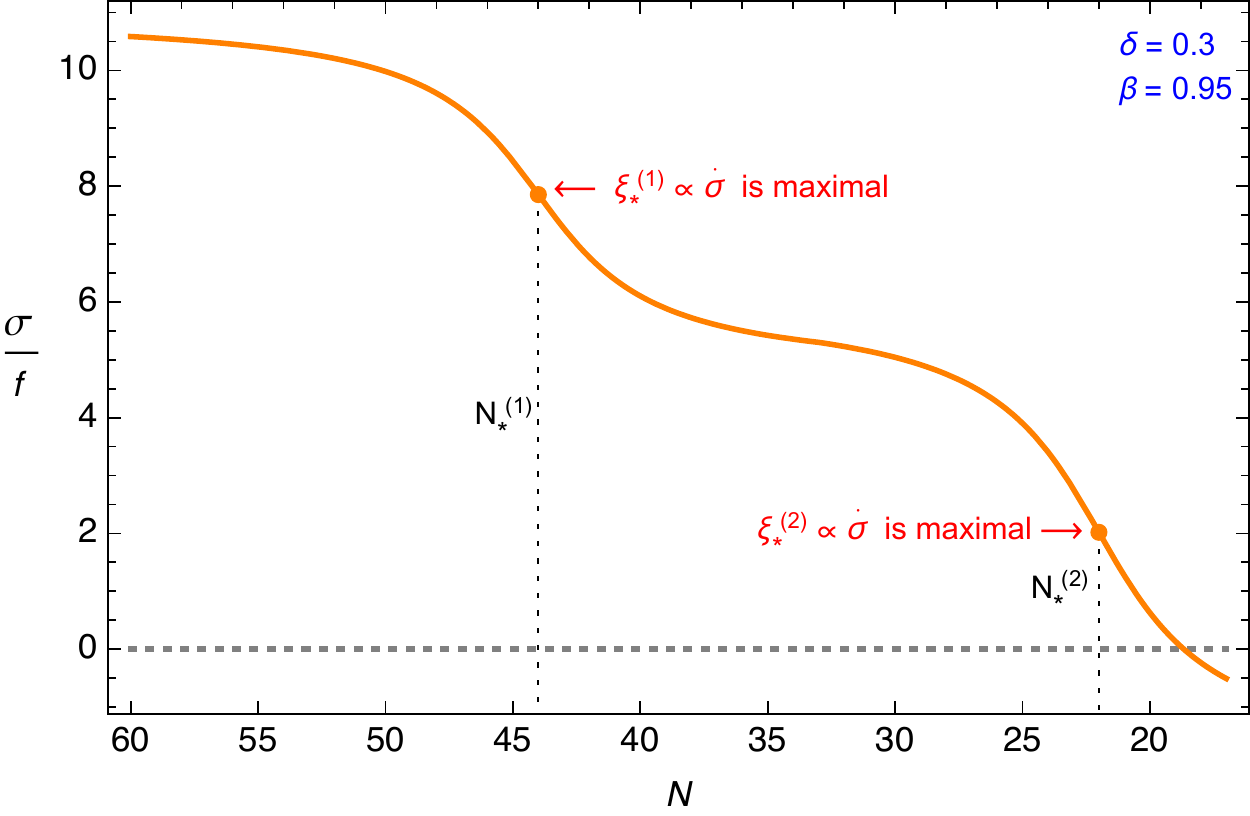}
\end{center}
\caption{ Evolution of $\sgm$ with respect to e-folds during inflation where we have used \eqref{san} to match two even branch solutions ($n=2$ and $n=0$) described in Appendix \ref{AppA}. In these plots, we choose the integration constants in \eqref{san} such that the peak in the velocity occurs at $N_*^{(1)} = 44$ and $N_*^{(2)} = 30$ (Left) and $N_*^{(1)} = 44$ and $N_*^{(2)} = 22$ (Right). For both field profiles, we assume that shortly after the field profile passes the horizontal dotted gray line $\sgm = 0$, the $\sgm$ settles to its minimum where $\dot{\sgm} \to 0$.\label{fig:2peaksgm}}
\end{figure}

\begin{figure}[t!]
\begin{center}
\includegraphics[scale=0.6]{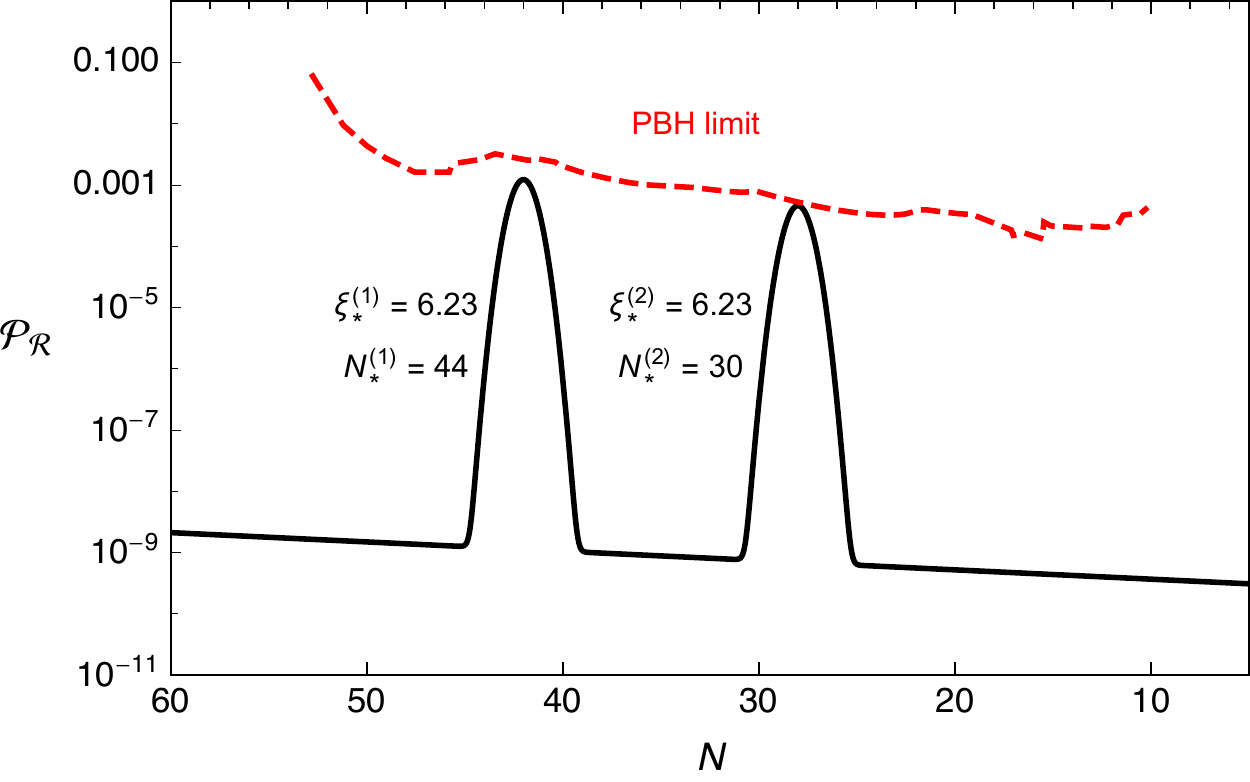}~~\includegraphics[scale=0.6]{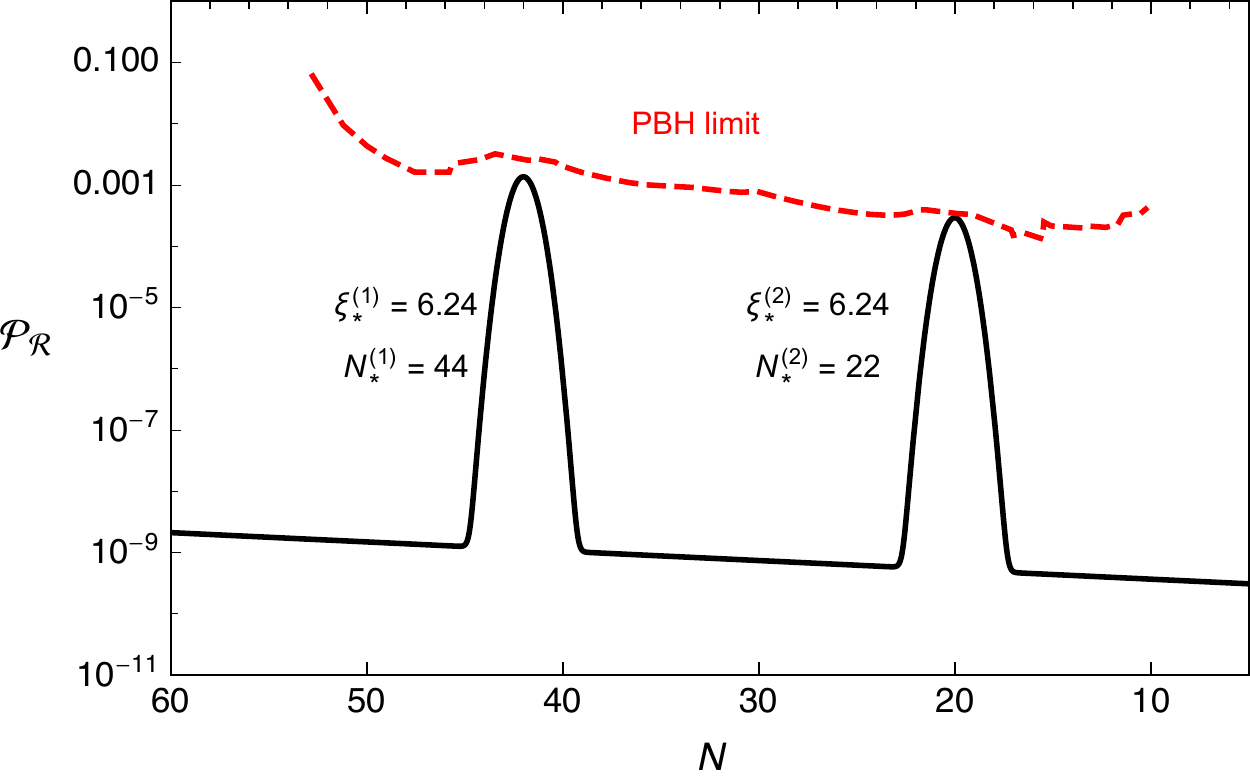}
\end{center}
\caption{ Scalar power spectrum in the two-field model \eqref{Lm} during inflation, where we assumed $\sgm$ field rolls through two successive cliff like regions in its potential, leading to a peaked signal at PTA-SKA and LISA scales (Left, Scenario 1) or at PTA-SKA and AdvLIGO scales (Right, Scenario 2). In these plots we choose a $\xi_*$ so that the peaked signal saturates the PBH bounds.      \label{fig:PBHL}}
\end{figure}

As an example, we consider two different scenarios where spectator axion $\sgm$ probe multiple cliff like regions in its potential before settling its global minimum at $\sgm =0$. For this purpose, we match two branch of solutions in eq. \eqref{san} (with $n=2$ and $n=0$) at an intermediate time. The resulting field profile(s) are shown in Figure \ref{fig:2peaksgm} in which we clearly indicated the e-folding number where the fields velocity is maximal. In these solutions, we choose the e-folds $N_*$ at which the motion of $\sgm$ is the fastest as $N^{(1)}_* = 44$ and $N^{(2)}_* = 30$ (Scenario 1) and $N^{(1)}_* = 44$ and $N^{(2)}_* = 22$ (Scenario 2) corresponding the optimal frequencies where the PTA-SKA, LISA and AdvLIGO experiments are sensitive to. In  making these choices we were guided by Table \ref{tab:GWW} and the relation between the e-folding number $N$ a given mode exits the horizon with respect ot the wave number $k = 2\pi f$ \cite{Garcia-Bellido:2016dkw}:
\beq\label{Nvsf}
N_{p} -N= 41.7 -\ln\left(\fr{k_{p}}{0.05~{\rm Mpc^{-1}}}\right)+\ln\left(\fr{f}{100~{\rm Hz}}\right)-\ln\left(\fr{H_{N}}{H_{p}}\right),
\eeq
where $N_p$ is the number of e-folds at which the pivot scale exits the horizon which we assume to be $ k_p = 0.05\, {\rm Mpc^{-1}}$. 
To study sub-CMB phenomenology in the multi-field scenarios we introduced, we need to specify the scalar potential $V_\phi(\phi)$ in the inflationary sector. Instead of fixing $V_\phi(\phi)$, we will take a phenomenological approach to determine the important set of parameters that characterize inflaton's dynamics. For this purpose, first notice that assuming the effects introduced by the rolling of $\sgm$ is negligible at CMB scales, we have $n_s -1 \simeq 2\eta_\phi -6\epsilon_\phi$ and $r \simeq 16\epsilon_\phi$. Therefore, using the results provided by CMB data, we can determine $\epsilon_\phi$ and $n_s -1$. In this regard,  we assume $r = 10^{-2}$ at CMB scales which is close to the current bound implied by Planck \cite{Akrami:2018odb}, to obtain $\epsilon_\phi \simeq 6.25 \times 10^{-4}$. On the other hand, the observed value of the spectral tilt gives $n_s -1 \simeq -0.035$ \cite{Akrami:2018odb}. 
Furthermore, neglecting higher order slow-roll parameters, we will assume that $\epsilon_\phi$ remain constant throughout the inflation. These simplifying approximations enable us the describe the resulting phenomenology at interferometer scales without affecting qualitative conclusions that can be drawn from the multi-field scenarios we described above. In light of this discussion, we note the total scalar and GW spectrum as
\begin{align}
\mathcal{P}_{\mathcal{R}} (k) &= \mathcal{P}^{(v)}_\mathcal{R}(k) + \left[ \epsilon_{\phi}\, \mathcal{P}_{\mathcal{R}}^{(v)}\left(k\right) \right]^{2} \sum_{i =1,2} f^{(i)}_{2,\mathcal{R}}(\xi_*,k/k_*,\delta),\\
\label{Pgw}\mathcal{P}_{h} (k) &= 16\epsilon_\phi \,\mathcal{P}^{(v)}_\mathcal{R}(k) + \left[ \epsilon_{\phi}\, \mathcal{P}_{\mathcal{R}}^{(v)}\left(k\right) \right]^{2} \sum_{i =1,2} f^{(i)}_{2,\mathcal{-}}(\xi_*, k/k_*, \delta),
\end{align}
where in the sourced terms above we sum over two sites of particle production. 
Noting the time dependence of Hubble parameter, $H(N)=H_{p}\, e^{-\epsilon_{\phi}\left(N_p-N\right)}$, we describe the vacuum scalar power spectrum as a function of e-folds as
\beq
\mathcal{P}^{(v)}_{\mathcal{R}}\left(k_{N}\right)=\mathcal{P}^{(v)}_{\mathcal{R}}\left(k_p\right) e^{-\left(1-\epsilon_{\phi}\right)\left(1-n_{s}\right)\left(N_p-N\right)},
\eeq
where $\mathcal{P}^{(v)}_{\mathcal{R}}\left(k_p\right) = \mathcal{A}_s = 2.1 \times 10^{-9}$. For $\delta =0.3$,  we then use the constraints\footnote{See also \cite{Domcke:2016bkh} for a discussion on the constraints that might arise through $N_{\rm eff}$. } on the scalar power spectrum from the PBH abundances\footnote{In the model we are considering, as the sourced signal originates from the convolution of two Gaussian modes and it obeys a $\chi^2$ statistics. Bounds on $\mathcal{P}_\mathcal{R}$ in this case is much stronger compared to the Gaussian fluctuatations  \cite{Lyth:2012yp,Byrnes:2012yx,Linde:2012bt}. We would like to thank Caner \"Unal for sharing the data on PBH limits.} as a function of e-folds to determine the limiting allowed value of $\xi_*$ at the peak of the sourced signal in both scenarios shown in Figure \ref{fig:2peaksgm}. The resulting peaks in the scalar power spectrum and the limiting $\xi_*$ are shown in Figure \ref{fig:PBHL}. We observe that as the PBH constraints\footnote{In Figure \ref{fig:PBHL} (Scenario 1), PBH limits  around the second peak originate from the disruption of stars by the capture of PBHs which destroy remnants like neutron stars \cite{Capela:2013yf,Capela:2014ita}. However, there are large uncertainties  \cite{Kawasaki:2016pql,Bartolo:2018rku} in these limits which opens up the possibility that  PBH abundance that can account for total DM density. In this case, the primordial GW spectrum (See \eg the left panel of Figure \ref{fig:GW}) we consider in this work will be accompanied by an induced component GW spectrum which originates from large scalar fluctuations at the horizon re-entry as first discussed in \cite{Garcia-Bellido:2017aan}. This implies that if primordial black holes $M_{\rm pbh} \sim 10^{-12} M_{\odot}$ generated by this mechanism account for the total dark matter density, LISA mission should be able detect its associated induced GW signal \cite{Garcia-Bellido:2017aan,Bartolo:2018evs}. } becomes tighter for smaller scales and the limit imposed on $\xi_*$ comes from the second peak of the sourced signal in both scenarios. We then use the limiting values of $\xi_*$ we obtained to determine the level of GW signal at the corresponding scales using
\beq\label{Omgw}
\Omega_{\rm GW}\, h^2 = \fr{\Omega_{r,0}\,h^2}{24} \mathcal{P}_{h}(k),
\eeq
where $\Omega_{r,0} \, h^2 \simeq 4.2 \times 10^{-5}$. The results are presented in Figure \ref{fig:GW} which shows that the model can simultaneously lead to an observable signal at PTA-SKA and LISA scales (Scenario 1). On the other hand, for the second scenario, the generated signal for GW's in the second bump barely overlaps with the sensitivity curve of future AdvLIGO setup.
The results we obtained so far are obtained under the assumption that i) the gauge field enhancement have negligible effects on the background dynamics ii) the calculations that leads to these results are under perturbative control. As it is clear from Figure \ref{fig:GW}, the gauge field amplification needs to be sufficiently strong to leave observable effects at interferometer scales. As a consistency check, one should therefore consider restrictions that might arise on the level of GW signal from i) and ii). In Appendix \ref{AppE}, we perform this analysis in detail to obtain the following bound in terms of the ratio $f/\Mp$:
\beq\label{ombrf}
0.07 \left(\fr{\Omega_{\rm GW}\, h^2}{10^{-9}}\right)_*^{1/4} e^{0.23\, \xi_*} <\fr{f}{\Mp} \lesssim 0.6.
\eeq
We see from \eqref{ombrf} that the spectator axion-gauge field dynamics is capable of producing visible GW signal that peaks at PTA-SKA, LISA and AdvLIGO scales as for all these probes, $\Omega_{\rm GW}\, h^2 \approx 10^{-9}$ is enough to generate an observable signal for a non-vanishing interval of $f/\Mp$, \ie for $\xi_* \lesssim 9.3$. For a GW signal at AdvLIGO scales (Scenario 2), a more demanding restriction is the constraints on PBH limits as shown by Figure \ref{fig:PBHL} (Right). In particular, the limits from PBH abundance requires $\xi_* \lesssim 6.2$, which puts the GW signal in the second scenario right on the AdvLIGO O-3 sensitivity curve \cite{TheLIGOScientific:2016wyq} (See \eg right panel of Figure \ref{fig:GW}).  
\subsection{Stringy parameter window for successful phenomenology}
In this section, we present the viable parameter space in the axion sector that leads to the observational results we derived in Sections \ref{S4p1p1} and \ref{S4p2}. For concreteness, we omit order one coefficients in the expressions we derive below by assuming $\delta\simeq \mathcal{O}(1)$ and focusing in the bumpy regime $\beta \to 1$.
From the definition \eqref{dapp} and $\beta = \Lambda^4/\mu^3 f =1$, we first note $\mu^3 \simeq H^2 f$ and $\Lambda^4 \simeq H^2 f^2$. In the following, we will use these expressions to determine the typical range of $\{\mu,\Lambda\}$ with respect to Planck scale $\Mp$. For this purpose, we can first use the normalization of the vacuum power spectrum at CMB scales using eq. \eqref{PSV} $\mathcal{P}^{(v)}_\mathcal{R} \equiv \mathcal{A}_s =  2.1 \times 10^{-9}$ to determine the Hubble rate as $H^2 \simeq \epsilon_\phi \times 10^{-7} \Mp^2$. Note that for scenarios we study in Sections \ref{S4p1p1} and \ref{S4p2}, this procedure is justified because the sourced scalar fluctuations are subdominant at CMB scales in both cases. Finally, using the bounds on $f$ from eq. \eqref{SoC}, we derived the following ranges:
\begin{align}\label{pscmb}
\nn 1.2\times 10^{-2}\, \epsilon_\phi^{1/2} & \lesssim \, \Lambda \,[M_{\mathrm pl}] \,\lesssim  7.5 \times 10^{-3} \,\,\epsilon_\phi^{1/4},\quad\quad\quad @\mathrm{CMB\,scales},\\
3.5 \times 10^{-3}\,\epsilon_\phi^{1/2} & \lesssim  \,\mu \,[M_{\mathrm pl}] \,\lesssim  2.6 \times 10^{-3}\,\, \epsilon_\phi^{1/3},
\end{align}
where we used a reference value of $\xi_* = 5$ for the CMB phenomenology we discussed in Section \ref{S4p1p1}. Similarly, for the sub-CMB phenomenology presented in Section \ref{S4p2}, we instead take $\xi_* = 6$ to obtain,
\begin{align}\label{psint}
\nn 4.5\times 10^{-2}\, \epsilon_\phi^{1/2} & \lesssim \, \Lambda \,[M_{\mathrm pl}] \,\lesssim  1.3 \times 10^{-2} \,\,\epsilon_\phi^{1/4},\quad\quad\quad @\mathrm{Interferometer\,scales},\\
 8.6 \times 10^{-3}\,\epsilon_\phi^{1/2} & \lesssim  \,\mu \,[M_{\mathrm pl}] \,\lesssim  4 \times 10^{-3}\,\, \epsilon_\phi^{1/3}.
\end{align}
On the other hand, from an effective field theory perspective, cut-off scale in the $\sgm$ sector is expected to be $M_{\rm cut} = f$ \footnote{In the $\sgm /f > 1$ regime, all interaction terms with $n>4$ will be important in the $\sgm$ sector, \ie $\mathcal{L}_{\rm int} \propto \sum_n V_{\sgm}^{(n)}(\bar{\sgm}) (\delta \sgm)^n$. Therefore the suppression scale of the most irrelevant operator ($n \to \infty$) sets the scale at which scattering amplitudes involving $\delta\sgm$ become non-unitary, implying a cut-off scale $M_{\rm cut} \simeq \lim_{n\to\infty} |V_{\sgm}^{(n)}|^{-1/(n-4)} \simeq f$  \cite{Bezrukov:2010jz}. From dimension 5 interaction \eqref{Lint}, demanding the unitarity of the 2-2 photon scattering also results with a similar cut-off: $M_{\rm cut} = 4\pi f/\alpha_{\rm c} \simeq f$ for $\alpha_{\rm c} = \mathcal{O}(10)$ values we consider in this work \cite{Burgess:2009ea}.} which is parametrically much larger than the inflationary Hubble scale $H \simeq 3 \times 10^{-4}\, \sqrt{\epsilon_\phi}\,\Mp$ considering the limits we imposed on $f$ in \eqref{SoC} for $\xi_* =\mathcal{O}(5-6)$. Therefore, the typical parameter space of the bumpy spectator axion model obeys the following hierarchy, $f \gg \Lambda \gtrsim \mu \gg H $. 

\section{Conclusions}\label{Sec5}

Forthcoming CMB experiments such as CMB-S4 \cite{Abazajian:2016yjj} and LiteBIRD \cite{Hazumi:2019lys} will measure the CMB B-mode polarization and its properties to an unprecedented accuracy. Given the expected improvements in the sensitivity of B-mode measurements, it is therefore important to explore alternative mechanisms to the standard scenario where GWs are produced through the enhancement of quantum vacuum fluctuations during inflation.

In this work, we have shown that the motion of a hidden sector axion-like field $\sgm$ in its wiggly potential (\ie eq. \eqref{Vs} with $\Lambda^4 \lesssim \mu^3 f$) can experience transient, relatively fast roll(s) (compared to smooth slow-roll) that can lead to significant amplification of gauge field fluctuations which in turn produce an additional component of tensor fluctuations whose amplitude is not proportional to the Hubble rate during inflation. In particular, this implies that, if the transient speeding up of $\sgm$ occurs while CMB scales leave the horizon, the model can generate an observable GW signal of primordial origin for an arbitrarily low energy scale of inflation (See \eg eq. \eqref{rpeak}) while respecting the limits on scalar non-Gaussianity at CMB scales (See \eg Figure \ref{fig:r}). 

The model we consider features a rich set of phenomenological signatures. First and foremost, the produced tensor fluctuations can exhibit strong scale dependence which could lead to a locally blue tilt for the tensor power spectrum (similar to the models studied in \cite{Mukohyama:2014gba,Namba:2015gja}). It is crucial to note that such a situation is typically considered as a smoking gun evidence to falsify inflationary paradigm. Moreover, at the peak of the signal, tensor power spectrum has a negative tilt which can be measured if the B-modes are observed for a range of CMB scales. Finally, the induced GWs has a significant departure from Gaussianity which can be detected by future CMB missions like LiteBIRD (See \eg Figure \ref{fig:r}). Importantly, together with the B-mode measurements, an observation of tensor non-Gaussianity would allow us to unambiguously determine vacuum vs non-vacuum nature of metric fluctuations as the vacuum part is expected to be nearly Gaussian. In other words, a departure from the standard vanilla scenario -- such as near scale invariance, near Gaussianity and parity invariance of tensor fluctuations-- will enable us to constraint the energy density contained in the hidden sector (See also \cite{Shiraishi:2016yun,Agrawal:2017awz,Agrawal:2018mrg}).

In addition to observable B-modes at CMB scales, we have shown that for a suitable choice of initial conditions and model parameters, the roll of $\sgm$ in its wiggly potential can result with significant enhancement of GWs on sub-CMB scales that can be detected at ground and spatial based interferometers. In particular, as an interesting application of our model, we showed that if the spectator axion $\sgm$ probes multiple cliff like regions of its potential during inflation, observably large GW signals can be generated both at scales probed by PTA-SKA and LISA missions without violating bounds from PBH abundance at the aforementioned scales. On the other hand, we found that the model (Scenario 2) can generate a GW signal right on the edge of the AdvLIGO O-3 sensitivity line \cite{TheLIGOScientific:2016wyq} while being consistent with bounds on the scalar fluctuations at those scales.   

In the model we studied in this paper, there remain several open problems to be investigated. The parity violating nature of particle production is expected to induce a characteristic mixed type 2-pt correlator (TB) \cite{Lue:1998mq} which might be detectable if the amplitude of sourced metric fluctuations is large enough \cite{Saito:2007kt,Gluscevic:2010vv,Namba:2015gja}. On the other hand, we have studied the phenomenology of the model only for a single choice of parameter $\delta$, which is proportional to axion mass in its global minimum. As we mentioned earlier in Section \ref{S4p1p1}, increasing $\delta$ reduces the amplitude of sourced scalar fluctuations more compared to the tensors and therefore a large sourced tensor component can be produced by increasing the axion mass (a similar situation appears in \cite{Namba:2015gja}). This situation is likely to give rise to larger TB and B-mode auto bispectrum (BBB) which could be detected by a Planck-like mission. We leave a comprehensive analysis on these interesting issues for future work. Finally, the axion fluctuations may result in a perturbation in the effective coupling $\delta \xi$ and since the amount of GWs sourced in this mechanism is controlled by $\xi$, this may lead to a scale dependent anisotropies in the GW signal at interferometer scales \cite{Bartolo:2019oiq}. We leave a detailed investigation of this matter for future analysis.  
\acknowledgments
I would like to thank Caner \"Unal for interesting discussions and especially for his help on numerics in the initial stages of this project. It is also a pleasure to thank Nicola Bartolo, Maresuke Shiraishi  and Scott Watson for useful conversations and comments pertaining to this work. The author acknowledges support by National Science Centre, Poland OPUS project 2017/27/B/ST2/02531, the European Structural and Investment Funds and the Czech Ministry of Education, Youth and Sports (Project CoGraDS-CZ.02.1.01/0.0/0.0/15003/0000437).
\begin{appendix}
\section{Background evolution of $\sgm$ and vector field production \label{AppA}}
Assuming a constant Hubble rate $H$, in the slow-roll regime, Klein-Gordon equation for the homogeneous background of the spectator $\sgm$ can be approximated as
\begin{align}\label{eomsapp}
\bar{\sgm}'(z)+ \bigg[ 1 +\beta \cos \left( \bar{\sgm}(z) \right) \bigg] = 0
\end{align}
where we defined a new time variable $\d z =  \mu^3/ (3H^2f) \d N$ with prime denotes differentiation with respect to the arguments, $\beta = \Lambda^4 /(\mu^3 f)$ and $\bar{\sgm} = \sgm/f -\pi/2$. Notice that the equation \eqref{eomsapp} is invariant under the discrete shift symmetry $\bar{\sgm} \to \bar{\sgm} + 2\pi n$ for arbitrary integer $n$. This implies that we can study the solution for \eqref{eomsapp} for any $2\pi$ interval in field space and the remaining regions of the solution can be found using the periodicity of the eq. \eqref{eomsapp}. For this purpose, we make a field redefinition to study the evolution of the scalar field within such an interval, \ie for even $n$, we define $\bar{\sgm}(z) = n\pi + 2 \arctan[y(z)]$ where the new variable $y(z)$ obeys,
\beq\label{eqy}
y'(z) + \fr{1}{2} \bigg[1 + \beta + (1-\beta) y(z)^2\bigg] = 0,
\eeq
and has the following solution:
\beq\label{y}
y(z) = \sqrt{\fr{1+ \beta}{1 - \beta}}\tan \left[\fr{\sqrt{1-\beta^2}}{2}(z_*-z)\right],
\eeq
where $z_*= \mu^3/ (3H^2f) N_*$ is an integration constant. Since we are interested in the bumpy regime, \ie $\beta \to 1$, one can further simplify the solution for $y$, which in turn simplifies the solution of the field profile as
\beq\label{san}
\sgm(N) = \left(n+\fr{1}{2}\right)\pi f + 2 f \arctan\left[\delta (N_* - N)\right],
\eeq
where we defined the following dimensionless parameter in terms of constant physical scales of the model: 
\beq\label{dapp}
\delta \equiv \fr{(1 + \beta)}{2} \fr{\mu^3}{3 H^2 f} = \left(1 + \fr{\Lambda^4}{\mu^3 f}\right) \fr{\mu^3}{6 H^2 f}.
\eeq

\begin{figure}[t!]
\begin{center}
\includegraphics[scale=0.61]{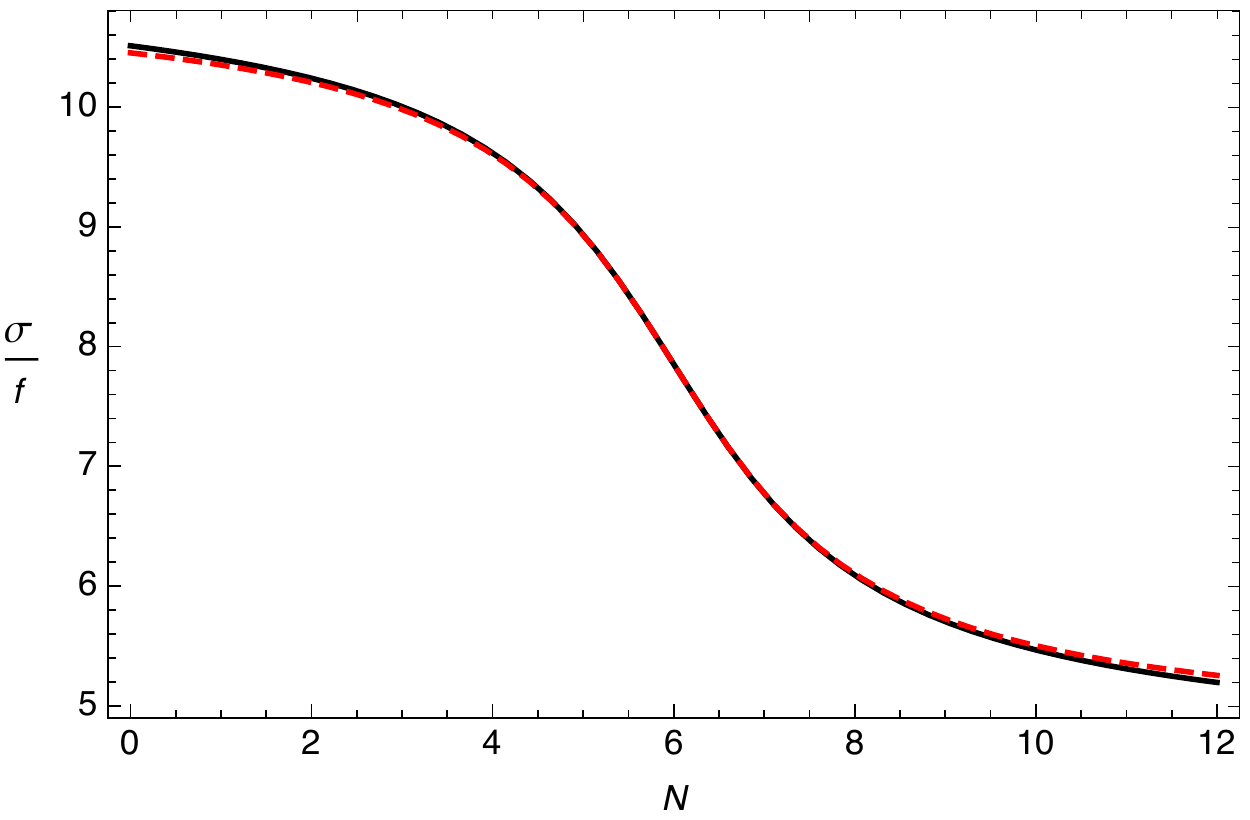}~\includegraphics[scale=0.64]{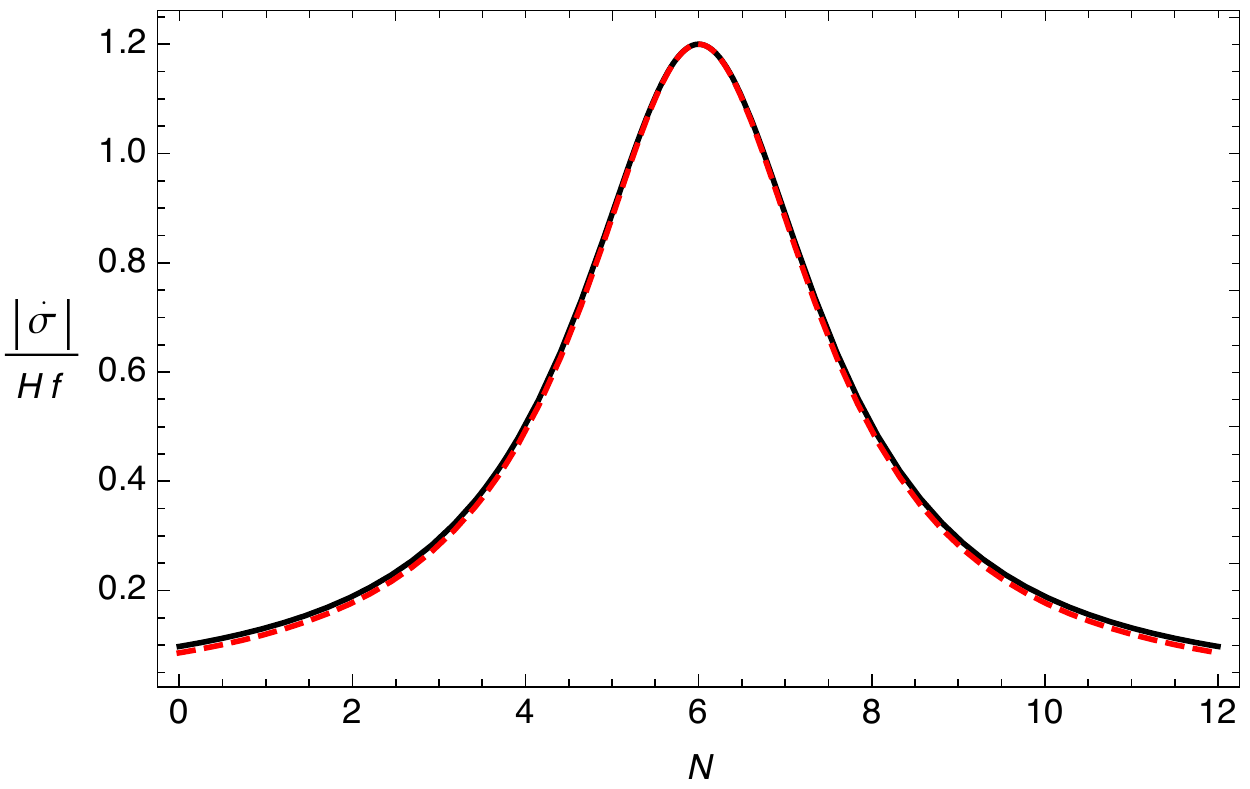}
\end{center}
\caption{Field profile $\sigma$ and the velocity $|\dot{\sgm}|$ as a function of e-folds $N$. In these plots, we have used $\delta= 0.6$, $\beta=\Lambda^4 / (\mu^3 f) =0.95$, $n = 2$ and $N_* = 6$. In both plots, dashed curves indicate the resulting profiles obtained via \eqref{san}.  \label{fig:sapp}}
\end{figure} 
In Figure \ref{fig:sapp}, the background evolution of axion within one bump of its potential is illustrated using the simplified profile \eqref{san} in comparison with the respective profile one can obtain using the full expression \eqref{y}. The accuracy of the approximation given by the solution (red dashed lines) in \eqref{san} is clearly visible. On the other hand, the expressions we derived so far assumes the slow-roll approximation:
\beq\label{dds}
\bigg|\fr{\ddot{\sgm}}{3H\dot{\sgm}}\bigg| = - \fr{2\delta^2 \Delta N }{3 \left[ 1+\delta^2  \Delta N^2 \right]}\ll 1,
\eeq
where we used \eqref{san} and defined $\Delta N = N-N_*$ with $N_*$ denoting the e-fold number where $\dot{\sgm}$ reaches its maximum value. From \eqref{dds} we realize that the slow-roll approximation is guaranteed both at the time $\dot{\sigma}$ peaks, \ie $\Delta N \to 0$ and asymptotically as $\Delta N \to \infty$. At intermediate times denoted by  $\Delta N = \mp \delta^{-1}$ however, it reaches its maximal values $\ddot{\sgm}/{(3H\dot{\sgm})} = \pm {\delta}/{3}$.
Therefore to ensure the slow-roll approximation and the validity of the formulas we derived so far, we require $\delta < 1$. 

{\bf Gauge field production:}
Now, we study gauge field amplification during the rollover of $\sgm$ through a single cliff like region of its potential which then can be identically used for scenarios where $\sgm$ traverses multiple cliff like regions during its entire evolution (See \eg Section \ref{S4p2}). We start our analysis from eq. \eqref{MEA2}, 
which describes a Schr\"odinger equation of the ``wave-function" $A_-$. Analytic solutions for $A_{-}$ can be derived by employing WKB approximation methods to study the analogous 1-D scattering problem in an effective potential that \eqref{MEA2} represents. For the sake of completeness, below we will outline the main steps of this procedure closely following \cite{QM,Namba:2015gja}. We begin by labeling the effective potential inside the brackets in eq. \eqref{MEA2} as $V_{\rm eff}(x)$ and note that there is a critical value (\ie turning point) of $x = x_{\rm c}$ where $V_{\rm eff}(x)$ vanishes, \ie  $V_{\rm eff}(x_{\rm c}) = 0$. On the other hand for on the opposite sides of $x_{\rm c}$, we define
\beq\label{Q}
V_{\rm eff}(x)=
 \begin{dcases} 
       \,\,\,\,p(x)^2, & x > x_{\rm c} \,\\
        -\kappa(x)^2, & x < x_{\rm c}  \,,
   \end{dcases}
\eeq 
where $p(x) = -i \kappa(x) = -i \sqrt{{2\xi_{*}}/{(x(1+\ln[(x_*/x )^{\delta})]^2))}-1}$. Sufficiently far away from the critical point $| x - x_{\rm c}| \gg 1$, the effective potential is varying adiabatically where the conditions $|p'(x)| \ll p(x)^2$ and $|\kappa'(x)| \ll \kappa(x)^2$ are satisfied. We label these asymptotic regimes as IN ($x \gg x_{\rm c}$) and OUT ($x \ll x_{\rm c}$) and write the general solutions for $A_-$ in these regimes as
\beq\label{Ain}
A_{-}(x\gg x_{\rm c}) \simeq \frac{\alpha}{\sqrt{p(x)}} \exp \left(-i\int_{x_{\rm c}}^{x} p(x') ~\d x'\right)+\frac{\beta}{\sqrt{p(x)}} \sin \left(\int_{x_{\rm c}}^{x} p(x') ~\d x'\right),
\eeq
and
\beq\label{Aout}
A_{-}(x\ll x_{\rm c}) \simeq \frac{\tilde{\alpha}}{\sqrt{\kappa(x)}} \exp \left(-\int_{x}^{x_{\rm c}} \kappa(x') ~\d x'\right)+\frac{\tilde{\beta}}{\sqrt{\kappa(x)}} \exp \left(\int_{x}^{x_{\rm c}} \kappa(x') ~\d x'\right),
\eeq
where $\alpha,\beta,\tilde{\alpha},\tilde{\beta}$ are complex constants. To obtain an expression for $A_{-}$ in the OUT region, we need to determine the coefficients $\tilde{\alpha},\tilde{\beta}$ in terms of the $\alpha,\beta$ of the IN state. For this purpose, we will match the solutions in these asymptotic regions with a solution obtained in the intermediate phase, \ie around the turning point $x = x_{\rm_c}$ where the WKB approximation ceases to hold. In this regime we can linearize the effective potential around the critical point $x = x_{\rm_c}$ to write the mode equation as $
A_{-}''(x) + V_{\rm eff}'(x_{\rm c})\,(x-x_{\rm c})\, A_{-}(x) \simeq 0$. 
This equation has two independent solutions in terms of Airy functions, ${\rm Ai}$ and ${\rm Bi}$ and by comparing their asymptotic behavior in the $x - x_{\rm_c} \to \pm \infty$ limits with the IN (eq. \eqref{Ain}) and OUT (eq. \eqref{Aout}) solutions, the WKB solutions on the opposite sides of the turning point $x = x_{\rm_c}$ can be described as \cite{QM},
\beq\label{Ainf}
A_{-}(x > x_{\rm c}) \simeq \frac{\alpha}{\sqrt{p(x)}} \cos \left(\int_{x_{\rm c}}^{x} p(x') ~\d x'-\frac{\pi}{4}\right)-\frac{\beta}{\sqrt{p(x)}} \sin \left(\int_{x_{\rm c}}^{x} p(x') ~\d x'-\frac{\pi}{4}\right),
\eeq
and
\beq\label{Aoutf}
A_{-}(x < x_{\rm c}) \simeq \frac{\alpha / 2}{\sqrt{\kappa(x)}} \exp \left(-\int_{x}^{x_{\rm c}} \kappa(x') ~\d x'\right)+\frac{\beta}{\sqrt{\kappa(x)}} \exp \left(\int_{x}^{x_{\rm c}} \kappa(x') ~\d x'\right).
\eeq
The complex coefficients of the solution (\eg eq. \eqref{Ainf})) in the IN region $x >  x_{\rm c}$ can be fixed by the requirement that the vector field modes are in their Bunch-Davies vacuum in the far past, implying $\alpha = 1/\sqrt{2k}$ and $\beta = -i/\sqrt{2k}$. 
To obtain an analytic expression for the mode functions in the OUT region, the final ingredient we require is the integrals that appear as arguments of the exponentials in eq. \eqref{Aoutf}. For a general set of model parameters, it is not possible to compute these integrals analytically to obtain definite expression for $A_-$. 
However, $x = -k\tau$ dependence of $A_-$ can be determined by noting
\beq\label{intkappa}
\int_{x}^{x_{\rm c}} \kappa(x')~ \d x' = \int_{0}^{x_{\rm c}} \kappa(x')~ \d x' - \int_{0}^{x} \kappa(x')~ \d x'
\eeq
so that only the second term has $x$ dependence. In the limit of sufficiently small $x$, we can simplify the integrand in the second integral of \eqref{intkappa} to get
\beq\label{intkappa2}
\int_{0}^{x} \kappa(x')~ \d x'  \simeq \int_{0}^{x} \sqrt{\fr{2\xi_*}{x'}}\fr{1}{\delta |\ln(x'/x_*)|}~ \d x'\underset{x\to0}{\simeq}\fr{2\sqrt{2\xi_* x} }{\delta |\ln(x/x_*)| }.
\eeq
We parametrize our ignorance of the first term in eq. \eqref{intkappa} by defining the normalization factor $\exp \left( \int_0^{x_{\rm c}} \kappa(x') \d x' \right) \equiv N(\xi_*,x_*,\delta)$. Therefore, recalling \eqref{Aoutf}, we have
\begin{align}\label{wkbsol}
\nn A_{-}(\tau,\vec{k}) &= A_{-}^{R} + i A_{-}^{I},\\ \nn
&= \fr{1}{\sqrt{2k}} \left(\fr{-k\tau}{2\xi(\tau)}\right)^{1/4}\Bigg\{ N(\xi_*,x_*,\delta) \exp \left[- \fr{2\sqrt{2\xi_*}~ (-k\tau)^{1/2} }{\delta |\ln(\tau/\tau_*)|}\right]\\
&\quad\quad\quad\quad\quad\quad\quad\quad\quad\quad+ \fr{i}{2N(\xi_*,x_*,\delta)}  \exp \left[ \fr{2\sqrt{2\xi_*}~ (-k\tau)^{1/2} }{\delta |\ln(\tau/\tau_*)|}\right]\Bigg\}~~~~~ \tau/\tau_* \ll 1,
\end{align}
where we chose an arbitrary initial phase to ensure that the growing part (first term in eq.  \eqref{wkbsol}) of the solution is real at late times $\tau/\tau_*\to 0$. To determine the normalization factors $N(\xi_*, x_*,\delta )$ we numerically solve \eqref{MEA2} and matching it to the WKB solution \eqref{wkbsol} at late times $-k\tau \ll 1$. In this way, we found that $N(\xi_*, x_*,\delta )$ can be accurately described by a log-normal shape,
\beq\label{Nform}
N\left(\xi_{*}, x_*, \delta\right) \simeq N^{c}\left[\xi_{*}, \delta\right] \exp \left(-\frac{1}{2 \sigma^{2}\left[\xi_{*}, \delta\right]} \ln ^{2}\left(\frac{x_*}{q^{c}\left[\xi_{*}, \delta\right]}\right)\right),
\eeq
where the functions $N^{c}, q^c$ and $\sgm$ is characterized by the background evolution of $\sgm$ and hence depend on $\xi_*$ and $\delta$. For an effective coupling to gauge fields within the range $3 \leq \xi_* \leq 6.5$, we found that these functions can be described accurately by a second order polynomial in $\xi_*$:
\begin{align}\label{fitN}
\nn N^{c} [\xi_*,\delta]&=\exp \left[0.325+2.72 \,\xi_{*}-0.00069 \,\xi_{*}^{2}\right], \quad\quad \delta=0.3, \quad 3 \leq \xi_{*} \leq 6.5, \\\nn
q^{c} [\xi_*,\delta]&= 0.013 + 0.710 \,\xi_{*}-0.00105\, \xi_{*}^{2},\\
\sigma [\xi_*,\delta]&=1.69-0.254\, \xi_{*} + 0.0164\, \xi_{*}^{2}.
\end{align} 
Armed with the knowledge of normalization factors, we verified numerically that the real part $A_{-}^{R}$ of eq. \eqref{wkbsol} dominates over $A_{-}^{R}$ in the late time limit $\tau/\tau_* \ll 1$. In this context, we would like to stress that the real part of the solution in eq. \eqref{wkbsol} represents the growing solution of $A_-$ and describes the physical amplification of the negative helicity gauge mode due to its coupling to $\sigma$. For the computation of cosmological correlators in this work (See \eg Appendices \ref{AppB} and \ref{AppC}), we will therefore drop the imaginary part of the gauge field mode functions, $A_- \simeq A_-^R$. On the other hand, $A_{-}^{I}$ in eq. \eqref{wkbsol} corresponds to the decaying mode and ensures that the Wronskian condition $A_{-}A^{\prime\,*}_{-} - c.c = i$ is satisfied. As we will show in detail in Appendix \eqref{AppD}, the presence of $A_{-}^{I}$ is important for understanding the smooth connection between the full solution in eq. \eqref{wkbsol} in the $\tau/\tau_* < 1$ regime with the UV solution of gauge modes (eq. \eqref{Ainf}) in the far past: 
\beq\label{wkbsolIN}
A_{-} \simeq \frac{1}{\sqrt{2k\,p(x)}} \exp \left(i\int_{x_{\rm c}}^{x} p(x') ~\d x'-i\frac{\pi}{4}\right)\propto \frac{1}{\sqrt{2k}} e^{-i k \tau} \quad\quad\quad \tau/\tau_* \gg 1,
\eeq
where we have used the values of the complex coefficients fixed by the adiabatic vacuum condition: $\alpha = 1/\sqrt{2k}$ and $\beta = -i/\sqrt{2k}$ and the fact that $p(x)\to 1$ for $ \tau/\tau_* \gg 1$. 

\section{Tensor correlators sourced by vector fields}\label{AppB}
In this Appendix, we will derive the tensor 2-pt and 3-pt correlators in the presence of gauge fields sources. For this purpose, we use \eqref{tdc} to note the following relation between the tensor mode operators $\hat{h}^{(s)}_\lambda$ and the canonical mode $\hat{Q}_\lambda$:
\beq\label{rqtoh}
\hat{h}_\lambda (\tau, k) = \Pi_{ij, \lambda} (\vec{k}) ~\hat{h}_{ij} (\tau, \vec{k}) = \fr{2}{\Mp a(\tau)} \hat{Q}_\lambda(\tau, \vec{k}).
\eeq
The equation of motion for the canonical operator  $\hat{Q}_\lambda$ is given in \eqref{ctme}. As we mentioned in the main text, we decompose the full solution into a homogeneous and particular one, corresponding to the modes generated by the vacuum and sourced fluctuations. The solution to the vacuum configuration can be approximated by the expression \eqref{VM} and we solve for the sourced contribution $\hat{Q}^{(s)}_\lambda$ using \eqref{QSL}.
Noting the relation \eqref{rqtoh}, we therefore have
\beq\label{sh}
\hat{h}^{(s)}_\lambda (0, k) = -\fr{2 H \tau}{\Mp} \int_{-\infty}^{0} \d\tau'~ G_k(\tau,\tau')~ \hat{J}_\lambda(\tau',\vec{k}),
\eeq
where the Green's function is given by \eqref{gf} and the source $\hat{J}_\lambda$ is defined in \eqref{ts}. Using the definitions \eqref{EBF} of vector fields, we can obtain an explicit expression for the source as
\begin{align}\label{JTs}
\nn\hat{J}_\lambda(\tau,\vec{k}) &\simeq-\fr{H}{\Mp} \sqrt{\fr{-\tau \xi(\tau)}{2}} \int \frac{\d^{3} p}{(2 \pi)^{3 / 2}} \,\epsilon_{\lambda}\left[\vec{k}, \vec{k}-\vec{p},\vec{p}\right]\, p^{1 / 4}\,|\vec{k}-\vec{p}|^{1 / 4}\left(1+\frac{-\tau}{2 \xi(\tau)} \sqrt{p|\vec{k}-\vec{p}|}\right) \\\nn\\
&\quad\quad\quad\quad\quad\quad\quad\quad\quad \times \tilde{A}(\tau,|\vec{k}-\vec{p}|)\, \tilde{A}(\tau, p)\, \hat{\mathcal{O}}_{-}(\vec{k}-\vec{p})\, \hat{\mathcal{O}}_{-}(\vec{p}),
\end{align}
where we defined the following shorthand notation for expressions involving annihilation and creation operators, 
\beq\label{Op}
\hat{\mathcal{O}}_\lambda(\vec{q}) \equiv \left[\hat{a}_{\lambda}(\vec{q})+\hat{a}_{\lambda}^{\dagger}(-\vec{q})\right],
\eeq
and for the products involving helicity vectors: 
\beq\label{eldef}
\epsilon_{\lambda} \left[ \vec{k}, \vec{k}-\vec{p},\vec{p}  \right] \equiv \epsilon_i^{\lambda} (\vec{k})^{*}\,  \epsilon_i^{-} (\vec{k}-\vec{p})\,  \epsilon_j^{\lambda} (\vec{k})^{*} \, \epsilon_j^{-} (\vec{p}).
\eeq
As we are interested in the phenomenology of tensor modes on super-horizon scales, $-k\tau \ll 1$, we can approximate the Green's function in \eqref{sh} as
\beq\label{gfapp}
G_{k}\left(\tau, \tau^{\prime}\right) \simeq \Theta\left(\tau-\tau^{\prime}\right) \sqrt{\fr{\pi}{2}} \fr{\sqrt{\tau \tau'}}{(-k\tau)^{3/2}} J_{3/2}(-k\tau')= \frac{\Theta\left(\tau-\tau^{\prime}\right)}{k^{3} \tau \tau^{\prime}}\left[k \tau^{\prime} \cos \left(k \tau^{\prime}\right)-\sin \left(k \tau^{\prime}\right)\right], \,\, -k \tau \ll 1.
\eeq
Combining \eqref{sh} with \eqref{JTs} and noting the approximation \eqref{gfapp}, we obtain
\begin{align}\label{hTs}
\nn\hat{h}^{(s)}_\lambda (0, k) &\simeq \sqrt{\fr{2}{k^{7}}}\left(\fr{H}{\Mp}\right)^2  \int \frac{\d^{3} p}{(2 \pi)^{3 / 2}} \epsilon_{\lambda}\left[\vec{k},\vec{k}-\vec{p}, \vec{p}\right]\, p^{1 / 4}\,|\vec{k}-\vec{p}|^{1 / 4} N\big(\xi_*, -|\vec{k}-\vec{p}|\tau_*,\delta\big)\\
&\quad\quad\quad\quad\quad\quad\quad\quad \times  N\big(\xi_*, -p\tau_*,\delta\big)\mathcal{I}\bigg[\xi_*,x_*,\delta,\fr{|\vec{k}-\vec{p}|}{k},\fr{p}{k}\bigg]\, \hat{\mathcal{O}}_{-}(\vec{k}-\vec{p})\, \hat{\mathcal{O}}_{-}(\vec{p}),
\end{align}
where we defined
\beq
\mathcal{I}\bigg[\xi_{*}, x_{*}, \delta, \tilde{p}, \tilde{q}\bigg] \equiv \mathcal{I}_{1}\bigg[\xi_{*}, x_{*}, \delta, \sqrt{\tilde{p}}+\sqrt{\tilde{q}}\bigg]+\frac{\sqrt{\tilde{p} \tilde{q}}}{2} \mathcal{I}_{2}\bigg[\xi_{*}, x_{*}, \delta, \sqrt{\tilde{p}}+\sqrt{\tilde{q}}\bigg]
\eeq
with $\mathcal{I}_1$ and $\mathcal{I}_2$ representing the time integral of the gauge field sources. They are defined as
\begin{align}
&\mathcal{I}_{1}\bigg[\xi_{*}, x_{*}, \delta, Q\bigg] \equiv \int_{0}^{\infty} \d x^{\prime}\left(x^{\prime} \cos x^{\prime}-\sin x^{\prime}\right) \sqrt{\frac{\xi\left(x^{\prime}\right)}{x^{\prime}}} \exp \left[-\frac{2 \sqrt{2\xi_{*}}}{\delta}\frac{x'^{1/2}}{|\ln(x'/x_*)\,|} Q\right]\\
&\mathcal{I}_{2}\bigg[\xi_{*}, x_{*}, \delta, Q\bigg] \equiv \int_{0}^{\infty} \d x^{\prime}\left(x^{\prime} \cos x^{\prime}-\sin x^{\prime}\right) \sqrt{\frac{x^{\prime}}{\xi\left(x^{\prime}\right)}} \exp \left[-\frac{2 \sqrt{2\xi_{*}}}{\delta}\frac{x'^{1/2}}{|\ln(x'/x_*)\,|} Q\right],
\end{align}
where $x' = -k \tau'$ and $x_* = - k\tau_*$ denotes the ratio of the physical momentum to the horizon side at the time when $\xi$ reaches its peak value $\xi_* = \alpha_{\rm c} \delta$. 
\\\noindent
{\bf Power Spectrum:} For each polarization, we define the total power spectrum as
\beq\label{DTPS}
\frac{k^{3}}{2 \pi^{2}}\left\langle\hat{h}_{\lambda}(0, \vec{k}) \hat{h}_{\lambda^{\prime}}(0, \vec{k}^{\prime})\right\rangle \equiv \delta_{\lambda \lambda^{\prime}}\, \delta \left(\vec{k}+\vec{k}^{\prime}\right) \,  \mathcal{P}_{\lambda}(k).
\eeq
Since the vacuum and the sourced mode are statistically uncorrelated, we separate the total power spectrum as $  \mathcal{P}_{\lambda}(k) 
=  \mathcal{P}^{(v)}_{\lambda}(k) +  \mathcal{P}^{(s)}_{\lambda}(k)$ where $ \mathcal{P}^{(v)}_{\lambda}(k) = H^2/\pi^2\Mp^2 $ using \eqref{VM}. For the sourced power spectrum, we use \eqref{hTs} in \eqref{DTPS} to obtain
\begin{align}\label{stps}
\nn \mathcal{P}_{\lambda}^{(s)}(k) \simeq & \frac{H^{4}}{8 \pi^{2} \Mp^{4} k^{4}} \int \frac{\d^{3} p}{(2 \pi)^{3}}(1-\lambda \hat{k} \cdot \hat{p})^{2}\left(1-\lambda \hat{k} \cdot \frac{\vec{k}-\vec{p}}{|\vec{k}-\vec{p}|}\right)^{2} \sqrt{p|\vec{k}-\vec{p}|} \\
&\quad\quad\quad\quad\quad\times N^2\bigg(\xi_*,-|\vec{k}-\vec{p}|\tau_*,\delta\bigg)N^2\bigg(\xi_*, -p\tau_*,\delta\bigg) \mathcal{I}^{2}\left[\xi_{*}, x_{*}, \delta, \frac{p}{k}, \frac{|\vec{k}-\vec{p}|}{k}\right],
\end{align}
where we have used the Wick's theorem to evaluate the correlators of the operators $\hat{\mathcal{O}}_{-}$ and used the following identity
\beq
\int \d \phi\,\, \epsilon_{\lambda}(\vec{k}, \vec{p},\vec{q}) \,\epsilon^{*}_{\lambda'}(\vec{k}, \vec{p},\vec{q})=\frac{\delta_{\lambda \lambda^{\prime}}}{16} \int \d \phi\, (1-\lambda \hat{k} \cdot \hat{p})^{2}(1-\lambda \hat{k} \cdot \hat{q})^{2}.
\eeq
To evaluate the integral over momentum in \eqref{stps}, we define $\tilde{p} = p/k$ and denote the cosine angle between $\hat{k}$ and $\hat{p}$ as $\eta$. In this way, we arrive at the final expression
\beq
\mathcal{P}_{\lambda}^{(s)}(k) \simeq \fr{H^4}{64 \pi^2 \Mp^4}\, f_{2,\lambda} (\xi_*,x_*,\delta),
\eeq
where 
\begin{align}\label{f2l}
\nn  f_{2,\lambda} (\xi_*,x_*,\delta) &=2 \int_{0}^{\infty} \d \tilde{p} \int_{-1}^{1} \d \eta \, \frac{\tilde{p}^{5 / 2}(1-\lambda \eta)^{2}\left[ \sqrt{1-2 \tilde{p} \eta+\tilde{p}^{2}}-\lambda (1-\tilde{p} \eta )\right]^{2}}{\left(1-2 \tilde{p} \eta+\tilde{p}^{2}\right)^{3 / 4}} \\
&\quad\quad \times  N^{2}\bigg(\xi_{*}, \sqrt{1-2 \tilde{p} \eta+\tilde{p}^{2}} x_{*}, \delta\bigg) N^{2}\bigg(\xi_{*}, \tilde{p} x_{*}, \delta\bigg)  \mathcal{I}^{2}\left[\xi_{*}, x_{*}, \delta, \tilde{p}, \sqrt{1-2 \tilde{p} \eta+\tilde{p}^{2}}\right].
\end{align}
In the presence of significant gauge field amplification, all the phenomenological features of tensor power spectrum in our model can be captured by the function $f_{2,\lambda}$ in \eqref{f2l} which can be evaluated numerically using the fitting functions we obtained for $N (\xi_*,x_*,\delta )$ in \eqref{fitN}. Note that since $A_{-}$ modes are amplified by the rolling $\sgm$, only negative helicity of tensor modes will be sourced efficiently, implying the hierarchy $f_{2,-} \gg f_{2,+}$.\\ 
{\bf Bispectrum:} We define the tensor bispectrum as 
\beq\label{tbsdef}
\left\langle\hat{h}_{\lambda_1}\left(0, \vec{k}_{1}\right) \hat{h}_{\lambda_2}\left(0, \vec{k}_{2}\right) \hat{h}_{\lambda_3}\left(0, \vec{k}_{3}\right)\right\rangle \equiv \mathcal{B}_{\lambda_1 \lambda_2 \lambda_3}\left(k_{1}, k_{2}, k_{3}\right) \delta\left(\vec{k}_{1}+\vec{k}_{2}+\vec{k}_{3}\right).
\eeq
As for all other correlators we consider in this work, tensor bispectrum takes contributions from the vacuum fluctuations of the metric and the sourced contributions due to enhanced gauge fields. In the presence of particle production in the gauge field sector, the latter gives the dominant contribution and hence, we will ignore vacuum fluctuations. More importantly, since only one polarization state of the gauge field is amplified, produced particles can efficiently source only one of the helicity state of tensor fluctuations. Keeping these in mind, in the following, we therefore focus on  $\mathcal{B}_{\lambda \lambda \lambda} \simeq \mathcal{B}^{(s)}_{\lambda \lambda \lambda}$. Using \eqref{hTs}, 3-pt correlator of $\hat{h}^{(s)}_{\lambda}$ is given by
\begin{align}
\nn &\left\langle\hat{h}_{\lambda}^{(s)}\left(0,\vec{k}_{1}\right) \hat{h}_{\lambda}^{(s)}\left(0,\vec{k}_{2}\right) \hat{h}_{\lambda}^{(s)}\left(0,\vec{k}_{3}\right)\right\rangle' \simeq  \left(\fr{H}{\Mp}\right)^6 \frac{2^{9 / 2}}{\left(k_{1} k_{2} k_{3}\right)^{7 / 2}} \int \frac{\d^{3} p}{(2 \pi)^{9 / 2}} \sqrt{p\left|\vec{p}+\vec{k}_{1}\right|\left|\vec{p}-\vec{k}_{3}\right|}\\\nn
&\quad\times \epsilon_{\lambda\lambda\lambda}\left[\vec{k}_1,\vec{k}_2,\vec{k}_3,\vec{p}\right] N^{2}\left(\xi_{*}, -p\tau_*, \delta\right) N^{2}\left( \xi_{*}, -|\vec{p}+\vec{k}_{1}|\tau_*, \delta\right) N^{2}\left(\xi_{*}, -|\vec{p}-\vec{k}_{3}|\tau_*, \delta\right)\\\nn
&\quad\times \mathcal{I}\left[\xi_{*}, -k_{1}\tau_{*}, \delta, \frac{p}{k_{1}}, \frac{\left|\vec{p}+\vec{k}_{1}\right|}{k_{1}}\right] \mathcal{I}\left[\xi_{*}, -k_{2}\tau_*, \delta, \frac{\left|\vec{p}+\vec{k}_{1}\right|}{k_{2}}, \frac{\left|\vec{p}-\vec{k}_{3}\right|}{k_{2}}\right] \mathcal{I}\left[\xi_{*}, -k_{3}\tau_{*}, \delta, \frac{\left|\vec{p}-\vec{k}_{3}\right|}{k_{3}}, \frac{p}{k_{3}}\right],
\end{align}
where prime denotes correlator without $\delta (\vec{k}_1+ \vec{k}_2+ \vec{k}_3)$ and the polarization products are defined as
\beq
\epsilon_{\lambda \lambda \lambda}\left[\vec{k}_{1}, \vec{k}_{2},\vec{k}_{3}, \vec{p} \right] \equiv \epsilon_{\lambda}\left[\vec{k}_{1},-\vec{p}, \vec{p}+\vec{k}_{1}\right] \epsilon_{\lambda}\left[\vec{k}_{2},-\vec{p}-\vec{k}_{1}, \vec{p}-\vec{k}_{3}\right] \epsilon_{\lambda}\left[\vec{k}_{3},-\vec{p}+\vec{k}_{3}, \vec{p}\right],
\eeq
where $\epsilon_\lambda$ is given in \eqref{eldef}. Noting \eqref{PSV}, we set $k_1 =k$ and define dimensionless variables $k x_2=k_2$, $k x_3 = k_3,$ $k\vec{\tilde{p}} = \vec{p}$ to obtain 
\beq
\mathcal{B}^{(s)}_{\lambda \lambda \lambda} \simeq \frac{\left[\epsilon_{\phi} \mathcal{P}_{\mathcal{R}}^{(v)}\right]^{3}}{k_{1}^{2} k_{2}^{2} k_{3}^{2}} f_{3, \lambda}\left(\xi_{*}, x_{*}, \delta, x_{2}, x_{3}\right),
\eeq
where we defined 
\begin{align}\label{f3l}
\nn&f_{3, \lambda}\left(\xi_{*}, x_{*}, \delta, x_{2}, x_{3}\right) =\frac{2^{27 / 2} \pi^{6}}{ (x_{2} x_{3})^{3/2}} \int \frac{d^{3} \tilde{p}}{(2 \pi)^{9 / 2}} \sqrt{\tilde{p}\left|\vec{\tilde{p}}+\hat{k}_{1}\right|\left|\vec{\tilde{p}}-x_{3} \hat{k}_{3}\right|} \epsilon_{\lambda \lambda \lambda}\left[\vec{k}_{1}, \vec{k}_{2},\vec{k}_{3},\vec{\tilde{p}}\right] N^{2}\left(\xi_{*}, \tilde{p}x_*, \delta\right)\\\nn
&\quad\quad\quad\quad\quad\quad\quad\quad\quad\times  N^{2}\left( \xi_{*}, |\vec{\tilde{p}}+\hat{k}_{1}|x_*, \delta\right) N^{2}\left(\xi_{*}, |\vec{\tilde{p}}-\hat{k}_{3}|x_*, \delta\right) \mathcal{I}\left[\xi_{*}, x_{*}, \delta, \tilde{p},\left|\vec{\tilde{p}}+\hat{k}_{1}\right|\right] \\
&\quad\quad\quad\quad\quad\quad\quad\quad\quad\times \mathcal{I}\left[\xi_{*}, x_{2}x_*, \delta, \fr{\left|\vec{\tilde{p}}+\hat{k}_{1}\right|}{x_{2}}, \frac{\left|\vec{\tilde{p}}-\hat{k}_{3}\right|}{x_{2}}\right] \mathcal{I}\left[\xi_{*}, x_{3}x_{*}, \delta, \frac{\left|\vec{\tilde{p}}-\hat{k}_{3}\right|}{x_{3}}, \frac{\tilde{p}}{x_{3}}\right].
\end{align}
In terms of the polarization vectors \eqref{eldef} the product $\epsilon_{\lambda\lambda\lambda}$ is given by
\begin{align}\label{epsf3l}
\nn \epsilon_{\lambda \lambda \lambda}\left[\vec{k}_{1}, \vec{k}_{2},\vec{k}_{3},\vec{\tilde{p}}\right] &=\epsilon_{i}^{\lambda}(\hat{k}_{1})^{*} \epsilon_{j}^{\lambda}(\hat{k}_{1})^{*}\,\, \epsilon_{j}^{-}(\vec{\tilde{p}}+\hat{k}_{1})\,\, \epsilon_{k}^{-}(\vec{\tilde{p}}+\hat{k}_{1})^{*} \,\,\epsilon_{k}^{\lambda}(\hat{k}_{2})^{*}\,\, \epsilon_{l}^{\lambda}(\hat{k}_{2})^{*}\\
&\times \epsilon_{l}^{-}(\vec{\tilde{p}}-x_{3} \hat{k}_{3})\,\, \epsilon_{m}^{-}(\vec{\tilde{p}}-x_{3} \hat{k}_{3})^{*}\,\epsilon_{m}^{\lambda}(\hat{k}_{3})^{*}\,\, \epsilon_{n}^{\lambda}(\hat{k}_{3})^{*}\,\, \epsilon_{n}^{-}(\vec{\tilde{p}}) \,\,\epsilon_{i}^{-}(\vec{\tilde{p}})^{*},
\end{align}
where we used the fact that $\epsilon^{\lambda} (\vec{a}/b) = \epsilon^{\lambda} (\vec{a}) $ (See \eg \eqref{epsdef}).
In order to evaluate momentum integrals in \eqref{f3l}, we align $\vec{k}_1$ along the z axis and write $\vec{k}_2$ and $\vec{k}_3$ in terms of $x_2$ and $x_3$,
\begin{align}\label{emc}
&\vec{k}_{1}=k\, (0,0,1)\\\nn
&\vec{k}_{2}= k\, x_2 \left(\fr{\sqrt{-\left(1-x_{2}+x_{3}\right)\left(1+x_{2}-x_{3}\right)\left(1-x_{2}-x_{3}\right)\left(1+x_{2}+x_{3}\right)}}{2 x_2},0, \fr{-1-x_{2}^{2}+x_{3}^{2}}{2 x_2}\right)\\\nn
&\vec{k}_{3}=k\, x_3 \left(-\fr{\sqrt{-\left(1-x_{2}+x_{3}\right)\left(1+x_{2}-x_{3}\right)\left(1-x_{2}-x_{3}\right)\left(1+x_{2}+x_{3}\right)}}{2 x_3},0, \fr{-1+x_{2}^{2}-x_{3}^{2}}{2 x_3}\right).
\end{align}
and define the polarization vector for a given momentum $\vec{q}$ in terms of its components as
\beq\label{epsdef}
\epsilon^{\lambda}(\vec{q})=\frac{1}{\sqrt{2}}\left(-\frac{\sqrt{q_{y}^{2}+q_{z}^{2}}}{|\vec{q}|},\frac{q_{x} q_{y} - i \lambda q_{z} |\vec{q}| }{ |\vec{q}|~\sqrt{q_{y}^{2}+q_{z}^{2}}}, \frac{q_{x} q_{z} + i \lambda q_{y} |\vec{q}|}{ |\vec{q}|~\sqrt{q_{y}^{2}+q_{z}^{2}}  }\right).
\eeq
One can immediately check that the definition in \eqref{epsdef} satisfies the desired relations listed below the eq. \eqref{DGF} for the vectors $k_1$, $k_2$ and $k_3$ in \eqref{emc}.
Using these explicit expressions, we can evaluate \eqref{f3l} numerically for a given set of parameters. Since only negative helicity mode of the gauge field is amplified, we have $f_{3,-} \gg f_{3,+}$.  On the other hand, since only gauge field modes that are approximately the size of the horizon are significantly amplified, we expect the bispectrum to be maximal at the equilateral configuration, $x_2 = x_3 = 1$ \footnote{See \eg the discussion in Appendix E of \cite{Namba:2015gja} where a model that shares very similar features is considered.}. 
\section{Sourced Scalar Fluctuations}\label{AppC}
In this appendix we present the derivation of the scalar 2-pt and 3-pt correlators in our model. We start from \eqref{CP} and seperate the canonical mode into its vacuum and sourced contribution, then using the solution \eqref{QS} for the sourced canonical mode, the sourced curvature perturbation is given by
\beq\label{CPSapp}
\hat{\mathcal{R}}^{(s)}(\tau, \vec{k}) \simeq \frac{3 \sqrt{2} H \tau}{\Mp} \int d \tau^{\prime} G_{k}\left(\tau, \tau^{\prime}\right) \frac{\sqrt{\epsilon_{\sigma}\left(\tau^{\prime}\right)}}{\tau^{\prime 2}} \int d \tau^{\prime \prime} G_{k}\left(\tau^{\prime}, \tau^{\prime \prime}\right) \hat{J}_{\sigma}\left(\tau^{\prime \prime}, \vec{k}\right)
\eeq
where the source is defined as in the right hand side of \eqref{usgm}. Using the definitions \eqref{EBF}, it is given by
\begin{align}\label{Jps}
\nn\hat{J}_\sgm(\tau'',\vec{k}) & =\fr{\alpha_{\rm c}}{4f a(\tau'')} \int \frac{\d^{3} p}{(2 \pi)^{3 / 2}} \,\epsilon^{-}_{i}(\vec{k}-\vec{p}) \epsilon^{-}_{i}(\vec{p})\,\, p^{1 / 4}\,|\vec{k}-\vec{p}|^{1 / 4}\left(p^{1/2}+ |\vec{k}-\vec{p}|^{1/2}\right) \\
&\quad\quad\quad\quad\quad\quad\quad\quad\quad \times \tilde{A}(\tau'',|\vec{k}-\vec{p}|)\, \tilde{A}(\tau'', p)\, \hat{\mathcal{O}}_{-}(\vec{k}-\vec{p})\, \hat{\mathcal{O}}_{-}(\vec{p}),
\end{align}
where we symmetrized the integrand with respect to $p$ and $|\vec{k}-\vec{p}|$ and $\mathcal{O}_{-}$ is defined as in \eqref{Op}. As we are interested in the correlators of $\mathcal{R}$ on super-horizon scales, we employ the  approximation \eqref{gfapp} in \eqref{CPSapp} for $G_k(\tau,\tau')$ while the same approximation does not hold for $G_k (\tau',\tau'')$. We therefore have
\begin{align}\label{CPSapp2}
\hat{\mathcal{R}}^{(s)}(\tau, \vec{k}) &\simeq \frac{3 \pi^{3 / 2} H}{2 \Mp \,k^{3 / 2}} \int_{-\infty}^{\tau} \frac{d \tau^{\prime}}{\tau^{\prime}} J_{3 / 2}\left(-k \tau^{\prime}\right) \sqrt{\epsilon_{\sigma}\left(\tau^{\prime}\right)} \int_{-\infty}^{\tau^{\prime}} d \tau^{\prime \prime} \sqrt{-\tau^{\prime \prime}}  \,\hat{J}_{\sigma}\left(\tau^{\prime \prime}, \vec{k}\right) \\\nn
&\quad\quad\quad\quad\quad\quad \times\left[J_{3 / 2}\left(-k \tau^{\prime}\right) Y_{3 / 2}\left(-k \tau^{\prime \prime}\right)-Y_{3 / 2}\left(-k \tau^{\prime}\right) J_{3 / 2}\left(-k \tau^{\prime \prime}\right)\right] .
\end{align}
Using \eqref{tA} in the source term \eqref{Jps}, we plug $\hat{J}_\sgm$ in \eqref{CPSapp2} to obtain
\begin{align}\label{sR}
\nn \hat{\mathcal{R}}^{(s)}(0,\vec{k}) &= \left(\fr{H}{\Mp}\right)^2 \fr{3\sqrt{2\pi^3}\xi_*}{8 k^4} \int \frac{\d^{3} p}{(2 \pi)^{3 / 2}} \,\epsilon^{-}_{i}(\vec{k}-\vec{p}) \epsilon^{-}_{i}(\vec{p})\,\, p^{1 / 4}\,|\vec{k}-\vec{p}|^{1 / 4}\left(p^{1/2}+ |\vec{k}-\vec{p}|^{1/2}\right) \\\nn
&\quad\quad\quad\quad\quad\quad\quad\quad\quad\quad\times  N\bigg(\xi_*, -|\vec{k}-\vec{p}|\tau_*,\delta\bigg)N\bigg(\xi_*, -p\tau_*,\delta\bigg)\, \hat{\mathcal{O}}_{-}(\vec{k}-\vec{p})\, \hat{\mathcal{O}}_{-}(\vec{p})\\
&\quad\quad\quad\quad\quad\quad\quad\quad\quad\quad \times  \mathcal{I}_{\mathcal{R}}\bigg[\xi_*,x_*,\delta,\sqrt{\fr{|\vec{k}-\vec{p}|}{k}} +\sqrt{\fr{p}{k}}\bigg],
\end{align}
where we have used $\alpha_{\rm c} \sqrt{\epsilon_{\sgm,*}}/f = \sqrt{2}\, \xi_*/ \Mp$ and we have defined the time integral of the sources as
\begin{align}\label{Rs}
\nn \mathcal{I}_{\mathcal{R}}\bigg[\xi_{*}, x_{*}, \delta, Q\bigg] \equiv \int_{0}^{\infty} \frac{d x^{\prime}}{x^{\prime}} J_{3 / 2}\left(x^{\prime}\right) \sqrt{\frac{\epsilon_{\sigma}\left(x^{\prime}\right)}{\epsilon_{\sigma, *}}} \int_{x^{\prime}}^{\infty} d x^{\prime \prime} x^{\prime \prime 3 / 2} \exp \left[-\frac{2 \sqrt{2\xi_{*}}}{\delta}\frac{x''^{1/2}}{|\ln(x''/x_*)\,|} Q\right] \\
\quad \times\left[J_{3 / 2}\left(x^{\prime}\right) Y_{3 / 2}\left(x^{\prime \prime}\right)-Y_{3 / 2}\left(x^{\prime}\right) J_{3 / 2}\left(x^{\prime \prime}\right)\right],
\end{align}
by sending the lower limit of the integral $-k\tau \to 0$. Using the standard definition of slow-roll parameters we note $
\sqrt{{\epsilon_{\sigma}(x^{\prime})}/{\epsilon_{\sigma, *}}} = {(1+\ln\left[(x_*/x' )^{\delta}\right]^2)^{-1}}$.\\
{\bf Power Spectrum:} We define the total scalar power spectrum as
\beq\label{DRPS}
\frac{k^{3}}{2 \pi^{2}}\left\langle\hat{\mathcal{R}}(0, \vec{k}) \hat{\mathcal{R}}(0, \vec{k}^{\prime})\right\rangle \equiv \, \delta \left(\vec{k}+\vec{k}^{\prime}\right) \,  \mathcal{P}_{\mathcal{R}}(k),
\eeq
where the total scalar power spectrum should be seperated as $  \mathcal{P}_{\mathcal{R}}(k) =  \mathcal{P}^{(v)}_{\mathcal{R}}(k) +  \mathcal{P}^{(s)}_{\mathcal{R}}(k)$ similar to the case with tensors. At leading order in slow-roll, using \eqref{vs}, the vacuum contribution is given in eq. \eqref{PSV}. Taking the 2-pt correlator of \eqref{sR} and using the Wick's theorem for the operators $\hat{\mathcal{O}}_{-}$, the sourced power spectrum can be extracted from the definition \eqref{DRPS} as
\begin{align}\label{PSS}
\nn   \mathcal{P}^{(s)}_{\mathcal{R}}(k) &= \left[\epsilon_\phi \mathcal{P}^{(v)}_\mathcal{R}\right]^2\fr{9\pi^5 \xi_*^2}{2 k^5} \int \frac{\d^{3} p}{(2 \pi)^{3}} \,\left(1- \fr{\vec{p}.\,(\vec{k}-\vec{p})}{p \, |\vec{k}-\vec{p}|}\right)^2\,\, p^{1 / 2}\,|\vec{k}-\vec{p}|^{1 / 2}\left(p^{1/2}+ |\vec{k}-\vec{p}|^{1/2}\right)^2 \\
&\quad\quad\quad\quad\quad\quad\quad\quad\times  N^2\bigg(\xi_*, -|\vec{k}-\vec{p}|\tau_*,\delta\bigg)N^2\bigg(\xi_*, -p\tau_*,\delta\bigg)\, \mathcal{I}^2_{\mathcal{R}}\bigg[\xi_*,x_*,\delta,\sqrt{\fr{|\vec{k}-\vec{p}|}{k}} +\sqrt{\fr{p}{k}}\bigg],
\end{align}
where we have used \eqref{PSV} to express the overall factors that appears in front of the integral in \eqref{PSS} and the identity $\big| \epsilon^{\lambda}_{i}(\vec{p}) \epsilon^{\lambda'}_{i}(\vec{q}) \big|^2 = (1 -\lambda\lambda' \hat{p}.\hat{q})^2/4$ between the polarization vectors. For the numerical integration of the momentum integral, we switch to dimensionless variable $\tilde{p} = p/k$ and denote by $\eta$ the cosine angle between $\vec{p}$ and $\vec{k}$. This gives $\mathcal{P}^{(s)}_{\mathcal{R}} = \left[\epsilon_\phi \mathcal{P}^{(v)}_\mathcal{R}(k)\right]^2 f_{2,\mathcal{R}}(\xi_*,x_*,\delta)$ where 
\begin{align}\label{f2Rf}
\nn f_{2,\mathcal{R}}(\xi_*,x_*,\delta) &= \fr{9\pi^3\, \xi_*^2}{8} \int_{0}^{\infty} \d \tilde{p} \int_{-1}^{1} \d \eta \,\,\, \tilde{p}^{5/2}\,\,\,(1-2 \tilde{p} \eta+\tilde{p}^{2})^{1 / 4}  \left[ \tilde{p}^{1/2}+ (1-2 \tilde{p} \eta+\tilde{p}^{2})^{1 / 4}\right]^2 \\\nn
&\quad\quad\quad\quad\times \left[1+ \fr{\tilde{p}-\eta}{(1-2 \tilde{p} \eta+\tilde{p}^{2})^{1 / 2}}\right]^2N^2\bigg(\xi_*, (1-2 \tilde{p} \eta+\tilde{p}^{2})^{1 / 2}\,x_*,\delta\bigg)N^2\bigg(\xi_*, \tilde{p}\,x_*,\delta\bigg)\,\\
&\quad\quad\quad\quad \times  \mathcal{I}^2_{\mathcal{R}}\bigg[\xi_*,x_*,\delta, (1-2 \tilde{p} \eta+\tilde{p}^{2})^{1 / 4}+\tilde{p}^{1/2}\bigg].
\end{align}
{\bf Bispectrum:} We define the bispectrum of comoving curvature perturbation as 
\beq\label{rbsdef}
\left\langle\hat{\mathcal{R}}\left(0, \vec{k}_{1}\right) \hat{\mathcal{R}}\left(0, \vec{k}_{2}\right) \hat{\mathcal{R}}\left(0, \vec{k}_{3}\right)\right\rangle \equiv \mathcal{B}_{\mathcal{R}}\left(k_{1}, k_{2}, k_{3}\right) \delta\left(\vec{k}_{1}+\vec{k}_{2}+\vec{k}_{3}\right).
\eeq
As in the case of 2-pt correlators, bispectrum consist of the vacuum and sourced part. In the presence of gauge field amplification, vacuum part is negligible (\ie slow-roll suppressed) and therefore we can mainly focus on the sourced contribution. Taking the 3-pt function of $\mathcal{R}^{(s)}$ in \eqref{sR}, we obtain
\begin{align}\label{3pR}
\nn &\left\langle\hat{\mathcal{R}}^{(s)}\left(0, \vec{k}_{1}\right) \hat{\mathcal{R}}^{(s)}\left(0, \vec{k}_{2}\right) \hat{\mathcal{R}}^{(s)}\left(0, \vec{k}_{3}\right)\right\rangle  = \left(\fr{H}{\Mp}\right)^6 \fr{27\,(2\pi^3)^{3/2} \,\xi_{*}^3}{2^9 \, k_1^4 k_2^4 k_3^4} \int \fr{\d^3 p_1 \d^3 p_2 \d^3 p_3}{(2\pi)^{9/2}} \prod_{i=1}^{3} \epsilon^{-}_k (\vec{k}_i-\vec{p}_i) \epsilon^{-}_k(\vec{p}_i) \\\nn
&\quad\quad\quad\quad\quad\quad\quad \times (p_i \, |\vec{k}_i - \vec{p}_i|)^{1/4} (p_i^{1/2} + |\vec{k}_i - \vec{p}_i|^{1/2}) N\bigg(\xi_*, -|\vec{k}_i-\vec{p}_i|\tau_*,\delta\bigg)N\bigg(\xi_*, -p_i \tau_*,\delta\bigg)\\
& \quad\quad\quad\quad\quad\quad\quad \times \ \mathcal{I}_{\mathcal{R}}\bigg[\xi_*, - k_i \tau_*,\delta,\sqrt{\fr{|\vec{k}_i-\vec{p}_i|}{k_i}} +\sqrt{\fr{p_i}{k_i}}\bigg] \langle \,  \hat{\mathcal{O}}_{-}(\vec{k}_i-\vec{p}_i)\, \hat{\mathcal{O}}_{-}(\vec{p}_i) \,\rangle.
\end{align}
Using Wick's theorem, we evaluate the product of expectation value in \eqref{3pR}. In this way, we found
\begin{align}
\nn\mathcal{B}^{(s)}_{\mathcal{R}}\left(k_{1}, k_{2}, k_{3}\right) &=  \left(\fr{H}{\Mp}\right)^6 \fr{27\,(2\pi^3)^{3/2} \,\xi_{*}^3}{2^6 \, k_1^4 k_2^4 k_3^4}\int \frac{\d^{3} p}{(2 \pi)^{9 / 2}} \,\epsilon\left[\vec{p}, \vec{p}+\vec{k}_{1}, \vec{p}-\vec{k}_{3}\right] \sqrt{p\left|\vec{p}+\vec{k}_{1}\right|\left|\vec{p}-\vec{k}_{3}\right|}\\\nn
&\quad\quad\quad\quad\quad \times (\sqrt{p}+\sqrt{\left|\vec{p}+\vec{k}_{1}\right|})(\sqrt{\left|\vec{p}+\vec{k}_{1}\right|}+\sqrt{\left|\vec{p}-\vec{k}_{3}\right|})(\sqrt{\left|\vec{p}-\vec{k}_{3}\right|}+\sqrt{p})\\\nn
&\quad\quad\quad\quad\quad \times N^{2}\left(\xi_{*}, -p\tau_*, \delta\right) N^{2}\left( \xi_{*}, -|\vec{p}+\vec{k}_{1}|\tau_*, \delta\right) N^{2}\left(\xi_{*}, -|\vec{p}-\vec{k}_{3}|\tau_*, \delta\right)\\\nn
&\quad\quad\quad\quad\times \mathcal{I}_{\mathcal{R}}\left[\xi_{*}, \frac{k_{1}}{k_{*}}, \delta, \frac{\sqrt{p}+\sqrt{| \vec{p}+\vec{k}_{1}} |}{\sqrt{k_{1}}}\right] \mathcal{I}_{\mathcal{R}}\left[\xi_{*}, \frac{k_{2}}{k_{*}}, \delta, \frac{\sqrt{\left|\vec{p}+\vec{k}_{1}\right|}+\sqrt{\left|\vec{p}-\vec{k}_{3}\right|}}{\sqrt{k_{2}}}\right]\\
&\quad\quad\quad\quad \times \mathcal{I}_{\mathcal{R}}\left[\xi_{*}, \frac{k_{3}}{k_{*}}, \delta, \frac{\sqrt{\left|\vec{p}-\vec{k}_{3}\right|}+\sqrt{p}}{\sqrt{k_{3}}}\right],
\end{align}
where we defined the product of polarization vectors as
\begin{align}\label{ppv}
\nn \epsilon\left[\vec{v}_{1}, \vec{v}_{2}, \vec{v}_{3}\right] &\equiv \epsilon_{i}^{-}\left(\vec{v}_{1}\right)^{*} \epsilon_{i}^{-}\left(\vec{v}_{2}\right) \epsilon_{j}^{-}\left(\vec{v}_{2}\right)^{*} \epsilon_{j}^{-}\left(\vec{v}_{3}\right) \epsilon_{k}^{-}\left(\vec{v}_{3}\right)^{*} \epsilon_{k}^{-}\left(\vec{v}_{1}\right) \\\nn
&= \frac{1}{8}\bigg[ \hat{v}_{1} \cdot \hat{v}_{2}+\hat{v}_{2} \cdot \hat{v}_{3}+\hat{v}_{3} \cdot \hat{v}_{1}+\left(\hat{v}_{1} \cdot \hat{v}_{2}\right)^{2}+\left(\hat{v}_{2} \cdot \hat{v}_{3}\right)^{2}+\left(\hat{v}_{3} \cdot \hat{v}_{1}\right)^{2} + \left(\hat{v}_{1} \cdot \hat{v}_{2}\right)\left(\hat{v}_{2} \cdot \hat{v}_{3}\right) \\\nn
& \quad\quad+\left(\hat{v}_{2} \cdot \hat{v}_{3}\right)\left(\hat{v}_{3} \cdot \hat{v}_{1}\right)+\left(\hat{v}_{3} \cdot \hat{v}_{1}\right)\left(\hat{v}_{1} \cdot \hat{v}_{2}\right)-\left(\hat{v}_{1} \cdot \hat{v}_{2}\right)\left(\hat{v}_{2} \cdot \hat{v}_{3}\right)\left(\hat{v}_{3} \cdot \hat{v}_{1}\right)\bigg]\\
&\quad+ \fr{i}{8}\,\, \hat{v}_1\cdot (\hat{v}_2 \times \hat{v}_3)(1 + \hat{v}_{1} \cdot \hat{v}_{2}+\hat{v}_{2} \cdot \hat{v}_{3}+\hat{v}_{3} \cdot \hat{v}_{1})
\end{align}
using the identity \cite{Barnaby:2012xt}:
\beq\label{phv}
\epsilon^{\pm}_{i} (\vec{q}) \epsilon_j^{\pm\,*} (\vec{q}) = \fr{1}{2} \left[\delta_{ij} - \hat{q}_i \hat{q}_j \mp i \epsilon_{ijk} \hat{q}_k\right].
\eeq
Since the bispectrum is real\footnote{See \eg the detailed discussion in the Appendix E of \cite{Namba:2015gja}.}, we disregard the imaginary part in \eqref{ppv} when computing the scalar bispectrum. Fixing $k_1 =k$, we define dimensionless variables $x_2,x_3$ and $\vec{\tilde{p}}$ as $
k\, x_2 = k_2, \,\, k\, x_3 = k_3, \,\, k\, {\vec{\tilde{p}}} = \vec{p}
$ to re-write the scalar bispectrum in terms of ratio of the external momenta $x_2 \equiv k_2/k_1$ and $x_3 \equiv k_3/k_1$ as
\beq
\mathcal{B}^{(s)}_{\mathcal{R}} \simeq \frac{\left[\epsilon_{\phi} \mathcal{P}_{\mathcal{R}}^{(v)}\right]^{3}}{k_{1}^{2} k_{2}^{2} k_{3}^{2}} f_{3, \mathcal{R}}\left(\xi_{*}, x_{*}, \delta, x_{2}, x_{3}\right),
\eeq
where we have used \eqref{PSV} and defined
\begin{align}\label{f3R}
\nn f_{3, \mathcal{R}}\left(\xi_{*}, x_{*}, \delta, x_{2}, x_{3}\right)&=\frac{27\, 2^{9 / 2} \pi^{21 / 2} \xi_{*}^{3}}{(x_{2} x_{3})^{2}} \int \frac{\d^{3} \tilde{p}}{(2 \pi)^{9 / 2}} {\rm Re}\left[\epsilon\left[\vec{\tilde{p}}, \vec{\tilde{p}}+\hat{k}_{1}, \vec{\tilde{p}}-x_{3} \hat{k}_{3}\right]\right] \sqrt{\tilde{p}\left|\vec{\tilde{p}}+\hat{k}_{1}\right|\left|\vec{\tilde{p}}-x_{3} \hat{k}_{3}\right|}\\\nn
&\quad\quad\times (\sqrt{\tilde{p}}+\sqrt{\left|\vec{p}+\hat{k}_{1}\right|})(\sqrt{\left|\vec{\tilde{p}}+\hat{k}_{1}\right|}+\sqrt{\left|\vec{\tilde{p}}-x_{3} \hat{k}_{3}\right|})(\sqrt{\left|\vec{\tilde{p}}-x_{3} \hat{k}_{3}\right|}+\sqrt{\tilde{p}})\\\nn
&\quad\quad\times N^{2}\left(\xi_{*}, \tilde{p} x_{*}, \delta\right) N^{2}\left(\xi_{*},\left|\vec{\tilde{p}}+\hat{k}_{1}\right| x_{*}, \delta\right) N^{2}\left(\xi_{*},\left|\vec{\tilde{p}}-x_{3} \hat{k}_{3}\right| x_{*}, \delta\right)\\\nn
&\quad\quad\times \mathcal{I}_{\mathcal{R}}\left[\xi_{*}, x_{*}, \delta, \sqrt{\tilde{p}}+\sqrt{\left|\vec{\tilde{p}}+\hat{k}_{1}\right|}\right] \mathcal{I}_{\mathcal{R}}\left[\xi_{*}, x_{2} x_{*}, \delta, \frac{\sqrt{\left|\vec{\tilde{p}}+\hat{k}_{1}\right|}+\sqrt{\left|\vec{\tilde{p}}-x_{3} \hat{k}_{3}\right|}}{\sqrt{x_{2}}}\right]\\
&\quad\quad\times \mathcal{I}_{\mathcal{R}}\left[\xi_{*}, x_{3} x_{*}, \delta, \frac{\sqrt{\left|\vec{p}-x_{3} \hat{k}_{3}\right|}+\sqrt{\tilde{p}}}{\sqrt{x_{3}}}\right].
\end{align}
For the numerical integration over $\d^3 \tilde{p}$, we align $\vec{k}_1$ with the z-axis and express $\vec{k}_2$ and $\vec{k}_3$ in terms of $x_2$ and $x_3$ as in \eqref{emc}.
As a result, using the normalization factors appearing inside the integrand in \eqref{f3R}, one can compute the integral numerically to understand the behavior of the bispectrum for general ratios of $x_2$ and $x_3$ which correponds to different deformations of the triangle formed by $\vec{k}_1$, $\vec{k}_2$ and $\vec{k}_3$.  Similar to the case with tensor fluctuations, equilateral configuration $x_2 = x_3 =1$ can be considered as a good measure of scalar non-gaussianity.
\section{Direct contribution to $\mathcal{R}$ from gauge fields}\label{AppD}
\begin{figure}[t!]
\begin{center}
\includegraphics[scale=0.89]{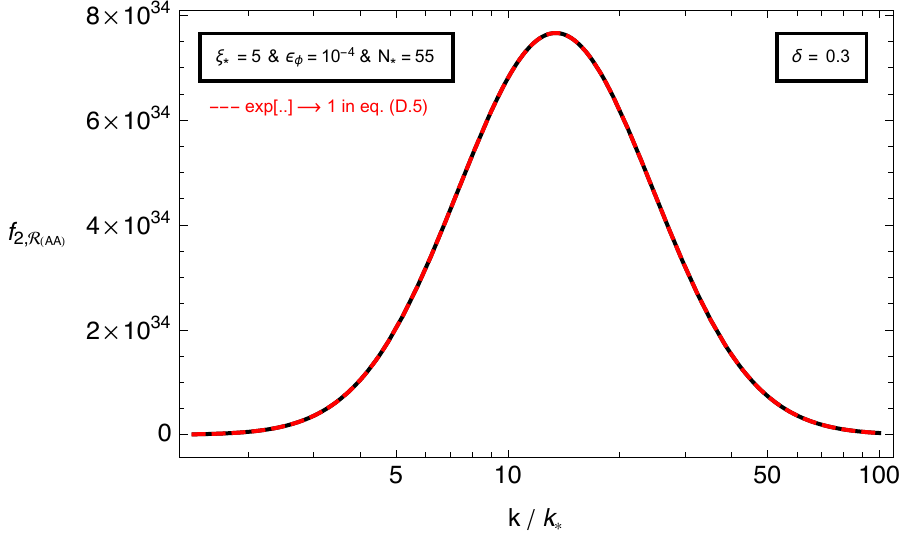}~\includegraphics[scale=0.89]{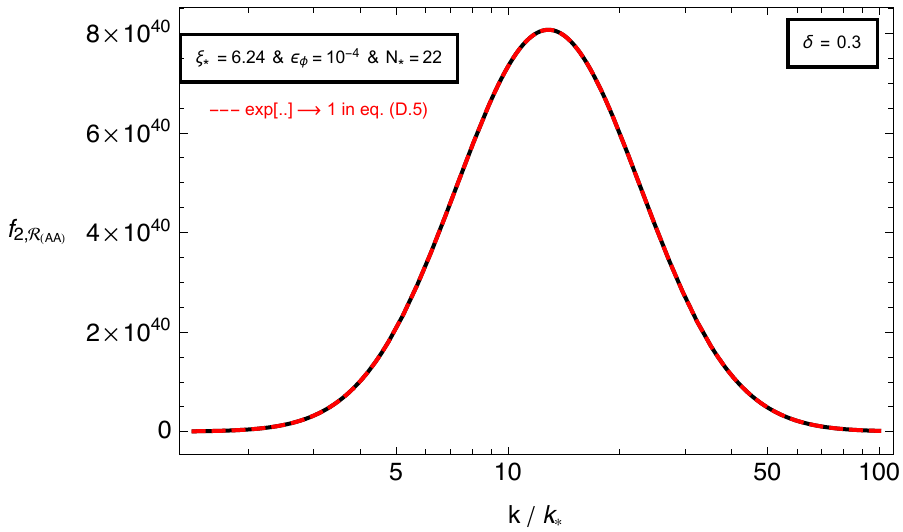}
\end{center}
\caption{Scale dependence of the power spectrum in eq. \eqref{PRGFF} for the same parameter choices $\xi_* = 5$ (left) and $\xi_* = 6.24$ (right) that can generate the CMB and sub-CMB phenomenology we present in Figures \ref{fig:r}, \ref{fig:PBHL} and \ref{fig:GW}. \label{fig:bf2RAA}}
\end{figure}

In this appendix, we calculate the power spectrum of the curvature perturbation induced by $\mathcal{R}_{(AA)}$ in \eqref{RAA} to decide if the latter contribution could alter the cosmological correlators computed using the standard relation \eqref{CP}. For this purpose, we begin by re-writing the expression \eqref{RAA},
\begin{align}\label{RGF}
\nn\hat{\mathcal{R}}_{(AA)}(\tau_{\rm end},\vec{k}) & =\fr{H}{4\dot{\phi}^2 a_{\rm end}^3}\fr{i\hat{k}_i}{k}\epsilon_{ijk} \int \frac{\d^{3} p}{(2 \pi)^{3 / 2}} \,\epsilon^{-}_{j}(\vec{k}-\vec{p}) \epsilon^{-}_{k}(\vec{p})\,\, p^{1 / 4}\,|\vec{k}-\vec{p}|^{1 / 4}\left(p^{1/2}- |\vec{k}-\vec{p}|^{1/2}\right) \\
&\quad\quad\quad\quad\quad\quad\quad\quad\quad\quad\quad\quad\, \times \tilde{A}(\tau_{\rm end},|\vec{k}-\vec{p}|)\, \tilde{A}(\tau_{\rm end}, p)\, \hat{\mathcal{O}}_{-}(\vec{k}-\vec{p})\, \hat{\mathcal{O}}_{-}(\vec{p}),
\end{align}
where we used the definitions \eqref{EBF} and anti-symmetrized the integrand with respect to $|\vec{k}-\vec{p}|$ and $p$. Using the standard definition of the power spectrum in \eqref{DRPS}, we take the 2-pt correlator of \eqref{RGF} to extract power spectrum of this contribution as
\begin{align}\label{PRGF}
\nn\mathcal{P}_{\mathcal{R}_{(AA)}}(k) & =\fr{k H^2\,  \hat{k}_i \hat{k}_l}{16\pi^2\dot{\phi}^4 a_{\rm end}^6}\epsilon_{ijk}\epsilon_{lmn} \int \frac{\d^{3} p}{(2 \pi)^{3}} \,\epsilon^{-}_{j}(\vec{k}-\vec{p}) \epsilon^{-}_{k}(\vec{p}) \epsilon^{-}_{m}(-(\vec{k}-\vec{p})) \epsilon^{-}_{n}(-\vec{p})\, p^{1 / 2}\,|\vec{k}-\vec{p}|^{1 / 2}\\
&\quad\quad\quad\quad\quad\quad\quad\quad\quad\quad\quad\quad\times \left(p^{1/2}- |\vec{k}-\vec{p}|^{1/2}\right)^2 \tilde{A}(\tau_{\rm end},|\vec{k}-\vec{p}|)^2\, \tilde{A}(\tau_{\rm end}, p)^2 ,
\end{align}
where we evaluated expectation values of involving $\hat{\mathcal{O}}_{-}$ using Wick's theorem. We can evaluate the products of helicity vectors in the first line of \eqref{PRGF} using the identity defined in \eqref{phv}. Proceeding this way, after a bit of algebra, one can show that contracted products of helicity vectors and $\hat{k}_i \hat{k}_l \epsilon_{ijk}\epsilon_{lmn}$ in the first line of \eqref{PRGF} is given by
\beq\label{pofhv}
\hat{k}_i \hat{k}_l \,\epsilon_{ijk}\epsilon_{lmn}\,\epsilon^{-}_{j}(\vec{k}-\vec{p}) \epsilon^{-}_{k}(\vec{p}) \epsilon^{-}_{m}(-(\vec{k}-\vec{p})) \epsilon^{-}_{n}(-\vec{p}) = \fr{1}{4} \left(\fr{\vec{k}\cdot(\vec{k}-\vec{p})}{k|\vec{k}-\vec{p}|}-\hat{k}\cdot\hat{p}\right)^2.
\eeq
Plugging the relation \eqref{pofhv} in \eqref{PRGF} and noting the mode functions \eqref{tA}, power spectrum can be shown to obtain the following form
\beq\label{PRGFF}
\mathcal{P}_{\mathcal{R}_{(AA)}}(k) = \left[\epsilon_\phi \mathcal{P}_{\mathcal{R}}^{(v)}\right]^2 \left(\fr{\tau_{\rm end}}{\tau_*}\right)^6 f_{2,\mathcal{R}_{(AA)}}(\xi_*,x_*,\delta),
\eeq
where we used power spectrum of vacuum fluctuations $\mathcal{P}_{\mathcal{R}}^{(v)} = H^4/(4\pi^2\dot{\phi}^2)$ to replace powers of $H^2/\dot{\phi}$ in \eqref{PRGF} in favor of $\mathcal{P}_{\mathcal{R}}^{(v)}$. The last factor in \eqref{PRGFF} parametrizes the scale dependence of the power spectrum \eqref{RGF} and is given by
\begin{align}\label{f2RAA}
\nn f_{2,\mathcal{R}_{(AA)}} &= \fr{x_*^{6}}{16\epsilon_\phi^2}\int_{0}^{\infty} \d \tilde{p} \int_{-1}^{1} \d \eta \,\,\, \tilde{p}^{5/2}\,\,\fr{(1-\tilde{p}\eta-\eta\sqrt{1-2 \tilde{p} \eta+\tilde{p}^{2}})^{2}}{(1-2 \tilde{p} \eta+\tilde{p}^{2})^{3/4}}  \left[\tilde{p}^{1/2}- (1-2 \tilde{p} \eta+\tilde{p}^{2})^{1/4}\right]^2 \\\nn
&\quad\quad\quad\quad\quad\quad\quad\quad\times N^2\bigg(\xi_*, (1-2 \tilde{p} \eta+\tilde{p}^{2})^{1 / 2}\,x_*,\delta\bigg)N^2\bigg(\xi_*, \tilde{p}\,x_*,\delta\bigg)\\
&\quad\quad\quad\quad\quad\quad\quad\quad \times \exp\bigg[-\fr{4\sqrt{2\xi_*x_*(\tau_{\rm end}/\tau_*)}((1-2 \tilde{p} \eta+\tilde{p}^{2})^{1 / 4}+\tilde{p}^{1/2})}{\delta|\ln(\tau_{\rm end}/\tau_*)|}\bigg],
\end{align}
\noindent
where we performed the trivial azimuthal integral in \eqref{PRGF} and switched to the dimensionless variable $\tilde{p} = p/k$. In \eqref{f2RAA}, we realize that argument of the exponential exhibit a factor of $\sqrt{\tau_{\rm end}/\tau_*} = e^{-N_*/2}$ (noting $N_{\rm end} = 0$) which result with significant suppression of its argument, in particular considering the phenomenological scenarios with $N_*\geq 22$ we consider in this work. To integrate the expression \eqref{f2RAA}, we can thus  set we can set $\exp[\dots] \to 1$, which in turn allow us universally capture the scale depence of the power spectrum for all the CMB and sub-CMB scenarios we study, irrespective of the location of the signal in terms of $N_*$. We illustrate these facts in Figure \ref{fig:bf2RAA} where \eqref{f2RAA} is plotted with respect to $x_* = k/k_*$. As expected, the power spectrum obtains a scale dependent bump due to the localized gauge field amplification by the rolling axion. Our main goal here is to determine whether this amplification is significant enough to influence the total late time power spectrum. For this purpose, we notice that the power spectrum \eqref{PRGFF} has an extra factor of $(\tau_{\rm end}/\tau_*)^{6} = e^{-6N_*}$ which give rise to an enourmous amount of suppression compared to power spectrum induced by the dominant channel we study (See \eg eq. \eqref{SC}). In particular, $(\tau_{\rm end}/\tau_*)^{6} = e^{-6N_*}$ ranges between $10^{-57}$ to $10^{-156}$ within the interval $22 \leq N_*\leq 60$ where all the phenomenological scenarios we consider in this work lives (See Section \ref{S4p1p1} and \ref{S4p2}) and comparing these values with the maximal amplitude that the scale dependent part of \eqref{PRGFF} can obtain in Figure \ref{fig:bf2RAA}, we can confidently conclude that contribution of $\mathcal{R}_{(AA)}$ on the scalar power spectrum can be completely ignored compared to the power spectrum that arise through the standard relation \eqref{CP}. We expect that the same conclusion to apply to the higher point auto/cross correlators of $\mathcal{R}_{(AA)}$. This is because, contrary to the expression \eqref{sR} we use to calculate the correlators of $\mathcal{R}$ in this work, the curvature perturbation induced directly by gauge fields in \eqref{RGF} include extra factors of $a_{\rm end}^{-3}$ that leads to a large supression factor that is much more dramatic than the scale dependent enhacement provided by the factors of $N(\xi_*,x_*,\delta)$. 
\section{Constraints on model building}\label{AppE}
In this appendix, we study the limitations on the phenomenological implications of our model focusing on the back-reaction effects of the produced particles on the background evolution and perturbativity considerations of scalar and vector fluctuations. To set the stage in this direction, we first provide a detailed account on the energy density of gauge field fluctuations.

\begin{figure}[t!]
\begin{center}
\includegraphics[scale=0.89]{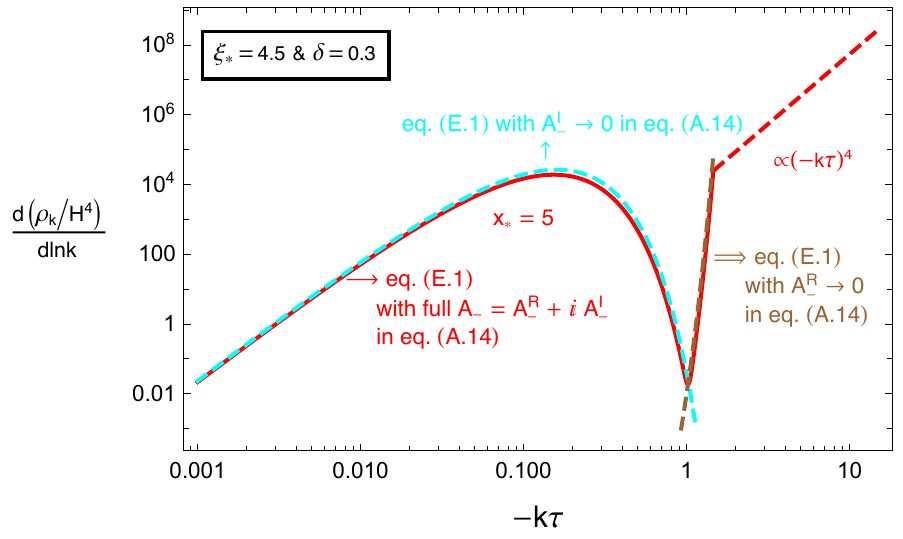}\includegraphics[scale=0.89]{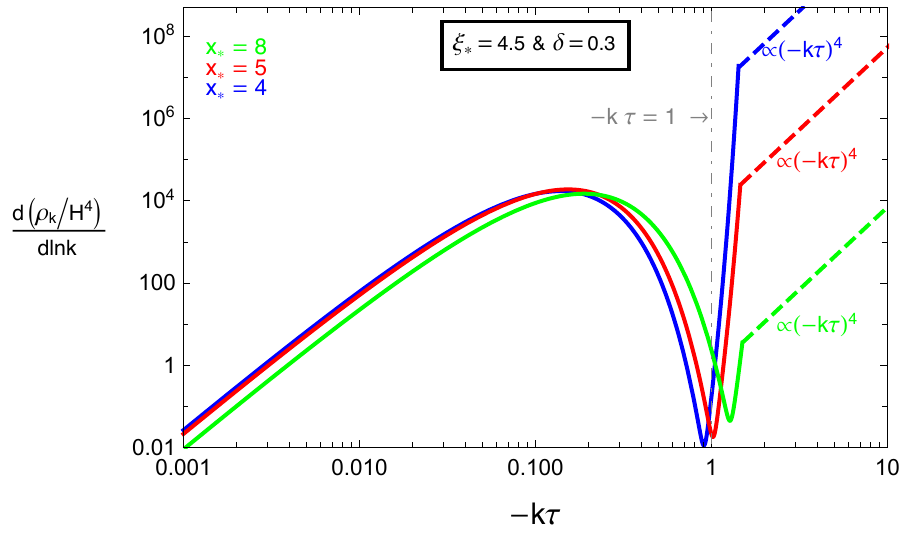}
\end{center}
\caption{Time evolution of the gauge field energy density for modes with a given comoving momentum $k$ where each mode is labeled by the ratio of the physical momentum and the Hubble rate at the time where axion's velocity peaks, $x_* = -k\tau_* = k/(a_*H_*)$. From the left panel, we observe that $A^{R}_{-}$ (dashed cyan curve) in \eqref{wkbsol} accurately describes the enhancement of the physical energy density (solid red curve) on super-horizon scales.\label{fig:rhoperk}}
\end{figure}

\noindent{\bf Gauge-field energy density:} Using the decomposition of gauge field in \eqref{DGF} and the definitions of ``electric" and ``magnetic" fields, we note the total energy density in the gauge field sector as
\beq\label{I}
\nn\rho_A \equiv \fr{1}{2}\langle\vec{E}^2+\vec{B^2}\rangle \simeq \fr{(H\tau)^4}{4\pi^2}\int \d k~ \Big\{ k^2 |A'_{-} (\tau,\vec{k})|^2 + k^4|A_{-}(\tau,\vec{k})|^2\Big\},
\eeq
where we take into account only negative helicity states as they contribute dominantly to the energy density. 
Switching to $x = -k\tau$ and defining the dimensionless mode functions via $\sqrt{2k} A_{-}(\tau,k) = \tilde{A}_{-}(x)$, we note the dimensionless measure of gauge field energy density per wavenumber as
\beq\label{rhopermode}
\frac{\d\left(\rho_{k, A} / H^{4}\right)}{\d \ln k}=\frac{x^{4}}{8 \pi^{2}}\left(\left|\frac{\d \tilde{A}_{-}}{\d x}\right|^{2}+|\tilde{A}_{-}|^{2}\right).
\eeq
In order to make a clear distinction between the physical enhancement by the rolling axion and the UV divergence piece of gauge field fluctuations, we use the mode functions derived in Appendix \ref{AppA} (see \eg \eqref{wkbsol} and \eqref{wkbsolIN}) in \eqref{rhopermode} to show in Figure \ref{fig:rhoperk} the time evolution (time flows from left to right) of energy density of a given wavenumber $k$. In particular, in the left panel we present the time evolution of the maximally amplified mode ($x_* =-k\tau_*=  5$, solid red curve) as the mode evolves from sub-horizon to super-horizon regime. From deep inside the horizon towards $-k\tau \sim 1$, we see that the vacuum energy density evolves from i) the standard phase associated with zero point fluctuations \eqref{wkbsolIN} (red dashed line with $(-k\tau)^4$) to ii) a phase where the energy density is dictated completely by the imaginary part $A^{I}_{-}$ of the solution \eqref{wkbsol} shown by the brown dotted line ($\propto (-k\tau)^p$ with $p > 4$.)\footnote{In the left panel of  Figure \ref{fig:rhoperk}, we illustrate this connection between the vacuum solution \eqref{wkbsolIN} and the WKB solution \eqref{wkbsol} by matching the energy density associated with these regimes slightly on the left hand side of the turning point, \ie at $x = x_{\rm c}/5$ where the solution in \eqref{wkbsol}(or equivalently in \eqref{Aoutf}) is still valid. Note that for each mode labeled by $x_*$, we obtained the critical point $x_{\rm c}$ by the condition $V_{\rm eff}(x_{\rm c})= 0$ using eq. \eqref{Q} at a fixed $\delta$ and $\xi_*$.}. The latter two regimes represent the UV divergent piece associated with vacuum energy density of $U(1)$ fields which must be renormalized away in an appropriate way. We would like to stress that this divergent piece is completely distinct from the physical enhancement of the gauge fields we study in this work and therefore when we calculate the observable effects induced by the gauge fields, we will ignore this UV regime $-k\tau > 1$. Note that this cut-off prescription corresponds to neglecting both the imaginary part $A^{I}_{-}$ of the late time solution we derived in \eqref{wkbsol} and the vacuum solution \eqref{wkbsolIN} as previously done in the literature \cite{Barnaby:2011vw,Barnaby:2012xt,Namba:2015gja,Peloso:2016gqs}. Following the decrease in the UV divergent piece, the growth in the energy density due to the rolling axion takes place for $-k\tau < 1$ followed by the subsequent dilution at late times $-k\tau \to 0$ due to the expansion of the universe.  As clearly visible by the dashed cyan curve in the left panel of Figure \ref{fig:rhoperk}, the resulting peak is completely dictated by the real part $A^{R}_{-}$ of the late time solution \eqref{wkbsol}. 

In the right panel of Figure \ref{fig:rhoperk}, we show the energy density of different modes around the maximally amplified mode $x_* = 5$ to confirm that the field amplification becomes smaller both for modes satisfying $x_* > 5$ and $x_* < 5$, in agreement with the profile of normalization factors we provided in eq. \eqref{Nform}. We see that the vertical $-k\tau \simeq \mathcal{O}(1)$ indicates a suitable position to distinguish between unphysical vacuum energy density and physical amplification caused by the coupling to $\sigma$.

\begin{figure}[t!]
\begin{center}
\includegraphics[scale=0.65]{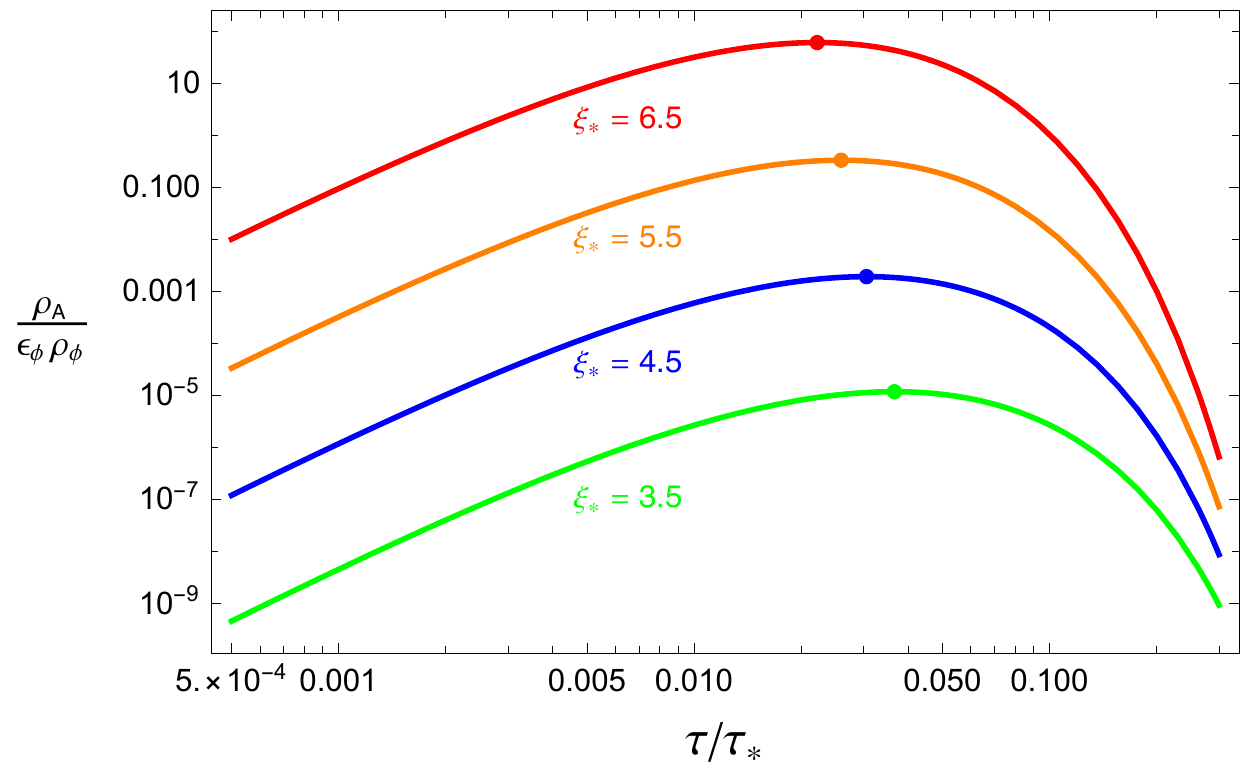}\includegraphics[scale=0.9]{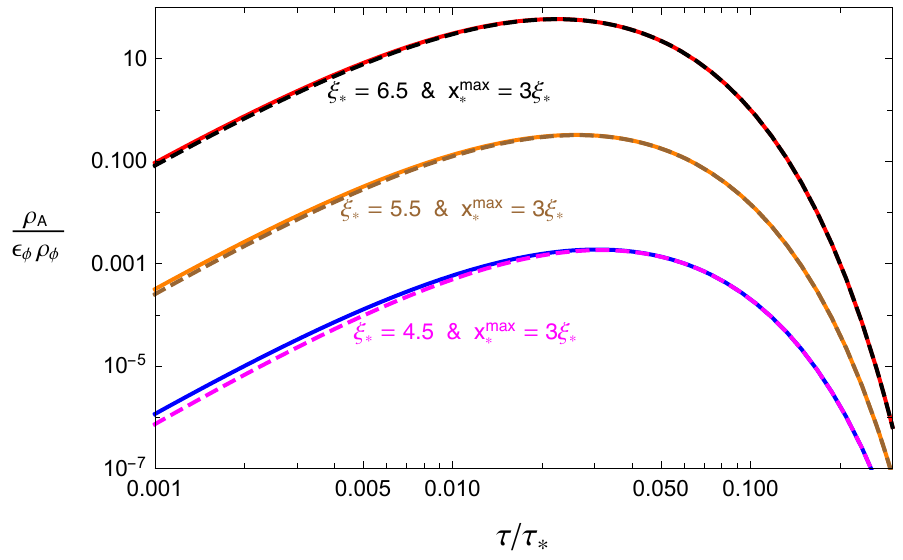}
\end{center}
\caption{The total energy density contained in the gauge field sector $\rho_A$ for $\delta = 0.3$ in units of the quantity $\epsilon_\phi \rho_\phi$ as a function of $y = \tau/\tau_*$ and for different values of $\xi_*$. The locations where the energy density reaches its maximum value are shown by colored points. In the right panel, we show the accuracy of the choice $x^{\rm max}_* = 3\xi_*$ (dashed colored curves) in the estimation of $\rho_A$ in \eqref{rhoA}.\label{fig:rhoA}}
\end{figure}

Following the discussion above, we disregard the modes that do not experience enhancement due to the roll of the axion to study the backreaction of the produced gauge field quanta. For this purpose, we focus on the real part $A^{R}_-$ of the gauge field amplitudes in \eqref{wkbsol} and noting the normalization factors \eqref{Nform}, $\rho_A$ is obtained by the following integral of \eqref{rhopermode} over the modes labeled by $x_*$:
\beq\label{rhoA}
\fr{\rho_{A}}{\epsilon_\phi \rho_\phi} = \fr{\mathcal{P}^{(v)}_{\mathcal{R}}y^{7/2} N^{c}[\xi_*,\delta]^2 \sqrt{2\xi(y)}}{3}\int_{0}^{x^{\rm max}_*}\d x_*\, x_*^{5/2} \exp\left[-\fr{4\sqrt{2\xi_*y}\,x_*^{1/2}}{\delta |\ln (y)|}-\fr{\ln^2(x_*/q_c)}{\sgm[\xi_*,\delta]^2}\right]\left(1 + \fr{x_*\,y}{2\xi(y)}\right),
\eeq
\noindent 
where we have defined $y \equiv \tau/\tau_*$ and used $\rho_\phi \simeq 3 H^2 \Mp^2$ together with \eqref{PSV} to eliminate $H$ factors. As can be also realized from the Gaussian profile of normalization factors \eqref{Nform} of mode functions and \eqref{fitN}, we expect that only modes with $x_* \sim q^c \sim \mathcal{O}(1)\xi_*$ to contribute significantly to the energy density. In fact, we verified that the upper limit of the integral \eqref{rhoA} can be extended to $x^{\rm max}_* \to \infty$ as the integrand of eq. \eqref{rhoA} rapidly decays outside the $x_* \sim q^c \sim \mathcal{O}(1)\xi_*$ region\footnote{We found that the choice $x^{\rm max}_* = 3\xi_*$ provides an accurate estimate for the total gauge field energy density around its peak although it makes small error at very late times especially for the smaller values of $\xi_*$ shown in the right panel of Figure \ref{fig:rhoA}. We note that to derive the back-reaction limits, we will instead use eq. \eqref{rhoAf} which assumes $x^{\rm max}_* \to \infty$.}. Proceeding in this way, in the left panel of Figure \ref{fig:rhoA}, we present the physical gauge field energy density as a function of $\tau/\tau_*$. The peak in total energy density $\tau/\tau_* = \mathcal{O}(0.01)$ and its decay as $\tau/\tau_* \to 0$ by the expansion of the universe can be clearly seen. At its maximum value, we studied $\xi_*$ dependence of $\rho_A / \epsilon_\phi \rho_\phi$ and found that it can be described very well by the following expression,
\beq\label{rhoAf}
\fr{\rho_{A,*}}{\epsilon_\phi \rho_\phi} \approx 3.4 \times 10^{-13} \,\,e^{1.54\pi \xi_*},\quad\quad\quad \delta = 0.3.
\eeq
Notice that \eqref{rhoAf} and \eqref{rpeak} nearly have the same $\xi_*$ dependence. This is expected as the main source of GWs emission is the energy density contained in the gauge field sector $\rho_A$.
\subsection{Perturbativity}\label{Spert}  The success of the CMB and sub-CMB phenomenology we presented in Sections \ref{S4p1p1} and \ref{S4p2} demands that tensor modes generated by the decay of gauge fields (\ie one loop computation we performed in Appendix \ref{AppB}) to be larger than the standard tree level expression arise from the vacuum fluctuations of the metric. The sourcing of such large tensor fluctuations require a sizeable amplitude (energy density) in the gauge field mode functions, \ie at the times/scales when the observable effects are produced. As a consequence, one may wonder if large amplitudes attained by the vector field mode functions can drive the system out of the perturbative regime which we base our analysis so far.  In what follows, our aim is therefore to establish the regime for which the results we derived in sections \ref{S4p1p1} and \ref{S4p2} are under perturbative control. In this context, we consider two main requirements that our model should fulfill \cite{Ferreira:2015omg,Peloso:2016gqs}: i) higher order loop effects do not spoil the leading order estimate in \eqref{wkbsol} for the gauge field modes amplified by the rolling spectator $\sgm$, ii) the fluctuations of $\sgm$ field do not induce a variance $\sqrt{\langle\delta \sgm^2\rangle}$ that is greater than the typical classical field excursion $\sgm_{\rm cl} = \mathcal{O}(1-10)\, f$ of the spectator axion. 

The former criterion is related to the renormalization of the gauge field wave function and is given by \cite{Peloso:2016gqs}
\beq\label{PA}
P_A \equiv \Bigg| \fr{\delta^{(1)}\big\langle \,\hat{A}_{-}(\tau, \vec{k})\hat{A}_{-}(\tau,\vec{k}') \,\big\rangle'}{\big\langle\,\hat{A}_{-}(\tau, \vec{k})\hat{A}_{-}(\tau, \vec{k}')\,\big\rangle'}\Bigg| \ll 1, 
\eeq
where $\hat{A}_{-} = A_\lambda(\tau,\vec{k})\hat{a}_\lambda(\vec{k}) + A^{*}_\lambda(\tau,-\vec{k})\hat{a}^{\dagger}_\lambda(-\vec{k})$ and the expression in the denominator is the tree level propagator without the corresponding $\delta$-function: $\big\langle\,\hat{A}_{-}(\tau, \vec{k})\hat{A}_{-}(\tau, \vec{k}')\,\big\rangle'  = A^{R}_{-}(\tau, \vec{k})^2 + A^{I}_{-}(\tau, \vec{k})^2$. The numerator in \eqref{PA} is the leading order loop contribution to the gauge field propagator which can be computed via the in-in formalism as \cite{Ferreira:2015omg,Peloso:2016gqs}, 
\beq\label{2ptPA}
\delta^{(1)}\langle \,\hat{A}_{-}(\tau, \vec{k})\hat{A}_{-}(\tau,\vec{k}') \,\big\rangle \simeq i^2 \int^{\tau} d \tau' \int^{\tau'} d \tau''\left\langle\left[\left[\hat{A}_{-}(\tau, \vec{k}) \hat{A}_{-}(\tau,\vec{k}'), \hat{H}_{\mathrm{int}}\left(\tau'\right)\right], \hat{H}_{\mathrm{int}}\left(\tau''\right)\right]\right\rangle,
\eeq
where
\beq\label{Hint}
\hat{H}_{\rm int}(\tau) =\fr{\alpha_c}{f} \int \d^3 x \,\, \delta\hat{\sgm}\,  \epsilon_{ijk}\, \hat{A}'_i \,\partial_j \hat{A}_k.
\eeq
On the other hand, the second criterion ii) stands to ensure that the interaction in \eqref{Hint} does not drive the amplitude of the axion perturbation $\delta \sgm$ to the non-linear regime, \ie we need to oblige that 
\beq\label{Ps}
P_\sgm \equiv \fr{\sqrt{\big\langle \delta \hat{\sgm}^{(1)}(\tau,\vec{x}) \delta \hat{\sgm}^{(1)}(\tau,\vec{x})\big\rangle}}{\sgm_{\rm cl}} = \fr{\sqrt{\int \d \ln k\, \mathcal{P}^{(1)}_{\sgm}(\tau, k)}}{\sgm_{\rm cl}} \ll 1,
\eeq
where we described the numerator as an integral of the leading order loop contribution to the axion's power spectrum $2\pi^2 \mathcal{P}^{(1)}_{\sgm}(\tau, k)/k^3 = \langle \delta \hat{\sgm}^{(1)} (\tau, \vec{k}) \delta\hat{\sgm}^{(1)} (\tau, -\vec{k})\rangle'$. Using the in-in formalism, the leading order loop correction to the 2-pt function of the axion fluctuations can be computed via
\beq\label{2ptPs}
\left\langle\delta \hat{\sgm}^{(1)}(\tau,\vec{k}) \delta \hat{\sgm}^{(1)}(\tau, \vec{k}')\right\rangle \simeq i^2\int^{\tau} d \tau' \int^{\tau'} d \tau''\left\langle\left[\left[\delta \hat{\sgm}(\tau,\vec{k}) \delta \hat{\sgm}(\tau, \vec{k}'), \hat{H}_{\rm int}(\tau')\right], \hat{H}_{\rm int}(\tau'')\right]\right\rangle,
\eeq
where $\delta \hat{\sgm} = \delta \sgm(\tau, \vec{k})\hat{b}(\vec{k}) +  \delta \sgm^{*}(\tau, \vec{k})\hat{b}^{\dagger}(\vec{k})$ with $\left[\hat{b}(\vec{k}),\hat{b}^\dagger(\vec{k}')\right] = \,\, \delta(\vec{k}-\vec{k}')$. Using \eqref{Hint}, perturbativity criterions in \eqref{PA} and \eqref{Ps} can be evaluated by carrying out explicitly the commutators and expectation values in \eqref{2ptPA} and \eqref{2ptPs}. This procedure generically leads to an integrand composed of many terms including the real and imaginary part of the gauge fields among which  the terms that have the highest possible power of $A^{R}_{-}$ dominates as it describes the late time growing part of the gauge field fluctuations. In this way, the expression on the left hand side of \eqref{PA} can be re-written as \cite{Peloso:2016gqs},
\begin{align}\label{PAF}
\nn \fr{\epsilon_{\sgm,*}}{\epsilon_\phi} P_A(\xi_*,\delta,x_*,x) &\simeq 16\pi^2 \mathcal{P}^{(v)}_{\mathcal{R}} \xi_{*}^{2}  \int \frac{d^{3} \tilde{p}}{(2 \pi)^{3}} \frac{\tilde{p}^{1 / 2}\left[1-\hat{k} \cdot \hat{\tilde{p}}\right]^{2}}{2\left|\hat{k}+\vec{\tilde{p}}\right|^{3}} \int_{x}^{x_{*}} d x' \int_{x'}^{x_{*}} d x'' \,\tilde{A}^{R}_{-}\left(\tilde{p}x'\right)\tilde{A}^{R}_{-}\left(\tilde{p} x''\right)\\\nn
& \quad\quad \quad\quad\quad \times \left[(\tilde{p}-1) \tilde{A}^{I}_{-}\left(x'\right)\tilde{A}^{R}_{-}\left(x''\right)+(1+\sqrt{\tilde{p}})^{2}\tilde{A}^{R}_{-}\left(x'\right)\tilde{A}^{R}_{-}\left(x''\right) \frac{ \tilde{A}^{I}_{-}(x)}{\tilde{A}^{R}_{-}(x)}\right]  \\
&\quad \quad\quad\quad\quad\times  {\rm Im}\left[\delta \tilde{\sigma}\left(|\hat{k}+\vec{\tilde{p}}| x'\right) \delta \tilde{\sigma}^{*}\left(|\hat{k}+\vec{\tilde{p}}| x''\right)\right],
\end{align}
where $\mathcal{P}^{(v)}_{\mathcal{R}}\simeq 2.1 \times 10^{-9}$, $- k \tau = x$ and we defined dimensionless mode functions of the gauge and axion fluctuations\footnote{For the evaluation of the perturbativity conditions in eq. \eqref{PAF} and \eqref{Psf}, we ignore the scale dependence that $\delta \sgm$ might posses due to $\mathcal{O}(\eta_\sgm)$ corrections in its effective mass, see \eg eq. \eqref{mm}. As far as \eqref{PAF} and \eqref{Psf} are concerned, this approximation is justified as $\eta_\sgm$ quickly vanishes at late times for $x > x_*$ proportional to $\Delta N^{-1}$. In other words, in the perturbativity calculations, we adopt the standard mode functions of a spectator scalar in a dS: $\delta\tilde{\sgm}(x) = i (1-ix) e^{ix} / \sqrt{2}$ (See eq. \eqref{dmfpert}). } as
\begin{align}\label{dmfpert}
\nn \tilde{A}^{R/I}_{-}(-k\tau) &= \sqrt{2k} \left(\fr{2\xi(\tau)}{-k\tau}\right)^{1/4} A^{R/I}_{-} (\tau,\vec{k}),\\
\delta \tilde{\sigma}(-k\tau) &\equiv \frac{k^{3 / 2}}{H} \delta \sigma\left(\tau, \vec{k}\right).
\end{align}
In \eqref{PAF}, time dependence of $P_A$ can be reformulated noting $x = x_* (\tau/\tau_*)$ to describe it as a function of the variable $y = \tau/\tau_*$, \ie $P_A = P_A(\xi_*,\delta,x_*,y)$. 

Similarly, the leading order loop contribution to the power spectrum of axion fluctuations can be described as
\begin{align}\label{Psf}
\nn \mathcal{P}_{\sigma}^{(1)}(\xi_*, \delta, x_*,y) &\simeq \frac{\alpha^{2} H^{4}}{64 \pi^{2} f^{2}} \int \frac{d^{3} \tilde{p}}{(2 \pi)^{3}}  \frac{\left(\tilde{p}^{1 / 2}+|\hat{k}-\vec{\tilde{p}}|^{1 / 2}\right)^{2}\left[(\tilde{p}+|\hat{k}-\vec{\tilde{p}}|)^{2}-1\right]^{2}}{\tilde{p}^{3 / 2}|\hat{k}-\vec{\tilde{p}}|^{3 / 2}} \\
&\quad\quad\quad\quad\times\left[\int_{x}^{x_{*}} d x' \, {\rm Im}\left[\delta \tilde{\sigma}(x) \,\delta \tilde{\sigma}^{*}\left(x'\right)\right] \tilde{A}^{R}_{-}\left(\tilde{p} x'\right) \tilde{A}^{R}_{-}\left(|\hat{k}-\vec{\tilde{p}}| x'\right)\right]^{2},
\end{align}
\begin{figure}[t!]
\begin{center}
\includegraphics[scale=0.88]{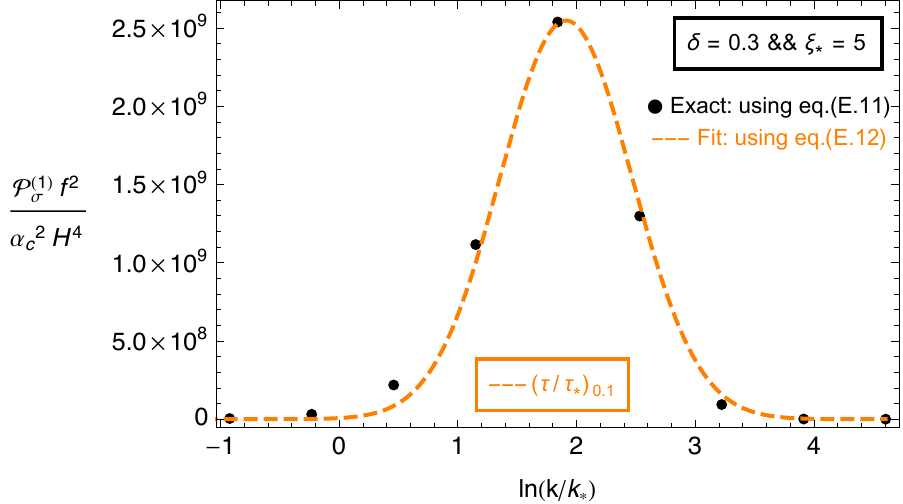}\includegraphics[scale=0.88]{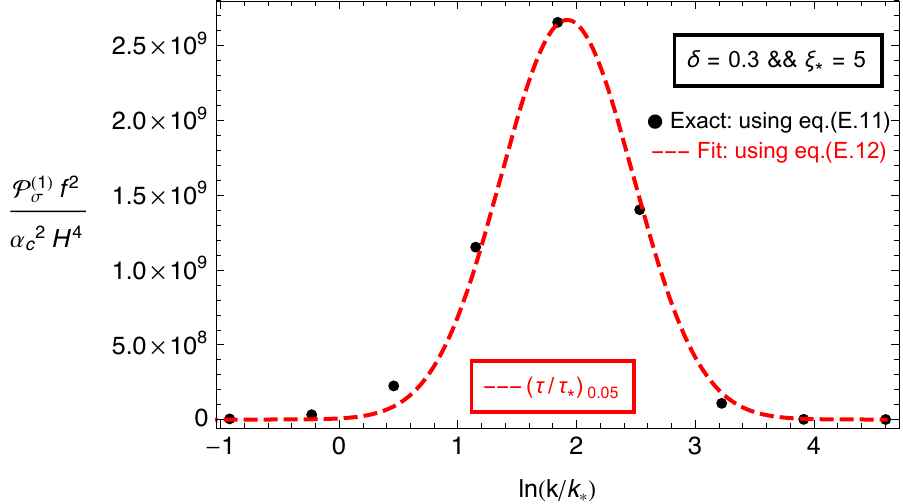}
\end{center}
\caption{ The power spectrum of axion fluctuations sourced by the gauge field evaluated at two different choices of time $y = \tau/\tau_*$, corresponding to the moments at which total energy density in the gauge field sector has decreased to $10\%$ (Left) and $5\%$ (Right) than the value it had at its peak. Between two panels, the slight change at the peak values of the power spectrum can be barely seen, signifying the threshold amplitude that $\mathcal{P}^{(1)}_{\sgm}$ obtains at late times $y \ll 1$.\label{fig:pspert}}
\end{figure} 
where apart from the scale ($x_*$) and time dependence ($y$), the functional dependence of $\mathcal{P}_{\sigma}^{(1)}$ on the background motion of the spectator is described by $\xi_*$ and $\delta$ as usual. Numerically evaluating \eqref{Psf} for different $x_*$ and at fixed $\xi_*$, $\delta$ and $y$, we found that its scale dependence is well fitted by a log-normal shape,
\beq\label{Psfit}
\mathcal{P}_{\sigma}^{(1)}\left(\xi_*, \delta, x_*,y\right) \simeq \frac{\alpha^{2} H^{4}}{f^{2}} \mathcal{A}_{\sgm}\left[\xi_{*}, \delta, y\right] \exp \left(-\frac{1}{2 \sigma^{2}_{\mathcal{A}_\sgm}\left[\xi_{*}, \delta,y\right]} \ln ^{2}\left(\frac{x_*}{x^{c}_{\mathcal{A}_\sgm}\left[\xi_{*}, \delta,y\right]}\right)\right).
\eeq
The explicit time dependence ($y$) of the amplitude $\mathcal{A}_\sgm$, the width of the peak $ \sigma_{\mathcal{A}_\sgm}$ and its location $x^{c}_{\mathcal{A}_\sgm}$ stems from the fact that the power spectrum in \eqref{Psfit} is sourced by the gauge fields which are clearly time dependent. Since the physical amplification of the real part of the gauge fields occur for $y < 1$, we are only interested in this regime. In particular we found that the scale dependent growth of the power spectrum eventually saturates to a large amplitude for $y \ll 1$.  To illustrate these facts and the accuracy of the expression  \eqref{Psfit} in describing the power spectrum in \eqref{Psf}, we plot  $P_{\sigma}^{(1)}$ for a fixed $\delta$ and $\xi_*$ for two different $y$ values corresponding to the times where the total energy density of the gauge field $\rho_A$ in \eqref{rhoA} reduces to $10 \%$ ($y_{0.1}$) and $5\%$ ($y_{0.05}$) of its value at the peak. Using the accurate expression in \eqref{Psfit}, we can then evaluate the integral in \eqref{Ps} analytically to re-write this expression as \cite{Peloso:2016gqs},
\beq\label{PsFin}
 \fr{\epsilon_{\sgm,*}}{\epsilon_\phi} P_\sgm(\xi_*, \delta, y) \simeq  5.25\times 10^{-7} \, \delta \xi_* \sqrt{\mathcal{A}_{\sgm}\left[\xi_{*}, \delta,y\right]\,  \sigma_{\mathcal{A}_{\sgm}}\left[\xi_{*}, \delta,y\right] }.
\eeq
Notice that contrary to \eqref{PAF},  the expression in \eqref{PsFin} is mode independent as it arise as a result of integration over modes through eq. \eqref{Ps}.
\begin{figure}[t!]
\begin{center}
\includegraphics[scale=0.90]{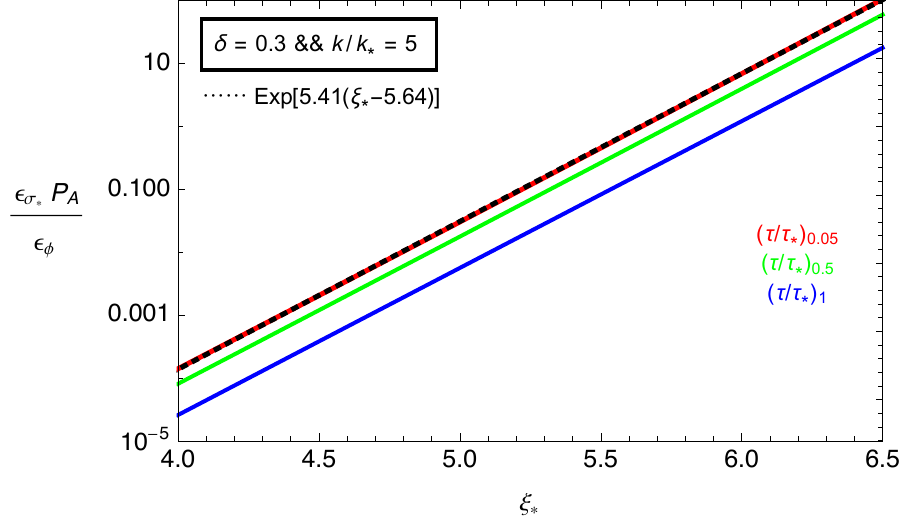}\includegraphics[scale=0.90]{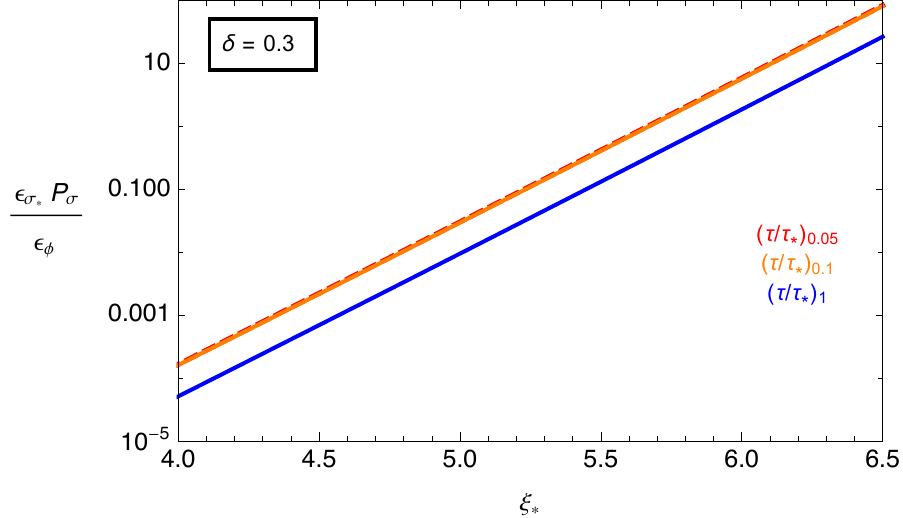}
\end{center}
\caption{ The expressions in eq.  \eqref{PAF} (left) and \eqref{Psf} (right) in terms of the effective coupling constant $\xi_*$ that controls the gauge field amplification. The different colored lines indicate different times at which these expressions are evaluated. \label{fig:pavsps}}
\end{figure} 

\noindent{\bf Summary of perturbativity constraints:} In Figure \ref{fig:pavsps}, we present the two expressions in eq. \eqref{PAF} (left panel) and \eqref{Psf} (right panel) as a function of the effective coupling $\xi_*$ at different times corresponding to the moments at which the energy density of the mode (left)/gauge field sector (right) reduces to a certain percentage of the its value at its peak due to the expansion of the universe. It is enough to study these conditions within these time limits as they converge to their maximal value at $(\tau/\tau_*)_{0.05}$. This is particularly clear in the right panel as $(\tau/\tau_*)_{0.1}$ line and  $(\tau/\tau_*)_{0.05}$ appear to be nearly superimposed on each other. 	Therefore, focusing on the maximal value among the lines shown, we found that their $\xi_*$ dependence can be well fitted by the following expressions
\beq\label{sumper}
\delta = 0.3 : \quad \fr{\epsilon_{\sgm,*}}{\epsilon_\phi} P_A \simeq e^{5.41(\xi_* - 5.64)},\quad\quad \fr{\epsilon_{\sgm,*}}{\epsilon_\phi} P_\sgm \simeq e^{5.23(\xi_* - 5.66)}.
\eeq
We find that in the parameter space that leads to interesting phenomenological results in our model, \ie $\xi_* \gtrsim 5$, the first expression is greater than the second in eq. \eqref{sumper}. Recalling, $\epsilon_{\sgm,*} = 2\delta^2 (f/\Mp)^2$, the strongest perturbativity condition $P_A \ll 1$ can be expressed as a lower bound on the axion decay constant $f$ as:
\beq\label{sumperf}
\boxed{5.6 \times 10^{-7} \sqrt{\epsilon_\phi}\,\, e^{2.71 \xi_*} < \fr{f}{\Mp},}
\eeq
where we replaced $\ll$ sign with $<$ due to exponential sensitivity to the parameter $\xi_*$.
\subsection{Back-reaction, spectral tilt and its running}\label{Sback}
In this subsection, we will i) study constraints on the parameter space of the model from back-reaction of the produced gauge quanta on the background dynamics (applicable to the both scenarios presented in Sections \ref{S4p1p1} and \ref{S4p2}) ii) investigate restrictions that might be imposed on the model from the scalar spectral tilt $n_s$ and its running $\alpha_s$ at CMB scales. 

\noindent{\bf An upper bound on $f/\Mp$:} We first need to make sure that $\sgm$ contributes negligible amount to the energy budget during inflation. To quantify this condition, we note the maximum value acquired by the slow-roll parameter $\epsilon_{\sgm,*} = 2 \delta^2 (f/\Mp)^2$ at $\tau = \tau_*$ where $\dot{\sgm}$ reaches its maximal value $\dot{\sgm}_*$ on the cliff like regions of its potential. Then plugging the field profile \eqref{san} in the potential \eqref{Vs}, at $\tau = \tau_*$, we obtain 
\beq\label{potvskin}
V_\sgm(\sgm_*) \simeq 3H^2\Mp^2\, \fr{\epsilon_{\sgm,*}}{\delta}\, \fr{f(n)}{2},\quad\quad \fr{\dot{\sgm}^2_*}{2} = 3 H^2 \Mp^2\, \fr{\epsilon_{\sgm,*}}{3},
\eeq
where $f(n) = 1+(n+1/2)\pi$ with $n=0,2 \dots$\,. In \eqref{potvskin}, the index $n$ can be seen as an indicator of the initial conditions one undertakes for $\sigma$, \ie for larger $n$ (and hence for larger $f(n)$) spectator axion starts its evolution higher up in its scalar potential and thus exhibits more potential energy. Noting this aside, we see from eq. \eqref{potvskin} that the potential energy always dominates over the kinetic energy of $\sigma$ for $\delta < 1$. Therefore the condition that $\sgm$ contributes negligibly to the total energy density during inflation, \ie $\rho_\sgm \ll 3H^2\Mp^2$ imposes an upper limit on $f/\Mp$ that depends on initial conditions of $\sgm$:
\beq\label{c1}
\epsilon_{\sgm,*} \ll \fr{2\delta}{f(n)} \quad \longrightarrow \quad \fr{f}{\Mp} < \fr{1}{\sqrt{\delta f(n)}}.
\eeq
For the sub-CMB scenarios we considered in Section \ref{S4p2}, the field excursion is as large as $\Delta\sgm/f \simeq \mathcal{O}(10)$ and $f(n=2)\simeq 8.9$ implying $f / \Mp \lesssim 0.6$ for $\delta =0.3$ whereas for the scenario we considered in Section \ref{S4p1p1}, $\sgm$ traverses a single cliff while the scales associated with CMB observations exit the horizon and hence probes the smallest possible distance in field space. In this case, we set $n =  0$ in eq. \eqref{c1}, implying a trivial condition $f /\Mp \lesssim 1$ for $\delta = 0.3$. To obtain a more restrictive upper bound, we consider the spectral tilt of the vacuum scalar power spectrum at CMB scales: $|n_s -1| \simeq 2\eta_\phi - 6\epsilon_\phi-4\epsilon_{\sgm} \sim 10^{-2}$. Assuming axion's velocity peaks (and hence the sourced GW signal) at around the CMB pivot scale, we may require $\epsilon_{\sgm,*} \ll 10^{-2}$ to avoid fine tuning through accidental cancellations between the terms appearing in $|n_s -1|$. We stress that one can not derive a tighter constraint on the ratio $f/\Mp$ considering the evolution of $\epsilon_\sigma$ away from its peak value as it quickly reduces to smaller values away from $\tau = \tau_*$. 
\begin{figure}[t!]
\begin{center}
\includegraphics[scale=0.84]{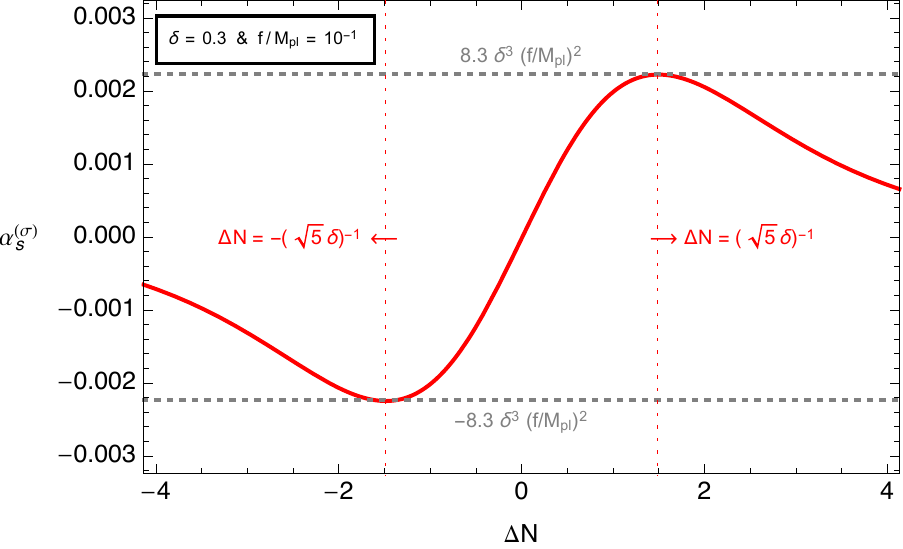}\includegraphics[scale=0.87]{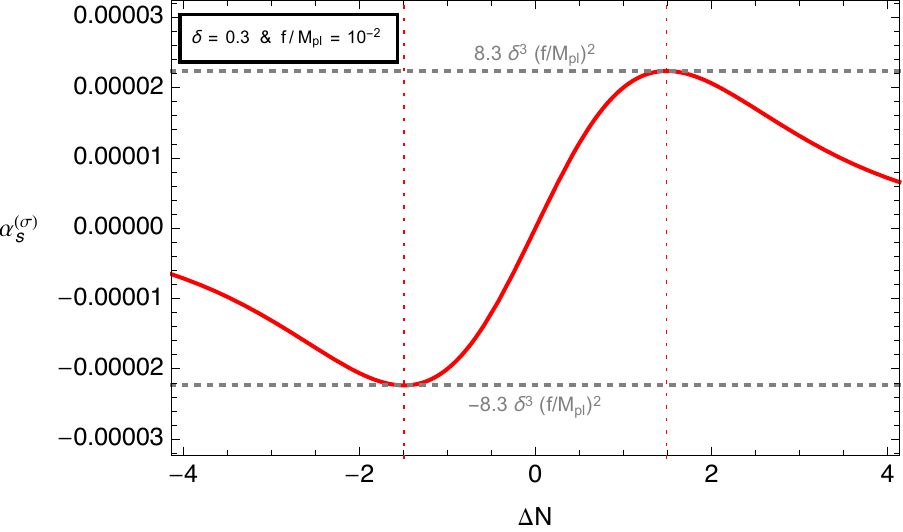}
\end{center}
\caption{Time evolution of $\alpha_s^{\sigma}$ in eq. \eqref{alphasgm} while $\sigma$ probes the step-like feature in its potential. In both plots $\epsilon_\phi = 10^{-4}$ is assumed. \label{fig:alphas}}
\end{figure} 

\noindent\underline{\emph{The running of the spectral index}}: The non-trivial structure of axion's wiggly potential may also influence the running of the spectral index $\alpha_s$ within an observable range of CMB scales\footnote{In the context of canonical single field inflation sizable modulations in the axion potential may also influence $n_s$ significantly  \cite{Kobayashi:2010pz}.}. In particular, $\alpha_s$ is sensitive to the higher derivates of the scalar potential \eqref{Vs} which may be large in the bumpy $\beta \lesssim 1$ regime we are operating. In what follows, we investigate the limitations that might arise on the parameter space of the model from the CMB constraints on the running of the spectral index. We begin by splitting $\alpha_s \equiv \d n_s / d\ln k = \alpha_s^{\phi} + \alpha_s^{\sigma}$ where $\alpha_s^{\phi}$ is the running due to the inflaton sector and is second order in slow-roll hierarchy \cite{Easther:2006tv}. Therefore for a smooth, flat enough inflaton potential we are assuming it is negligible compared the running caused by the rolling $\sigma$:
\beq\label{alphasgm}
 \alpha_s^{\sigma} 
 \equiv -4 \fr{\d \epsilon_{\sigma}}{\d \ln k} \simeq 8\epsilon_{\sigma} \left( \eta_\sigma - \epsilon_\phi- \epsilon_\sigma\right),
\eeq
where $\eta_\sigma = \Mp^2 V''_{\sigma}/V$.  Using the field profile \eqref{san} and the definition $\epsilon_\sigma = \dot{\sigma}^2/(2H^2\Mp^2)$, in Figure \ref{fig:alphas}, we present the time evolution (left to right) of $\alpha_s^{\sigma}$ while $\sigma$ rolls through a step-like feature in its potential for two different choices of the ratio $f/\Mp$ and for $\delta=0.3$. We observe that in the region outside the red vertical dotted lines denoted by $\Delta N = \mp (\sqrt{5}\delta)^{-1}$, as $\sigma$ probes flat plateau regions of the step-like feature, $\alpha_s^{\sigma}$ tends to be vanishing at very early and late times whereas inside this region it evolves from negative to positive values when $\sigma$ probes the cliff-like region on its potential. In particular, for $-(\sqrt{5}\delta)^{-1} \leq \Delta N \leq (\sqrt{5}\delta)^{-1}$, $\alpha_s^{\sigma}$ evolves linearly $\Delta N$  from its maximally negative to positive value and therefore vanishes at $N=N_*$ where axion acquires its maximal velocity. In Figure \ref{fig:alphas}, the envelopes shown by the horizontal line are good indicators of the maximal values that running can obtain. In terms of model parameters, we found that these maximal values are reached at $\Delta N = \mp (\sqrt{5}\delta)^{-1}$ and are given by $\alpha_s^{\sigma} \simeq \mp 8.3\, \delta^3\, (f/\Mp)^2$, indicating their sensitivity on the ratio $f/\Mp$. Considering $2\sigma$ limits on the running from the Planck 2018 data \cite{Akrami:2018odb}, we have $-0.0206 < \alpha_s < 0.0074$ ($\alpha_s = -0.0066 \pm 0.0070$ at $68\, \%\,{\rm CL}$) which implies that imposing the upper bound $f/\Mp < 0.18$ allow us to keep these envelopes ($\alpha_s^{\sigma} \simeq \mp 8.3\, \delta^3\, (f/\Mp)^2$) within $2\sigma$ observational limits. The magnitude of $\alpha_s^{\sigma}$ shown in the right panel of Figure \ref{fig:alphas} clearly confirms this result where we adopted $f/\Mp = 10^{-2}$. Therefore, the running induced by the spectator axion can be kept below the current observational limits at CMB scales as far as we assume $f/\Mp < 0.18$ which is parametrically very close to the upper bound we found in eq. \eqref{LBf} from the considerations on the spectral index. Summarizing our findings above, for the two scenarios we consider in Sections \ref{S4p1p1} and \ref{S4p2}, we can impose the following conservative upper bounds on the axion decay constant
\beq\label{LBf}
\boxed{\fr{f}{\Mp} \lesssim 0.6\,\,\, @\mathrm{Interferometer\, scales}, \quad\quad \fr{f}{\Mp} \lesssim 0.18\,\,\, @\mathrm{CMB\,scales}\,.}
\eeq
{\bf A lower bound on $f/\Mp$:} Next, we need to make sure that the gauge field amplification does not significantly alter the motion of $\sgm$. For this purpose we consider the equation of motion of $\sgm$:
$$\ddot{\sgm} + 3H\dot{\sgm} + V'_{\sgm}(\sgm) = \alpha_c \langle \vec{E} \cdot \vec{B} \rangle /f,$$
and impose $  \alpha_c \langle \vec{E} \cdot \vec{B} \rangle /f \ll 3 H \dot{\sgm}$. Notice that $|\vec{E}|/|\vec{B}| \simeq \sqrt{\xi/x} \sim \xi$ (see \eg \eqref{EBF}),  where we have used $x \sim \xi^{-1}$ for an optimal estimate on the latter ratio since for modes that satisfy $x \gg \xi^{-1} \sim \mathcal{O}(10^{-1})$, amplitude of mode functions is suppressed further (see eq. \eqref{tA} and  Figure \ref{fig:rhoA}). Therefore, the second backreaction condition can be re-written as
\beq\label{c2}
\fr{\alpha_c \langle \vec{E} \cdot \vec{B} \rangle}{f} \ll 3 H \dot{\sgm} \quad\quad \longrightarrow\quad\quad \rho_ A \ll 3\, \fr{\dot{\sgm}^2}{2},
\eeq
where we used the fact that $\vec{E}$ fields contribute dominantly to the energy density of the gauge fields in \eqref{I}. In light of the expression \eqref{c2}, at the maximum of gauge field energy density, a simpler conservative criterion is therefore given by $\rho_ {A,*} \ll  \dot{\sgm}_*^2/{2} \simeq \epsilon_{\sgm,*} \,{\rho_{\phi}}/{3}$ where we made use of \eqref{potvskin}. Recalling the result \eqref{rhoAf} we derived earlier, we can derive an upper on $f$ as
\beq\label{BRC}
\boxed{2.4 \times 10^{-6} \sqrt{\epsilon_\phi} \,e^{2.42\,\xi_*}<\fr{f}{\Mp}. }
\eeq
\subsection{Summary of perturbativity and back-reaction limits}
Comparing the perturbativity constraint in \eqref{sumperf} with the one derived from back-reaction consideration in \eqref{BRC}, we found that the former is more restrictive at fixed $\epsilon_\phi$ for $\xi_* \gtrsim 5$ corresponding to the parameter space that leads to the phenomenology we presented in Section \ref{S4p1p1} and \ref{S4p2}. Therefore, combining the lower bound in \eqref{sumperf} with the upper bounds we derived in \eqref{LBf}, we arrive at
\beq\label{SoC}
\boxed{
5.6 \times 10^{-7} \sqrt{\epsilon_\phi}\,\, e^{2.71 \xi_*} < \fr{f}{\Mp} \lesssim \{0.18, 0.6\},}
\eeq
where the upper bound changes depending on the scales we are considering as in \eqref{LBf}.
Finally, for sources that peaks at CMB scales (interferometer scales), we can use $r_*^{1/2} \simeq 2.8 \times 10^{-8} \epsilon_\phi\, e^{4.955\,\xi_*}$ (\eqref{Pgw} and \eqref{Omgw}) to eliminate $\epsilon_\phi$ in terms of $r_*$ ($\Omega_{\rm GW,*}\, h^2$) to re-write these limits as
\begin{align}\label{BRF}
\nn 0.0017 \left(\fr{r_*}{0.063}\right)^{1/4} e^{0.23\, \xi_*} &< \fr{f}{\Mp} \lesssim 0.18, \quad\quad @\mathrm{CMB\,scales},\\
0.07 \left(\fr{\Omega_{\rm GW}\, h^2}{10^{-9}}\right)_*^{1/4} e^{0.23\, \xi_*} &<\fr{f}{\Mp} \lesssim 0.6, \quad\quad@\mathrm{Interferometer\,scales}.
\end{align}
Considering $\xi_* = \mathcal{O}(5-6)$ we adopt in this work, from \eqref{BRF} one can verify that there is a sizeable portion of parameter space available (in terms of $f/\Mp$) in which limits from perturbativity considerations (Section \ref{Spert}) and back-reaction (Section \ref{Sback}) are satisfied while gauge field sources amplified by the transient roll of axion can produce observable GWs at CMB and sub-CMB scales as we show in Sections \ref{S4p1p1} and \ref{S4p2}.
\end{appendix}

\addcontentsline{toc}{section}{References}
\bibliographystyle{utphys}
\bibliography{paper2}
\end{document}